\documentclass[10pt,twocolumn,letterpaper]{article}

\usepackage[pagenumbers]{cvpr} % To force page numbers, e.g. for an arXiv version

\usepackage[dvipsnames]{xcolor}

\usepackage{caption}
\usepackage{algorithm}
\usepackage{algorithmic}
\usepackage{xcolor}
\usepackage{bm}
\usepackage{amsmath}
\usepackage{multirow}
\usepackage{booktabs}
\usepackage{float}
\usepackage{amssymb}
\usepackage{subcaption}

\usepackage{color, colortbl}
\definecolor{Gray}{gray}{0.9}

\usepackage{soul}
\soulregister{\cite}7 % 注册\cite命令
\soulregister{\citep}7 % 注册\citep命令
\soulregister{\citet}7 % 注册\citet命令
\soulregister{\ref}7 % 注册\ref命令
\soulregister{\pageref}7 % 注册\pageref命令

\definecolor{cvprblue}{rgb}{0.21,0.49,0.74}
\usepackage[pagebackref,breaklinks,colorlinks,citecolor=cvprblue]{hyperref}

\def\paperID{14093} % *** Enter the Paper ID here
\def\confName{CVPR}
\def\confYear{2024}

\title{DFDG: Data-Free Dual-Generator Adversarial Distillation for One-Shot Federated Learning}

\author{Kangyang Luo$^{1}$, Shuai Wang$^{1}$, Yexuan Fu$^{1}$, Renrong Shao$^{3}$, \\ Xiang Li$^{1}$\thanks{Corresponding author}, Yunshi Lan$^{1}$, Ming Gao$^{1}$, Jinlong Shu$^{2}$\\
{\tt\small East China Normal University$^{1}$, Shanghai, China}\\
{\tt\small Shanghai Normal University$^{2}$, Shanghai, China}\\
{\tt\small Naval Medical University$^{3}$, Shanghai, China}\\
{\tt\small \{52205901003, 51215903058, 51215903042\}@stu.ecnu.edu.cn,} \\ 
{\tt\small roryshaw6613@smmu.edu.cn,} \\ 
{\tt\small \{xiangli, yslan, mgao\}@dase.ecnu.edu.cn,}\\
{\tt\small jlshu@shnu.edu.cn}
}

\begin{document}
\maketitle
\begin{abstract}
Federated Learning~(FL) is a distributed machine learning scheme in which clients jointly participate in the collaborative training of a global model by sharing model information rather than their private datasets. 
In light of concerns associated with communication and privacy, one-shot FL with a single communication round has emerged as a de facto promising solution.
However, existing one-shot FL methods either require public datasets, focus on model homogeneous settings, or distill limited knowledge from local models, making it difficult or even impractical to train a robust global model.
To address these limitations, we propose a new data-free dual-generator adversarial distillation method~(namely DFDG) for one-shot FL, which can explore a broader local models' training space via training dual generators.
DFDG is executed in an adversarial manner and comprises two parts: \textit{dual-generator training} and \textit{dual-model distillation}.
In \textit{dual-generator training}, we delve into each generator concerning fidelity, transferability and diversity to ensure its utility, and additionally tailor the cross-divergence loss to lessen the overlap of dual generators' output spaces.
In \textit{dual-model distillation}, the trained dual generators work together to provide the training data for updates of the global model.
At last, our extensive experiments on various image classification tasks show that DFDG achieves significant performance gains in accuracy compared to SOTA baselines. 
We provide our code here: \url{https://anonymous.4open.science/r/DFDG-7BDB}.
% $\mathcal{F}$
\end{abstract}

\section{Introduction}
\label{sec:intro}

Federated Learning~(FL)~\cite{McMahan2017Communication} is a privacy-preserving distributed machine learning scheme that enables multiple clients to collaboratively train a global model without compromising their private data, and thus has received extensive attention from academia and industry.
In recent years, FL has shown promise in many practical application domains, such as healthcare~\cite{jiang2022harmofl}, finance~\cite{yang2019ffd}, IoT~\cite{nguyen2021federated}, and autonomous driving~\cite{li2021privacy}, among others. %financial institutions
Despite its remarkable success, most existing FL methods suffer from 
high communication cost and high risk of being attacked. 
Specifically, FedAvg and its variants~\cite{McMahan2017Communication, karimireddy2020scaffold, Acar2021Federated, li2021model, luo2023gradma, luo2024dfrd, kim2022multi} require clients to frequently communicate with the central server to exchange local model parameters or their gradients. % require
The resulting high communication cost may be intolerable and impractical in real-world FL.
Furthermore, frequent communication has many known security weaknesses.
For example, frequent communication is at high risk of being intercepted by attackers, who can make eavesdropping or man-in-the-middle attacks~\cite{wang2020man, huang2021evaluating} or even reconstruct training data from gradients ~\cite{geiping2020inverting, wu2023learning}.
To get around the aforementioned issues in mainstream FL, an intuitively promising solution, one-shot FL~\cite{guha2019one}, which only allows a single communication round between clients and the server, came into being.

\begin{figure}[t]
    \centering
    \includegraphics[width=0.47\textwidth]{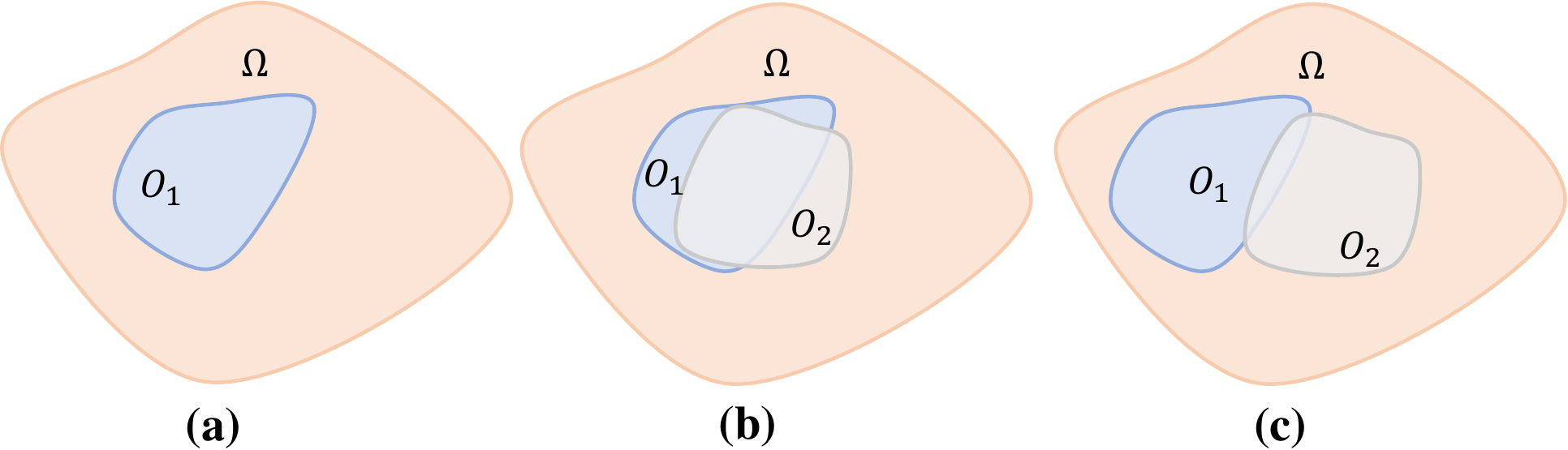} % Reduce the figure size so that it is slightly narrower than the column.
    \caption{A sketch of real data space~($\Omega$) as well as output spaces~($O_1$, $O_2$) of dual generators~($G_1$, $G_2$) that mimic $\Omega$ using data-free knowledge distillation.}
    \vspace*{-3ex}
    \label{dual_gene_analysis_pic:}
\end{figure}
% practical
A practical scenario for one-shot FL is model markets~\cite{vartak2016modeldb}, where pre-trained models purchased by users do not undergo iterative communication during training.
However, in some scenarios (such as healthcare and finance), the individual circumstances of data generation and computing resources vary greatly among different institutions~(i.e., clients). 
This results in diverse data distribution~(i.e., data heterogeneity)~\cite{li2020federated} and varying model capacity~(i.e., model heterogeneity)~\cite{diao2020heterofl} among distinct clients.
There have been a panoply of efforts in one-shot FL struggle with the above-mentioned scenarios.
For example, \cite{guha2019one, li2020practical, gong2021ensemble, gong2022preserving} utilize a public dataset for training to enhance the global model.
But the desired public data is not always available in practice and the performance may decrease dramatically if the apparent disparity in distributions exists between public data and clients' private data~\cite{zhu2021data, tan2022fedproto}. 
Moreover, these methods only engage in model homogeneous settings. 
Recently, DENSE~\cite{Zhang2022DENSE} and FedFTG~\footnote{For clarity, we only consider FedFTG with a single communication round in our work. Also, FedFTG can be directly applied to model heterogeneous scenarios.}~\cite{Zhang2022Fine} aggregate knowledge from local models with data-free knowledge distillation~(DFKD)~\cite{chen2019data,  yoo2019knowledge} to train a global model of one-shot FL, thereby bypassing public data and adapting to model heterogeneous setting.
Therefore, we focus on how to train a robust global model in one-shot FL with the help of DFKD. % bypassing / getting rid of
% of in this paper,

\begin{figure*}[t]
  \centering
  \includegraphics[scale=1.15]{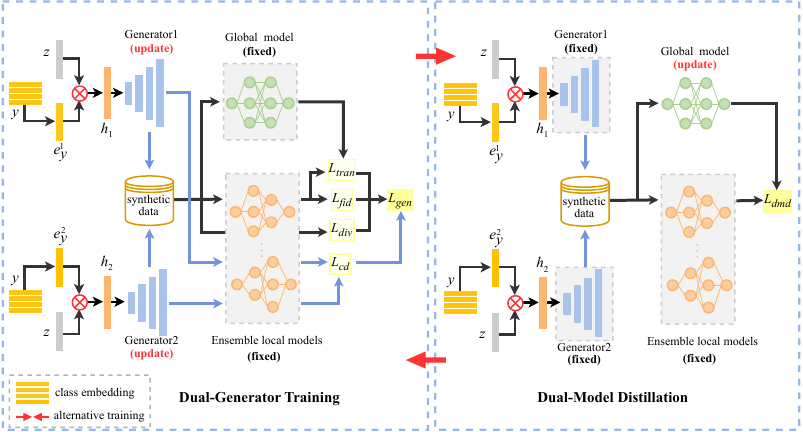}
  \caption{Illustration of DFDG. DFDG works on the server and contains two phases, {\it dual-generator training} and {\it dual-model distillation}, where $\mathcal{L}_{gen}$ and $\mathcal{L}_{dmd}$ are the loss objectives of the dual conditional generators and the global model, respectively.}
  \label{Framework:}
  \vspace*{-3ex}
\end{figure*}

Existing methods DENSE and FedFTG equip the server with a generator alone to approximate the training space of the local models, and train the generator and the global model in an adversarial manner.
We revisit them and find that the output space~($O_1$) of a single generator~($G_1$) can only cover a limited part of the real data space~($\Omega$)~\cite{beetham2022dual, do2022momentum}, as shown in Fig.~\ref{dual_gene_analysis_pic:}~(a).
Intuitively, we reckon that training dual generators~($G_1$, $G_2$) can expand the space covering~$\Omega$, i.e., $O_1 \subset O_1\cup O_2$~(see Fig.~\ref{dual_gene_analysis_pic:}~(b)).
But if the overlap of $O_1$ and $O_2$ is large, the gain is not noticeable.
As such, a key goal of this paper is to reduce the overlap of $O_1$ and $O_2$, thus enabling $O_1\cup O_2$ to cover a wider space in real data space (see Fig.~\ref{dual_gene_analysis_pic:}~(c)).
Besides, the quality of individual generators can greatly affect the extent that $O_1\cup O_2$ can cover. 
Thus, training well-performing individual generators is also a crucial goal of our work.

Built upon the said observations, we propose a new \underline{\textbf{D}}ata-\underline{\textbf{F}}ree \underline{\textbf{D}}ual-\underline{\textbf{G}}enerator adversarial distillation method~(termed as DFDG) for one-shot FL. %Based on 
% To be specific 
Concretely, DFDG equips the server with dual conditional generators to generate synthetic data after the server collects local models~(teachers) from clients.
% simulate
To effectively emulate the local models' training space, we investigate each generator in terms of fidelity, transferability, and diversity to ensure its utility. 
% Also, we carefully craft a cross-divergence loss to reduce the overlap of the dual generators' output space.
Also, we carefully craft a cross-divergence loss to guide the generators to explore different regions of the local models' training space, thereby reducing the overlap of the dual generators' output spaces.
The trained dual generators then work together to provide the training data for updates of the global model~(student). % to
Repeating the above processes alternately, DFDG trains the dual generators and the global model in an adversarial manner.
We overview the schematic of DFDG in Fig.~\ref{Framework:}.

To sum up, we highlight our contributions as follows:
\begin{itemize}
    \item  We bring forward a novel one-shot FL method, DFDG, which can explore a broader  local models' training space by training dual generators, thereby enabling a robust global model.
    To the best of our knowledge, we are the first effort in one-shot FL to attempt to train dual generators to emulate the local models' input space.
    
    \item To achieve well-trained dual generators, we look into each generator concerning fidelity, transferability and diversity to ensure its utility on the one hand, and on the other hand, we elaborate a cross-divergence loss to lessen the overlap of dual generators' output space. %space

    \item 
    % We conduct extensive experiments on four commonly used image classification datasets~(i.e., FMNIST, CIFAR-10, SVHN and CINIC-10) to show that DFDG is highly competitive compared with other state-of-the-art baselines.
    % Meanwhile, ablation studies demonstrate efficacy and indispensability for core modules and key parameters. 
    We conduct extensive experiments on multiple popular image classification datasets to show that DFDG is highly competitive compared with other state-of-the-art baselines. Ablation studies further validate core modules and key parameters' effectiveness.  %(i.e., FMNIST, CIFAR-10, SVHN, and CINIC-10)
\end{itemize}

\section{Related Work}
\label{sec:related_work}

\textbf{Data-Free Knowledge Distillation~(DFKD).} DFKD methods are promising, which transfer knowledge from the teacher model to another student model without any real data.
The generation of synthetic data that facilitates knowledge transfer is crucial in DFKD methods.
Existing DFKD methods can be broadly classified into non-adversarial and adversarial training methods.
Non-adversarial training methods
~\cite{chen2019data, yoo2019knowledge, luo2020large, nayak2019zero, wang2021data, luo2024privacy} exploit certain heuristics to search for synthetic data similar to the original training data.
For example, DAFL~\cite{chen2019data} and ZSKD~\cite{nayak2019zero} regard the predicted probabilities and confidences of the teacher model as heuristics.
% Adversarial-based training DFKD methods
Adversarial training methods
~\cite{fang2021contrastive, yin2020dreaming, do2022momentum} take advantage of adversarial learning to explore the distribution space of the original dataset.
They take the quality and/or diversity of the synthetic data as important objectives.
For example, CMI~\cite{fang2021contrastive} improves the diversity of synthetic data by leveraging inverse contrastive loss.
% DeepInversion~\cite{yin2020dreaming} augments synthetic data by regularizing the distribution of Batch Norm to ensure visual interpretability.
DeepInversion~\cite{yin2020dreaming} improves synthetic data with Batch Norm distribution regularization for visual interpretability.
% MAD~\cite{do2022momentum} mitigates the generator's forgetting of previous knowledge by maintaining an exponential moving average copy of the generator.
MAD~\cite{do2022momentum} prevents generator forgetting by using an exponential moving average copy.

\textbf{One-shot Federated Learning.} % ~(FL)
One-shot FL, which aims to address issues stemming from communication and security in standard FL~\cite{McMahan2017Communication, karimireddy2020scaffold, Acar2021Federated, li2021model, luo2023gradma, kim2022multi}, was first proposed by~\cite{guha2019one}.
Although the original one-shot FL provides a practical and simple solution for aggregation, it still suffers from performance deterioration or even fails to run due to the vast data or model heterogeneity among real-world clients~\cite{li2020federated, diao2020heterofl}.
Shortly thereafter, a plethora of modifications have been proposed to handle the mentioned issue. 
For example, 
\cite{guha2019one, li2020practical, gong2021ensemble, gong2022preserving} perform knowledge transfer on a public dataset to mitigate data heterogeneity, augmenting the global model.
% Furthermore, these methods only focus on model homogeneous setting.
But the desired public data is not always available in practice and the performance may decrease dramatically if the apparent disparity in distributions exists between public data and clients' private data~\cite{zhu2021data, tan2022fedproto}. 
Besides, these methods only focus on model-homogeneous settings, which greatly limits their application domain expansion. 
To get rid of public data and accommodating to model heterogeneous settings,
DENSE~\cite{Zhang2022DENSE} aggregates knowledge from heterogeneous local models based on DFKD to train a global model for one-shot FL.  
FedFTG~\cite{Zhang2022Fine} leverages DFKD to fine-tune the global model in model-homogeneous FL to overcome data heterogeneity.
Note that FedFTG with a single communication round can be regarded as a one-shot FL method and directly applied to model heterogeneous scenarios.
However, they can merely extract limited knowledge from local models due to the fact that a generator alone is equipped on the server,
and do not carefully frame a learning objective for enhancing the generator.
On this account, our study is inspired by the aforementioned pitfalls.
Going beyond the said methods, there are a handful of recent studies~\cite{zhou2020distilled, shin2020xor, dennis2021heterogeneity, heinbaugh2022data,  diao2022towards, humbert2023one} on one-shot FL.
% % For example, 
% ~\cite{wang2018dataset} dataset distillation
% ~\cite{dennis2021heterogeneity} clustering
% ~\cite{heinbaugh2022data} FEDCVAE-KD
% ~\cite{shin2020xor, diao2022towards} data augmentation technique and open-set recognition
% ~\cite{humbert2023one} Conformal Prediction
\section{Notations}
\label{sec:notations}
This section describes the notations used in this paper.
We focus on the centralized setup that consists of a central server and $N$ clients owning private labeled datasets  $\{(\bm{X}_i, \bm{Y}_i)\}_{i=1}^N$, where $\bm{X}_i=\{\bm{x}_i^b\}_{b=1}^{n_i}$ follows the data distribution $\mathcal{D}_i $ over feature space $\mathcal{X}_i$, i.e., $\bm{x}_i^b \sim \mathcal{D}_i$, and $\bm{Y}_i=\{y_i^b\}_{b=1}^{n_i}$ ($y_i^b \in [C]:=\{1, \cdots, C\}$) denotes the ground-truth labels of $\bm{X}_i$. 
% We denote total number of labels as $C$.
And $C$ refers to the total number of labels.
Each client $i$ holds an on-demand local model $f_i$ parameterized by $\bm{\theta}_i$. 
On the server, the global model $f$ is parameterized by $\bm{\theta}$.
% Additionally, we consider generating synthetic data using conditional generator
% % .
% % Specifically,
% % we consider conditional generator 
% $G(\cdot)$  parameterized by $\bm{w}$. 
Additionally, we consider conditional generator $G(\cdot)$ parameterized by $\bm{w}$ to generate synthetic data.
It takes as input a random noise $\bm{z} \in \mathbf{R}^d$ sampled from standard normal distribution $\mathcal{N} (\bm{0}, \bm{I})$,
and a random label $y \in [C]$ sampled from label distribution $p(y)$, i.e., the probability of sampling $y$,
thus generating the synthetic data $\bm{s}=G(\bm{h}=o(\bm{z}, y), \bm{w})$. Note that $o(\bm{z}, y)$ represents the merge operator of $\bm{z}$ and $y$. 
\section{The Proposed Method}
\label{sec:propsed_method}
In this section, we detail our method DFDG.
Figure~\ref{Framework:} visualizes the training process of DFDG on the server, which occurs after the server receives the trained local models uploaded by all clients, and consists of two components: \textit{dual-generator training} and \textit{dual-model distillation}.
% Fig.~\ref{Framework:} visualizes the training procedure of DFDG on the server, conducted independently of local training, and consisting of two components: \textit{dual-generator training} and \textit{dual-model distillation}.
Thereafter, we detail the two components.
Moreover, we present pseudocode for the training process of DFDG in Appendix~\ref{sup_sec:pseudo}.

% In this paper, we work on improving the performance of global model in one-shot FL 
% % with both data and model heterogeneity
% with the help of data-free knowledge distillation~\cite{chen2019data, yoo2019knowledge}.
% Note that the strategy of integrating DFKD to FL is not unique to us.

% Our work points out that existing methods combining FL and DFKD do not thoroughly study the training of the generator, and neglect the catastrophic forgetting of the global model caused by large distribution shifts of the generator in the scenarios of coexisting data and model heterogeneity. 

\subsection{Dual-Generator Training}
In this subsection, we consider thoroughly training dual conditional generators.
Simply put, in addition to ensuring the utility of the individual generators, we lessen the overlap in their output spaces.
For a single generator, high-quality synthetic data should satisfy several key characteristics: \textit{fidelity}, \textit{transferability}, and \textit{diversity}~\cite{Zhang2022DENSE, Zhang2022Fine}.
Further, on account of ensuring the utility of a single generator, the small overlap of dual generators' output spaces can notably enlarge the covered manifold in the real data space, thus capturing more knowledge of local models.
Thereafter, we construct the loss objective from the mentioned aspects to ensure the quality and utility of the dual generators.
To differentiate the dual generators, we set $\bm{s}_k=G_k(\bm{h}_k=o^k(\bm{z}, y), \bm{w}_k)$, $k\in \{1, 2\}$.
 % $G_1(\cdot)$ and $G_2(\cdot)$.

\textbf{Fidelity.} 
To commence, we study the fidelity of the synthetic data for a single generator $G_k(\cdot)$. 
Specifically, we expect
$G_k(\cdot)$
to simulate the training space of the local models to generate the synthetic dataset with a similar distribution to the original dataset. 
To put it differently, we want the synthetic data $\bm{s}_k$ to approximate the training data with label $y$ without access to clients' training data. 
To achieve it, we form the fidelity loss $\mathcal{L}_{fid, k}$ at logits level:  
\begin{align}
    \label{L_fidelity:}
     \mathcal{L}_{fid,k}=CE(\sum_{i \in [N]}\tau_{i, y} f_i(\bm{s}_k, \bm{\theta}_i), y),
\end{align} %\tau_{i, y}
where $CE$ denotes the cross-entropy function, $f_i(\bm{s}_k, \bm{\theta}_i)$ is the logits output of the local model from client $i$ when $\bm{s}_k$ is given, $\tau_{i, y}$ dominates the weight of logits from different clients $\{i | i \in [N]\}$ when $y$ is given. 
And on top of $G_k(\cdot)$, $\mathcal{L}_{fid, k}$ is the cross-entropy loss between the weighted average logits $\sum_{i \in [N]}\tau_{i, y} f_i(\bm{s}_k, \bm{\theta}_i)$ and the label $y$.
By minimizing $\mathcal{L}_{fid,k}$, $\bm{s}_k$ is enforced to be classified into label $y$ with a high probability, thus facilitating the fidelity of $\bm{s}_k$. 

% \begin{wrapfigure}{r}{0cm}
%   \vspace*{-6ex}
%   \centering
%   \includegraphics[scale=0.3]{Distribution_diagram.pdf}
%   \caption{The visualization of synthetic data and decision boundaries of global model~(student) and ensemble model~(teacher). \textit{Left panel}: synthetic data~(red circles) are far away from the decision boundary $d_T$. \textit{Middle panel}: synthetic data~(black circles) near the decision boundaries $d_T$.
%   \textit{Right panel}: synthetic data~(yellow and purple circles) cross over the decision boundary $d_T$.}
%   \label{tran_analysis_pic:}
%   \vspace*{-3ex}
% \end{wrapfigure}

\begin{figure}[t]
    \centering
    \includegraphics[width=0.47\textwidth]{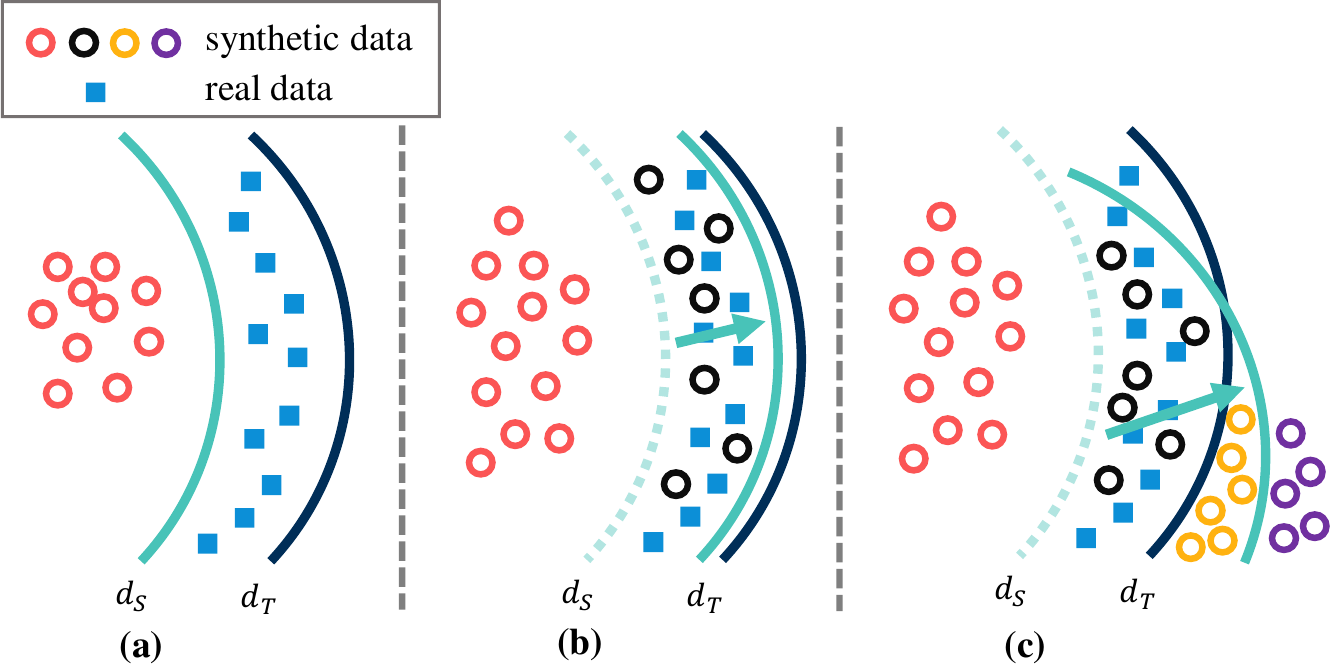} % Reduce the figure size so that it is slightly narrower than the column.
    \caption{The sketch of synthetic data and decision boundaries of global model~(student) and ensemble model~(teacher). (a): synthetic data~(red circles) are far away from the decision boundary $d_T$. (b): synthetic data~(black circles) near the decision boundaries $d_T$.
  (c): synthetic data~(yellow and purple circles) cross over the decision boundary $d_T$.}
    \vspace*{-3ex}
    \label{tran_analysis_pic:}
\end{figure}

In reality, the conditional generator $G_k(\cdot)$ easily generates synthetic data with low classification errors~(i.e. $\mathcal{L}_{fid, k}$ close to $0$) as the training proceeds.
This may cause the synthetic data to fall into a space far from the decision boundary of the ensemble model~(i.e., $\sum_{i \in [N]}\tau_{i, y} f_i(\cdot, \bm{\theta}_i)$) if only $\mathcal{L}_{fid, k}$ is optimized, as shown in the synthetic data represented by red circles in Fig.~\ref{tran_analysis_pic:} (a).
Note that $d_S$ and $d_T$ denote the decision boundaries of the global model~(student) and the ensemble model~(teacher), respectively.
An obvious observation is that the red circles are correctly classified on the same side of the two decision boundaries~(i.e., $d_S$ and $d_T$), making it difficult to transfer knowledge from teacher to student.
We next explore how to augment the transferability of the synthetic data to ameliorate this downside.

\textbf{Transferability} 
is intended to guide $G_k(\cdot)$ in generating
synthetic data that moves the decision boundary of the global model towards that of the ensemble model, such as synthetic data with black circles in Fig.~\ref{tran_analysis_pic:}~(b). 
However, during the training of $d_S$ to approach $d_T$, we find that $G_k(\cdot)$ can easily generate two other types of synthetic data, the yellow and purple circles in Fig.~\ref{tran_analysis_pic:}~(c).  Both of them are misclassified by the ensemble model~($d_T$), while the yellow circles are correctly classified and the purple circles are misclassified by the global model~($d_S$).
For the conditional generator $G_k(\cdot)$ that takes label information as one of the inputs, 
yellow and purple circles can mislead the generator, thereby leading to $d_S$ approximating $d_T$ with a large deviation, as shown in Fig.~\ref{tran_analysis_pic:}~(c).
Based on the above observation, we deem that the synthetic data $\bm{s}_k=G_k(\bm{h}_k=o^k(\bm{z}, y), \bm{w}_k)$ is useful if it is classified as $y$ by the ensemble model but classified not as $y$ by the global model.
To achieve this, we maximize the logits discrepancy between the global and ensemble models utilizing Kullback-Leibler divergence loss on synthetic data with black circles, as follows:
\begin{align}
     \label{L_tran:}
     \mathcal{L}_{tran,k}=- \varepsilon \cdot KL(\sum_{i \in [N]}\tau_{i, y} f_i(\bm{s}_k, \bm{\theta}_i), f(\bm{s}_k, \bm{\theta})),
\end{align}
where $KL$ is Kullback-Leibler divergence function and $f(\bm{s}_k, \bm{\theta})$ denotes the logits output of the global model on $\bm{s}_k$ with label $y$. Note that
$\varepsilon = 1$ if $\arg \max f(\bm{s}_k, \bm{\theta}) \neq y$ and $\arg \max \sum_{i \in [N]}\tau_{i, y} f_i(\bm{s}_k, \bm{\theta}_i) = y$ hold, otherwise $\varepsilon = 0$.~($\Diamond$)

We would like to point out that the existing works~\cite{Zhang2022Fine} and~\cite{Zhang2022DENSE} align with our research perspective on the transferability of the generator, which aims to generate more synthetic data with black circles.
% We note that the existing works~\cite{Zhang2022Fine} and~\cite{Zhang2022DENSE} align with our research perspective on generator transferability, aiming to generate more synthetic data with black circles.
However, they do not carefully frame the learning objective for enhancing the transferability of the generator.
Concretely, \cite{Zhang2022Fine} does not consider the type of synthetic data, i.e., $\varepsilon = 1$ always holds, thus inducing the generation of synthetic data with yellow and purple circles.~($\bigtriangleup$)
% Compared with \cite{Zhang2022Fine}, 
\cite{Zhang2022DENSE} focuses on synthetic data satisfying $\arg \max f(\bm{s}, \bm{\theta}) \neq \arg \max \sum_{i \in [N]}\tau_{i, y} f_i(\bm{s}, \bm{\theta}_i)$, but enables the generation of synthetic data with purple circles yet.~($\bigtriangledown$)  
\footnote{
Note that $\bigtriangleup$, $\bigtriangledown$ and $\Diamond$ denote the strategies for the transferability of generator  in~\cite{Zhang2022Fine}, in~\cite{Zhang2022DENSE} and in this paper, respectively.
}

\textbf{Diversity.} 
Although we enable $G_k(\cdot)$ to generate synthetic data that falls around the real data by optimizing $\mathcal{L}_{fid,k}$ and $\mathcal{L}_{tran,k}$, the diversity of synthetic data is insufficient.
Due to the fact that the generator may get stuck in \textit{local equilibria} as the training proceeds, model collapse occurs~\cite{odena2017conditional, kodali2017convergence}.
In this case, the generator may produce similar data points for each class with little diversity. 
Also, the synthetic data points may not differ significantly among classes.
This causes the empirical distribution estimated by $G_k(\cdot)$ to cover only a small manifold in the real data space, and thus only partial knowledge of the ensemble model is extracted.
% To alleviate this issue, we introduce a diversity loss $\mathcal{L}_{div,k}$ with label information to increase the diversity of synthetic data as follows:
To mitigate this, we introduce a diversity loss $\mathcal{L}_{div,k}$ with label information to enhance synthetic data diversity:
% \begin{align}
%      \label{L_div:}
%      \mathcal{L}_{div,k}= e^{\left(-\sum\limits_{j,l\in [B]}\|\bm{s}_k^j-\bm{s}_k^l\|_2\cdot\|\bm{h}_k^j-\bm{h}_k^l\|_2/B^2\right)},
% \end{align}
\begin{align}
     \label{L_div:}
     \mathcal{L}_{div,k}= \exp(-\sum\limits_{j,l\in [B]}\|\bm{s}_k^j-\bm{s}_k^l\|_2\|\bm{h}_k^j-\bm{h}_k^l\|_2/B^2),
\end{align}
where $B$ denotes the batch size and $\bm{s}_k^{j/l}=G_k(\bm{h}_k^{j/l}=o^k(\bm{z}^{j/l}, y^{j/l}), \bm{w}_k)$. 
Intuitively, $\mathcal{L}_{div,k}$ takes  
% the pairwise distance 
$\|\bm{h}_k^j-\bm{h}_k^l\|_2$ as a weight, and then multiplies it by the corresponding 
% pairwise distance 
$\|\bm{s}_k^j-\bm{s}_k^l\|_2$ in each batch $B$, 
thus imposing a larger weight on the synthetic data points pair~($\bm{s}_k^j$ and $\bm{s}_k^l$) at the more distant input pair~($\bm{h}_k^j$ and $\bm{h}_k^l$).
Notably, we merge the random noise $\bm{z}$ with label $y$ as the input of $G_k(\cdot)$ to overcome spurious solutions~\cite{do2022momentum}.
Further, we propose a multiplicative merge operator, i.e., $o^k(\bm{z}, y)= \bm{z} \times \mathcal{E}^k(y)$, where $\mathcal{E}^k$ is a trainable embedding and $\times$ means vector element-wise product. 
We empirically find that our merge operator enables the global model with better performance compared to others, possibly because the label information is effectively absorbed into the stochasticity of $\bm{z}$ by multiplying them when updating $\mathcal{E}^k$.
See the ablation study of Experiments section for more details and empirical justification.

\textbf{Cross-Divergence.} 
While we have systematically investigated how to guide a single generator $G_k(\cdot)$, $k\in[2]$ in generating valid synthetic data, training a single generator alone can only encompass a limited portion of the real data space~(as shown in Fig.~\ref{dual_gene_analysis_pic:}~(a)), thus failing to effectively explore the knowledge of the local models.
As such, we opt for training dual generators simultaneously to capture a broader space of real data. 
Nevertheless, if the dual generators are solely optimized in terms of \textit{fidelity}, \textit{transferability} and \textit{diversity}, the overlap of their output spaces may be substantial~(see Fig.~\ref{dual_gene_analysis_pic:}~(b)).
This motivates us to reduce or even eliminate the overlap, thereby covering a wider extent of real data space, see Fig.~\ref{dual_gene_analysis_pic:}~(c).
In other words, given $(\bm{z}, y)$, we aim for $G_1(\cdot)$ and $G_2(\cdot)$ to generate discrepant synthetic data $\bm{s}_1$ and $\bm{s}_2$, capturing the distinct valid information from local models.
%解释 我们希望s1和s2不同当y1=y2时，即抓取不同的有效信息。
To this end, we propose to maximize the logits discrepancy of the ensemble model between $\bm{s}_1$ and $\bm{s}_2$ by employing Kullback-Leibler divergence loss.
We call it cross-divergence loss.
Note that in practical training, we train $G_1(\cdot)$ and $G_2(\cdot)$ alternately, see Algorithm~\ref{server_update:} in Appendix~\ref{sup_sec:pseudo} for details.
Thus, the cross-divergence loss of $G_k(\cdot)$ takes the form:
\begin{align}
     \label{L_cd:}
      \mathcal{L}_{cd, k}  =
     -  & KL(\sum_{i \in [N]}\tau_{i, y} f_i(\bm{s}_k, \bm{\theta}_i), \notag \\ 
     & \quad \quad \quad \quad \quad \sum_{i \in [N]} \tau_{i, y} f_i(\bm{s}_{[2]/k}, \bm{\theta}_i)),
\end{align}
where $[2]/k = 1$ if $k=1$, otherwise $[2]/k = 2$.

Combining $\mathcal{L}_{fid,k}$, $\mathcal{L}_{tran,k}$, $\mathcal{L}_{div,k}$ and $\mathcal{L}_{cd,k}$, the overall objective of a single generator $G_k(\cdot)$ can be formalized as follows:
\begin{align}
     \label{L_gen_k:}
     \mathcal{L}_{gen, k}=\mathcal{L}_{fid,k} + & \beta_{tran} \cdot \mathcal{L}_{tran,k} + \notag \\ 
     & \quad \quad\beta_{div} \cdot \mathcal{L}_{div,k} + \beta_{cd} \cdot \mathcal{L}_{cd,k},
\end{align}
where $\beta_{tran}$, $\beta_{div}$ and $\beta_{cd}$ are tunable hyper-parameters. 
Further, the loss objective of the dual generators is $\mathcal{L}_{gen}=\mathcal{L}_{fid} + \beta_{tran}\mathcal{L}_{tran} + \beta_{div}\mathcal{L}_{div}+\beta_{cd}\mathcal{L}_{cd}$, where $\mathcal{L}_{fid}=\mathcal{L}_{fid,1}+\mathcal{L}_{fid,2}$. Also, $\mathcal{L}_{tran}$, $\mathcal{L}_{div}$ and $\mathcal{L}_{cd}$ are obtained in the same way.
% Of note, the synthetic data generated by a well-trained generator should be visually distinct from the real data for privacy protection, while it can capture the common knowledge of the local models to ensure similarity to the real data distribution for utility. 
% More privacy protection is discussed in Appendices~\ref{Vis_syn_samp:} and~\ref{app_discussion:}.

\subsection{Dual-Model Distillation}
Now we update the global model with the dual generators~(discussed in the previous section) and the ensemble model~(i.e., $\sum_{i \in [N]}\tau_{i, y} f_i(\cdot, \bm{\theta}_i)$).
Particularly, we compute the Kullback-Leibler divergence between logits of the ensemble model and the global model on the synthetic data points $\bm{s}_k=G_k(\bm{h}_k=o^k(\bm{z}, y), \bm{w}_k)$, $k\in [2]$, 
which is formulated as follows:
\begin{align}
    \label{L_dmd:}
     \mathcal{L}_{dmd}=\sum_{k\in[2]} KL(f(\bm{s}_k, \bm{\theta}), \sum_{i \in [N]}\tau_{i, y} f_i(\bm{s}_k, \bm{\theta}_i)).
\end{align}

So far, the appropriate $\tau_{i, y}$ and $p(y)$ are essential for the effective extraction of knowledge from local models. Here, we set $\tau_{i, y} = n_i^{y} / n^{y}$ and $p(y)=n^{y} / \sum_{y\in[C]}n^{y}$, where $n^{y} = \sum_{ j\in [N]}n_j^{y}$ and $n_i^{y}$ denotes the number of data points with label $y$ on the $i$-th client~\cite{Zhang2022Fine}.

\section{Experiments}
\subsection{Experimental Settings}
\textbf{Datasets and Baselines.} %In this paper,
We evaluate different methods with multiple image classification task-related datasets, namely Fashion-MNIST~\cite{xiao2017fashion}~(FMNIST in short), CIFAR-10~\cite{Krizhevsky2009Learning}, SVHN~\cite{Netzer2011Reading}, CINIC-10~\cite{darlow2018cinic}, CIFAR-100~\cite{Krizhevsky2009Learning}, Tiny-ImageNet~\footnote{http://cs231n.stanford.edu/tiny-imagenet-200.zip} and FOOD101~\cite{bossard2014food}.
We detail the said datasets in Appendix~\ref{sup_sec:dataset}. 
% ~\ref{dataset_app:}.
To gauge the effectiveness of DFDG with dual generators, we design a baseline~(marked as DFAD), which considers DFKD with a single generator based on Eq.~(\ref{L_fidelity:})-(\ref{L_div:}).
Also, we compare DFDG against FedAvg~\cite{McMahan2017Communication}, FedFTG~\cite{Zhang2022Fine} and DENSE~\cite{Zhang2022DENSE}, which are the most relevant methods to our work.

\textbf{Configurations.} 
Unless otherwise stated, all experiments are performed on a centralized network with $N=10$ clients.
To mimic data heterogeneity across clients, as in previous works~\cite{luo2023decentralized, luo2023gradma}, we use Dirichlet process $Dir(\omega)$ to partition the training set for each dataset, thereby allocating local training data for each client.  
It is worth noting that $\omega$ is the concentration parameter and smaller $\omega$ corresponds to stronger data heterogeneity. We set $\omega=0.5$ as default.
To simulate model-heterogeneous scenarios, we formulate exponentially distributed resource budgets for a given $N$: $R_i = [\frac{1}{2}]^{\min\{\sigma, \lfloor\frac{\rho \cdot i}{N}\rfloor\}} (i \in [N])$, where $\sigma$ and $\rho$ are both positive integers. 
See Appendix~\ref{Budget_Distribution_app:} for more details. 
We set both $\sigma$ and $\rho$ to $0$ by default.
%~\ref{Budget_Distribution_app:}
Unless otherwise specified, we set $\beta_{tran}$, $\beta_{div}$ and $\beta_{cd}$ to $1$ in \textit{dual-generator training}.
Also, all baselines utilize the same settings as ours.
Due to space limitations, see Appendix~\ref{Com_Exp_Setup:} for the full experimental setup. % ~\ref{Com_Exp_Setup:}

\textbf{Evaluation Metrics.} 
We evaluate the performance of different one-shot FL methods by global test accuracy~(\textit{G.acc} for short). 
To be specific,
we employ the global model on the server to evaluate the global performance of different one-shot FL methods with 
the original test set.
To ensure reliability, we report the average for each experiment over $3$ different seeds.

\begin{table*}[t]
  \centering
  \caption{Top $G.acc$~(\%) of distinct methods across $\omega \in \{0.1, 0.5, 1.0\}$ on different datasets~(with $\rho=0$).}
  \resizebox{2.0\columnwidth}{!}{
    \begin{tabular}{cp{4.19em}p{4.19em}p{4.19em}|p{4.19em}p{4.19em}p{4.19em}|p{4.19em}p{4.19em}p{4.19em}|p{4.19em}p{4.19em}p{4.19em}}
    \toprule
    \multirow{2}[0]{*}{Alg.s} & \multicolumn{3}{c|}{FMNIST} & \multicolumn{3}{c|}{CIFAR-10} & \multicolumn{3}{c|}{SVHN} & \multicolumn{3}{c}{CINIC-10} \\
    \cmidrule{2-13}
          & \multicolumn{1}{c}{$\omega=1.0$} & \multicolumn{1}{c}{$w=0.5$} & \multicolumn{1}{c|}{$w=0.1$} & \multicolumn{1}{c}{$w=1.0$} & \multicolumn{1}{c}{$w=0.5$} & \multicolumn{1}{c|}{$w=0.1$} & \multicolumn{1}{c}{$w=1.0$} & \multicolumn{1}{c}{$w=0.5$} & \multicolumn{1}{c|}{$w=0.1$} & \multicolumn{1}{c}{$w=1.0$} & \multicolumn{1}{c}{$w=0.5$} & \multicolumn{1}{c}{$w=0.1$} \\
    \midrule
    % \rowcolor{Gray}
    FedAvg & 10.34$\pm$0.57 & 11.41$\pm$1.90 & 10.38$\pm$0.66 & 13.83$\pm$0.59 & 9.71$\pm$0.42 & 10.56$\pm$0.76 & 18.05$\pm$2.39 & 16.52$\pm$4.73 & 10.03$\pm$5.13 & 9.52$\pm$0.50 & 12.56$\pm$2.23 & 9.76$\pm$0.23 \\
    \midrule
    DENSE & 40.98$\pm$6.18 & 33.93$\pm$3.90 & 14.09$\pm$2.48 & 24.26$\pm$2.75 & 23.55$\pm$2.27 & 12.39$\pm$2.28 & 23.76$\pm$4.22 & 20.05$\pm$4.42 & 11.03$\pm$4.33 & 14.75$\pm$1.28 & 14.54$\pm$1.81 & 11.14$\pm$0.82 \\
    % \rowcolor{Gray}
    FedFTG & 21.37$\pm$3.45 & 18.88$\pm$5.45 & 16.12$\pm$3.50 & 17.91$\pm$1.94 & 16.02$\pm$2.24 & 12.96$\pm$2.99 & 18.94$\pm$0.90 & 17.60$\pm$3.71 & 11.51$\pm$3.95 & 13.48$\pm$2.51 & 13.33$\pm$0.93 & 10.28$\pm$0.31 \\
    DFAD  & 54.44$\pm$14.6 & 48.90$\pm$6.77 & 25.46$\pm$0.54 & 32.45$\pm$5.58 & 29.78$\pm$0.72 & 15.04$\pm$0.77 & 24.58$\pm$1.23 & 21.21$\pm$1.13 & 20.92$\pm$1.41 & 20.10$\pm$1.60 & 15.80$\pm$2.18 & 13.80$\pm$0.69 \\
    % \rowcolor{Gray}
    DFDG  & \textbf{63.25$\pm$3.05} & \textbf{62.00$\pm$1.60} & \textbf{36.69$\pm$4.97} & \textbf{36.72$\pm$3.92} & \textbf{36.56$\pm$3.77} & \textbf{19.04$\pm$3.93} & \textbf{25.89$\pm$2.30} & \textbf{24.54$\pm$4.63} & \textbf{22.74$\pm$1.99} & \textbf{22.94$\pm$0.39} & \textbf{18.57$\pm$2.27} & \textbf{15.32$\pm$0.64} \\
    \bottomrule
    \end{tabular}}%
  \label{tab_data_heter:}%
\end{table*}%

\begin{figure}[b]
    \centering
    \includegraphics[width=0.47\textwidth]{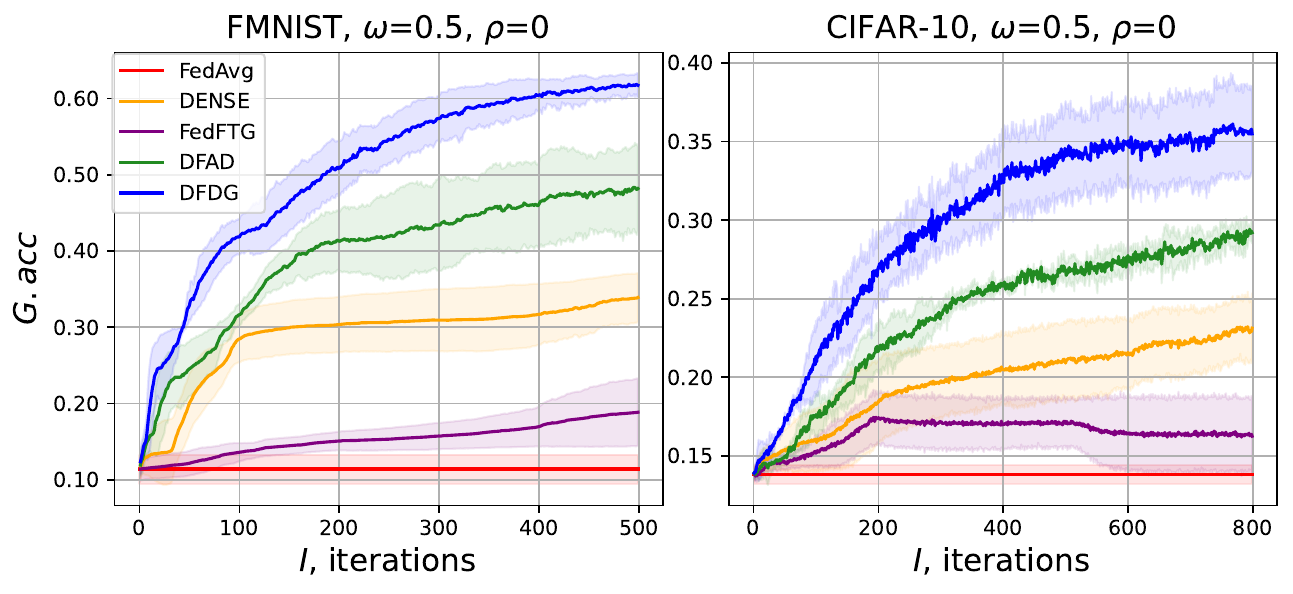} 
    \caption{Accuracy curves selected of DFDG and baselines on FMNIST and CIFAR-10.}
    \label{fmnist_cifar10_iter:}
\end{figure}

\subsection{Results Comparison}

%这个textbf因为在开头，和别的textbf没有对齐，需要调整吗
\textbf{Data Heterogeneity.}
To demonstrate the efficacy of our method in different data heterogeneity scenarios, we set $\omega \in \{0.1, 0.5, 1.0\}$. 
Also, we fix $\rho=0$, i.e., the model-homogeneous scenario. % that is, model-homogeneous scenario. 
We report the global performance on different datasets~(FMNIST, CIFAR-10, SVHN and CINIC-10) with different methods in Table~\ref{tab_data_heter:}.
The results show that: 
1) The global performances of all methods uniformly degrade with decreasing $\omega$ over all datasets, with DFDG being the only method that remarkably surpasses
other baselines against all scenarios.
Specifically, 
DFDG improves $G.acc$ by an average of $11.05\%$, $5.02\%$, $2.15\%$ and $2.38\%$ on FMNIST, CIFAR-10, SVHN and CINIC-10 respectively, compared to the best baseline DFAD.
This suggests that training dual generators simultaneously enables a more effective and extensive exploration of the local models' training space as opposed to training a single generator.
Also, Fig.~\ref{fmnist_cifar10_iter:} shows that the learning efficiency of DFDG consistently beats other baselines~(see Fig.~\ref{fig:data_full_res} in Appendix~\ref{sup:Add_Ex_Re} for complete curves).
2) FedAvg underperforms significantly all other methods w.r.t. $G.acc$, which indicates that merely averaging the model parameters cannot yield a robust global model in one-shot FL with the non-i.i.d setting (even with mild data heterogeneity setting, i.e., $\omega = 1.0$).
3) DFAD uniformly leads FedFTG and DENSE in terms of $G.acc$, revealing that a single generator trained with our proposed Eq.~(\ref{L_fidelity:})-(\ref{L_div:}) generates more effective synthetic data.
% , thus enhancing the global model more effectively.

\textbf{Model Heterogeneity.} 
We explore the impacts of different model heterogeneity distributions on different methods with FMNIST, CIFAR-10, SVHN and CINIC-10 datasets. 
We fix $\sigma=2$ and choose $\rho$ from $\{2, 3, 4\}$.
It is remarkable that a higher $\rho$ means more clients with $\frac{1}{4}$-width capacity w.r.t. the global model.
From Table~\ref{tab_model_heter:}, we can clearly observe that the performance of all methods drops uniformly with increasing $\rho$~(except FedFTG on CINIC-10), where DFDG consistently and overwhelmingly dominates other baselines in terms of $G.acc$.
For FMNIST~(CIFAR-10, SVHN, CINIC-10), DFDG improves accuracy by an average of $7.74\%$~($3.97\%$, $2.01\%$, $2.59\%$) over the best baseline DFAD.
%图里用的是curves，是不是统一比较好？
The selected learning curves shown in Fig.~\ref{svhn_cinic10_iter:} also verifies the said statement~(see Fig.~\ref{fig:model_full_res} in Appendix~\ref{sup:Add_Ex_Re} for more results). 
The above empirical results validate the superiority of our proposed method in one-shot FL with model heterogeneity. 

\begin{table*} %[htbp]
  \centering
  \caption{Top $G. acc$~(\%) of distinct methods across $\rho \in \{2, 3, 4\}$ on different datasets~(with $\omega=0.5$).}
  \resizebox{2.0\columnwidth}{!}{
    \begin{tabular}{cp{4.19em}p{4.19em}p{4.19em}|p{4.19em}p{4.19em}p{4.19em}|p{4.19em}p{4.19em}p{4.19em}|p{4.19em}p{4.19em}p{4.19em}}
\midrule    \multirow{2}[4]{*}{Alg.s} & \multicolumn{3}{c|}{FMNIST} & \multicolumn{3}{c|}{CIFAR-10} & \multicolumn{3}{c|}{SVHN} & \multicolumn{3}{c}{CINIC-10} \\
\cmidrule{2-13}          & \multicolumn{1}{c}{$\rho=2$} & \multicolumn{1}{c}{$\rho=3$} & \multicolumn{1}{c|}{$\rho=4$} & \multicolumn{1}{c}{$\rho=2$} & \multicolumn{1}{c}{$\rho=3$} & \multicolumn{1}{c|}{$\rho=4$} & \multicolumn{1}{c}{$\rho=2$} & \multicolumn{1}{c}{$\rho=3$} & \multicolumn{1}{c|}{$\rho=4$} & \multicolumn{1}{c}{$\rho=2$} & \multicolumn{1}{c}{$\rho=3$} & \multicolumn{1}{c}{$\rho=4$} \\
    \midrule
    DENSE & 29.09$\pm$2.85 & 27.47$\pm$5.91 & 25.90$\pm$4.12 & 17.67$\pm$0.81 & 16.51$\pm$0.37 & 15.84$\pm$0.81 & 19.67$\pm$0.13 & 19.08$\pm$0.88 & 18.93$\pm$1.14 & 19.57$\pm$0.69 & 19.45$\pm$0.29 & 17.92$\pm$0.29 \\
    % \rowcolor{Gray}
    FedFTG & 17.67$\pm$4.26 & 16.81$\pm$2.99 & 15.56$\pm$2.28 & 13.21$\pm$1.68 & 11.98$\pm$0.75 & 11.45$\pm$0.34 & 12.28$\pm$4.22 & 11.83$\pm$3.81 & 10.98$\pm$4.27 & 11.57$\pm$0.68 & 11.87$\pm$1.21 & 12.04$\pm$1.45 \\
    DFAD  & 57.82$\pm$3.66 & 53.84$\pm$9.01 & 51.83$\pm$14.1 & 19.52$\pm$1.45 & 18.02$\pm$1.35 & 17.80$\pm$0.32 & 20.78$\pm$1.81 & 20.48$\pm$1.18 & 20.23$\pm$1.36 & 22.93$\pm$0.79 & 21.80$\pm$0.39 & 18.95$\pm$1.48 \\
    % \rowcolor{Gray}
    DFDG  & \textbf{65.26$\pm$5.04} & \textbf{62.35$\pm$1.38} & \textbf{59.10$\pm$1.08} & \textbf{22.89$\pm$0.64} & \textbf{22.63$\pm$1.43} & \textbf{21.73$\pm$0.84} & \textbf{23.19$\pm$0.77} & \textbf{22.62$\pm$0.87} & \textbf{21.71$\pm$1.22} & \textbf{26.58$\pm$1.82} & \textbf{23.92$\pm$1.68} & \textbf{20.94$\pm$1.23} \\
    \bottomrule
    \end{tabular}}%
  \label{tab_model_heter:}%
  \vspace*{-3ex}
\end{table*}%

\begin{figure}[b]
    \centering
    \includegraphics[width=0.47\textwidth]{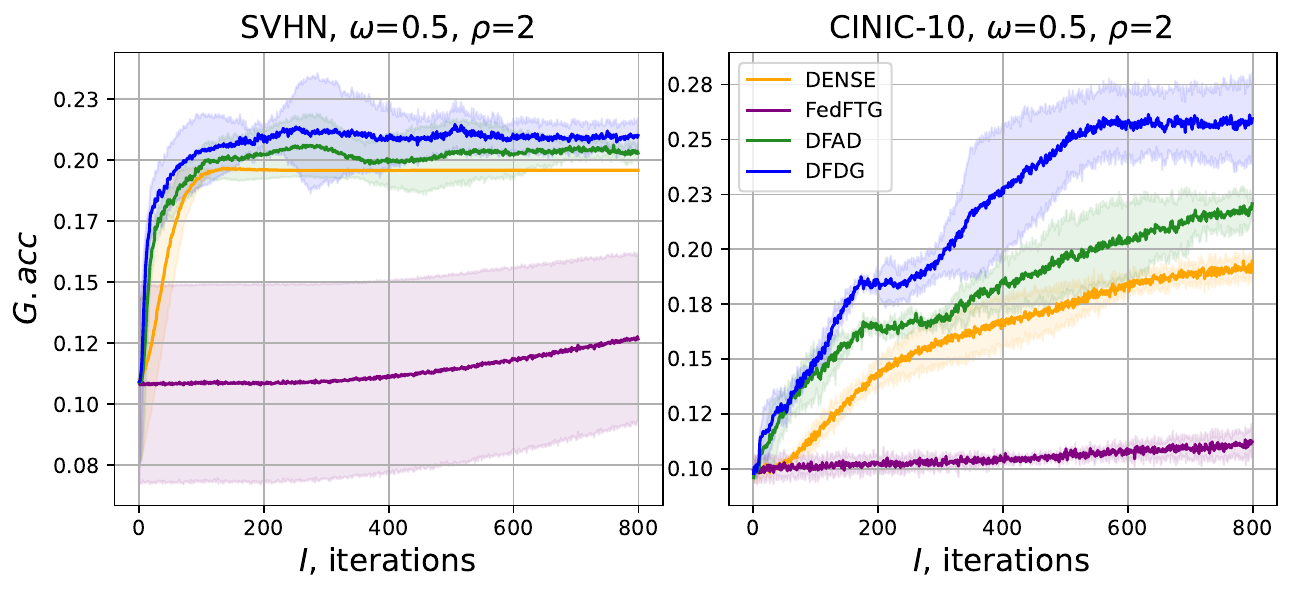} 
    \caption{Accuracy curves selected of DFDG and baselines on SVHN and CINIC-10.}
    \label{svhn_cinic10_iter:}
\end{figure}

\textbf{Difficult Image Classification Tasks.}
We further delve into the effectiveness of DFDG with different $N$ values in difficult image classification tasks~(CIAR-100, Tiny-ImageNet and FOOD101).
% We further delve into the effectiveness of DFDG over difficult image classification tasks~(CIAR-100, Tiny-ImageNet and FOOD101) w.r.t. different the number of clients in one-shot FL with homogeneous setting~(i.e., $\rho=0$).
We set $\omega=0.5$, $\rho=0$ and $N \in \{10, 50, 100\}$. 
% See Appendix~\ref{Com_Exp_Setup:} for detailed experimental setup.
Due to space limitations, we only report the global test accuracy results corresponding to $N = \{10, 100\}$ in Table~\ref{di_im_task:}. 
See Table~\ref{full_di_im_task:} in Appendix~\ref{sup:Add_Ex_Re} for full results.
% We report the results with respect to the global test accuracy in Table~\ref{di_im_task:}.
One can observe that DFDG consistently beats the baselines, suggesting that DFDG with dual generators exemplifies a robust performance gain on difficult image classification tasks.
Specifically, compared to FedAvg, DFDG has shown an average G.acc improvement of $9.62\%$.
In addition, DFAD uniformly outstrips DENSE and FedFTG, which further validates the utility of our proposed Eq.~(\ref{L_fidelity:})-(\ref{L_div:}).
% 感觉the可以去掉 而且这里和前文的difficult image classification tasks也不统一
With the exception of FedAvg and FedFTG, the $G.acc$ of the other methods consistently improves as $N$ increases. 
We conjecture that the increase in the number of clients makes the generator(s) extract more knowledge, enhancing the global model's performance.
Notably, DFDG does not meet the practical application requirements on the difficult image classification task, although it dominates in terms of $G.acc$.
We argue that adopting more powerful global model and generator can alleviate this issue.

% \begin{table}[htbp]
%   \centering
%   \caption{Top $G. acc$~(\%) of distinct methods over $(\omega, \rho)=(0.5, 0)$ on CIFAR-100, Tiny-ImageNet and FOOD101 datasets.}
%   \resizebox{0.8\columnwidth}{!}{
%     \begin{tabular}{cccc}
%     \toprule
%           & CIFAR-100 & Tiny-ImageNet & FOOD101\\
%     \midrule
%     FedAvg & 8.62$\pm$0.89 & 3.58$\pm$0.26 & 4.35$\pm$0.54\\
%     \midrule
%     DENSE    & 12.55$\pm$1.35 & 9.62$\pm$2.11 & 6.14$\pm$0.24\\
%     FedFTG      & 10.92$\pm$0.74 & 9.06$\pm$1.88 & 7.16$\pm$0.36\\
%     DFAD   & 14.24$\pm$1.61 & 11.52$\pm$1.36 & 8.68$\pm$0.55\\
%     % \rowcolor{Gray}
%      DFDG  & \textbf{17.05$\pm$1.62} & \textbf{13.97$\pm$1.75} & \textbf{10.98$\pm$0.67}\\
%     \bottomrule
%     \end{tabular}}%
%   \label{di_im_task:}%
%   % \vspace*{-1ex}
% \end{table}%

\begin{table}[htbp]
  \centering
  \caption{Top $G. acc$~(\%) of different methods over $N\in\{10, 100\}$ on CIFAR-100, Tiny-ImageNet and FOOD101 datasets with $(\omega, \rho)=(0.5, 0)$.}
  \resizebox{1.0\columnwidth}{!}{
    \begin{tabular}{ccc|cc|cc}
    \toprule
    \multicolumn{1}{c}{\multirow{2}[4]{*}{Alg.s}} & \multicolumn{2}{c|}{CIFAR-100} & \multicolumn{2}{c|}{Tiny-ImageNet} & \multicolumn{2}{c}{FOOD101} \\
\cmidrule{2-7}          & $N=10$  & $N=100$ & $N=10$  & $N=100$ & $N=10$  & $N=100$ \\
    \midrule
    FedAvg & 8.62$\pm$0.89 & 7.98$\pm$0.65 & 3.58$\pm$0.26 & 3.87$\pm$0.67 & 4.35$\pm$0.54 & 3.73$\pm$0.65 \\
    \midrule
    DENSE & 12.55$\pm$1.35 & 14.66$\pm$0.77 & 9.62$\pm$2.11 & 10.06$\pm$1.01 & 6.14$\pm$0.24 & 8.97$\pm$0.28 \\
    FedFTG & 10.92$\pm$0.74 & 9.01$\pm$0.38 & 9.06$\pm$1.88 & 8.24$\pm$0.78 & 7.16$\pm$0.36 & 6.33$\pm$0.47 \\
    DFAD  & 14.24$\pm$1.61 & 16.39$\pm$0.88 & 11.52$\pm$1.36 & 13.29$\pm$1.34 & 8.68$\pm$0.55 & 10.57$\pm$0.26 \\
    DFDG  & \textbf{17.05$\pm$1.62} & \textbf{18.98$\pm$0.66} & \textbf{13.97$\pm$1.75} & \textbf{15.99$\pm$1.32} & \textbf{10.98$\pm$0.67} & \textbf{12.88$\pm$0.36} \\
    \bottomrule
    \end{tabular}}%
  \label{di_im_task:}%
  \vspace*{-3ex}
\end{table}%

\subsection{Ablation Study}
In this section, we carefully demonstrate the efficacy and indispensability of core modules and key parameters in our method DFDG on FMNIST and CIFAR-10.

\begin{figure}[h]
    \centering
    \includegraphics[width=0.47\textwidth]{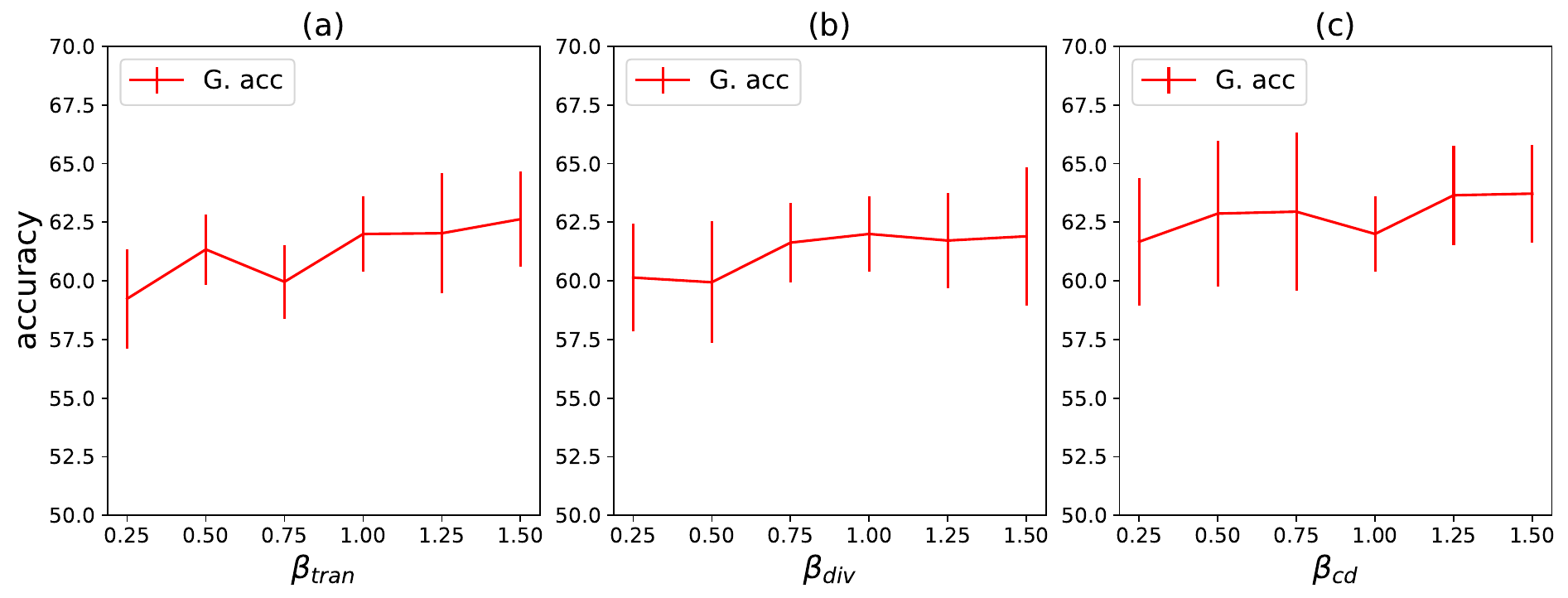} 
    \caption{Test accuracy~(\%) with varying $\beta_{tran}$, $\beta_{div}$ and $\beta_{cd}$ on FMNIST.}
    \label{varying_para_FMNIST:}
    \vspace*{-3ex}
\end{figure}

\textbf{Varying $\beta_{tran}$, $\beta_{div}$ and $\beta_{cd}$.} 
We explore the impacts of $\beta_{tran}$, $\beta_{div}$ and $\beta_{cd}$ and select them from $\{0.25, 0.50, 0.75, 1.00, 1.25, 1.50\}$.
% From Fig.~\ref{varying_para_FMNIST:}, we can see 
Fig.~\ref{varying_para_FMNIST:} shows that $G.acc$ of DFDG fluctuates slightly with the increases of $\beta_{tran}$, $\beta_{div}$ and $\beta_{cd}$ over FMNIST and CIFAR-10~(see Fig.~\ref{fig:model_full_res} in Appendix~\ref{sup:Add_Ex_Re}).
% From Fig.~\ref{varying_para_FMNIST:}, we can see that DFDG maintains stable test performance among all selections of $\beta_{tran}$, $\beta_{div}$ and $\beta_{cd}$ over FMNIST. 
% Meanwhile, $G.acc$ fluctuates slightly with the increases of $\beta_{tran}$, $\beta_{div}$ and $\beta_{cd}$ on CIFAR-10~(see Fig.~\ref{fig:model_full_res} in Appendix).
% 感觉这里说的有点不明确，the worst $G.acc$说明一下是我们的比较好，同理，Fig.12以及其他很多图片的说明也没有说明是DFDG的方法，感觉是不是写清楚好一点？
Besides, we observe that the worst $G.acc$ in Fig.~\ref{fig:model_full_res} outperforms the baseline with the best $G.acc$ in Table~\ref{tab_data_heter:}.
The above results indicate that DFDG is not sensitive to choices of $\beta_{tran}$, $\beta_{div}$ and $\beta_{cd}$ over a wide range.

\textbf{Necessity of each component for DFDG.}
We perform the leave-one-out test to explore the contributions of $\mathcal{L}_{tran}$, $\mathcal{L}_{div}$ and $\mathcal{L}_{cd}$ to DFDG separately, and further display the test results by simultaneously discarding them.
From Table~\ref{tab:impact_component}, removing
either $\mathcal{L}_{tran}$, $\mathcal{L}_{div}$ or $\mathcal{L}_{cd}$ adversely affects the performance of DFDG.
In addition, the multiple losses' absence further exacerbates the degradation of $G.acc$,
demonstrating that $\mathcal{L}_{tran}$, $\mathcal{L}_{div}$ and $\mathcal{L}_{cd}$ are vital for the training of the dual generators.
Interestingly, we observe that removing $\mathcal{L}_{tran}$ or $\mathcal{L}_{div}$ negatively impacts the global model more than removing $\mathcal{L}_{cd}$.
Meanwhile, we find that DFDG without $\mathcal{L}_{cd}$ still outperforms DFAD in terms of global performance~(see Table~\ref{tab_data_heter:}).
% 这个-是本来就需要用的吗？
Note that $-\mathcal{L}_{cd}$ means training dual generators simultaneously and independently only with Eq.~(\ref{L_fidelity:})-(\ref{L_div:}).
The mentioned empirical results indicate that in dual-generator training, the quality of the synthetic data from a single generator is crucial for obtaining superior performance of DFDG and the dual generators further augment the global model.
Besides, compared to a single generator,
training dual generators simultaneously and independently can improve the global model's performance, and introducing $\mathcal{L}_{cd}$ can result in a better $G.acc$.
We posit that training dual generators built upon $\mathcal{L}_{cd}$ can effectively explore a wider range of local models' training space under the premise of ensuring high-quality synthetic data from a single generator.
\begin{table}[h]
  \centering
  \caption{Impact of each component w.r.t test accuracy~($\%$).}
  \resizebox{0.8\columnwidth}{!}{
    \begin{tabular}{cll}
    \toprule
          & FMNIST & CIFAR-10 \\
    \midrule
    baseline & \textbf{62.00$\pm$1.60} & \textbf{36.56$\pm$3.77} \\
    \midrule
    $-\mathcal{L}_{tran}$ & 47.81$\pm$10.41 & 29.26$\pm$4.35 \\
    % \rowcolor{Gray}
    $-\mathcal{L}_{div}$ & 57.44$\pm$3.49 & 28.13$\pm$2.33 \\
    $-\mathcal{L}_{cd}$ & 60.78$\pm$3.10 & 32.85$\pm$1.88 \\
    \midrule
    $-\mathcal{L}_{tran}$, $-\mathcal{L}_{div}$ & 43.49$\pm$3.78 & 22.17$\pm$2.33 \\
    % \rowcolor{Gray}
    $-\mathcal{L}_{tran}$, $-\mathcal{L}_{cd}$ & 45.61$\pm$9.76 & 26.21$\pm$3.60 \\
    $-\mathcal{L}_{div}$, $-\mathcal{L}_{cd}$ & 50.99$\pm$0.97 & 24.96$\pm$7.54 \\
    \midrule
    $-\mathcal{L}_{tran}$, $-\mathcal{L}_{div}$, $-\mathcal{L}_{cd}$ & 37.45$\pm$5.14 & 20.11$\pm$4.88 \\
    \bottomrule
    \end{tabular}}%
  \label{tab:impact_component}%
  \vspace*{-2ex}
\end{table}%

\textbf{Impacts of different transferability constraints and merge operators.}
To investigate the utility of the merge operator, we consider multiple merge operators, including $mul$, $add$, $cat$, $ncat$ and $none$.
Among them, $mul$ is $o(\bm{z}, y)= \bm{z} \times \mathcal{E}(y)$,  $add$ is $o(\bm{z}, y)= \bm{z} + \mathcal{E}(y)$, $cat$ is $o(\bm{z}, y)= [\bm{z}, \mathcal{E}(y)]$, $n\-cat$ is $o(\bm{z}, y)= [\bm{z}, y]$ and $none$ is $o(\bm{z}, y)= \bm{z}$.
It can be observed from Table~\ref{ablation_noise_tran:}  that our proposed transferability constraint $\Diamond$ uniformly beats $\bigtriangleup$ and $\bigtriangledown$ w.r.t. $G.acc$ over two datasets for given $mul$.
This means that $\Diamond$ can guide the generator to generate more effective synthetic data~(like those with the black circles see Fig.~\ref{tran_analysis_pic:}), thereby improving the test performance of the global model.
Additionally, from Table~\ref{ablation_noise_tran:}, we can see that our proposed $mul$ significantly outperforms other competitors in terms of $G.acc$ for given $\Diamond$.
This suggests that $mul$ can better exploit the diversity constraint to make the generator generate more effective and diverse synthetic data, thus enhancing the global model.
% Also, the visualization of the synthetic data in Appendix~\ref{Vis_syn_samp:} validates this statement.

% Also, we conjecture that generating more synthetic data may positively impact on the performance of local models, since $\Diamond$ also slightly and consistently trumps other competitors w.r.t. $L.acc$.

\begin{table}[htbp]
  \centering
  \caption{Test accuracy~(\%) comparison among different merger operators and transferability constraints.}
  \resizebox{0.8\columnwidth}{!}{
    \begin{tabular}{ccc}
    \toprule
          & FMNIST & CIFAR-10 \\
    \midrule
    baseline~($mul$, $\Diamond$) & \textbf{62.00$\pm$1.60} & \textbf{36.56$\pm$3.77} \\
    \midrule
    ($mul$, $\bigtriangledown$)    & 60.37$\pm$3.95 & 32.39$\pm$2.00 \\
    ($mul$, $\bigtriangleup$)      & 59.42$\pm$0.85 & 31.59$\pm$2.96 \\
    \midrule
    ($add$, $\Diamond$)   & 44.26$\pm$2.01 & 21.81$\pm$4.92 \\
    % \rowcolor{Gray}
     ($cat$, $\Diamond$)  & 44.95$\pm$3.15 & 20.86$\pm$2.86 \\
     ($ncat$, $\Diamond$) & 55.74$\pm$6.62 & 30.92$\pm$5.46 \\
     % \rowcolor{Gray}
     ($none$, $\Diamond$)  & 56.55$\pm$4.71 & 32.89$\pm$2.69 \\
    \bottomrule
    \end{tabular}}%
  \label{ablation_noise_tran:}%
  \vspace*{-3ex}
\end{table}%

\section{Discussion}
For a one-shot FL system, there are many trade-offs, including utility, privacy protection, computational efficiency and communication cost, etc~\cite{li2020federated1}.
Thereafter, we discuss DFDG in terms of the said trade-offs.
Notably, our method DFDG mainly focuses on improving the utility of the global model in one-shot FL.

\begin{figure}[h]
  \centering
  \begin{subfigure}{1.0\linewidth}
    \centering
    \includegraphics[width=1.0\linewidth]{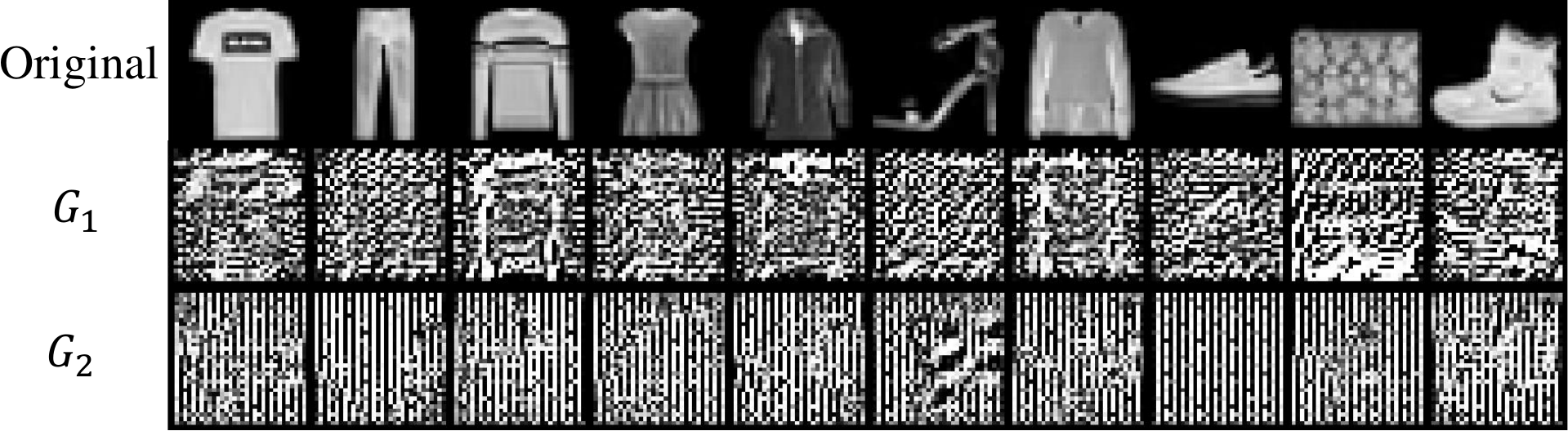}
        \caption{FMNIST}
  \end{subfigure}
  \\
  \centering
  \begin{subfigure}{1.0\linewidth}
    \centering
    \includegraphics[width=1.0\linewidth]{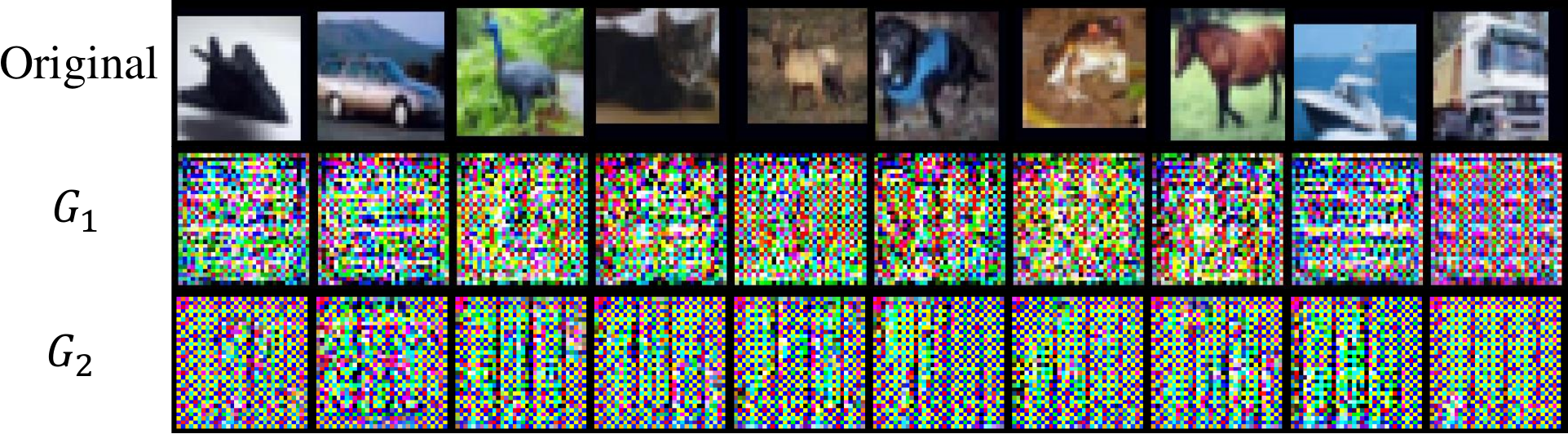}
        \caption{CIFAR-10}
  \end{subfigure}
  \caption{Visualization of synthetic data on FMNIST and CIFAR-10 by using DFDG with diversity constraint based on \textit{mul} for $G_1$ and $G_2$. Note that original data is selected from FMNIST and CIFAR-10 datasets, and we select $1$ image from each class in each dataset as a row. } % See Fig.~\ref{sup:raw_data_data_gen_mul} in Appendix~\ref{sup:Add_Ex_Re} for more visualizations.
  \label{raw_data_data_gen_mul:}
  \vspace*{-2.5ex}
\end{figure}

\textbf{Privacy Protection.} 
We acknowledge that it is difficult for DFDG to strictly guarantee privacy without privacy protection.
Since DFDG generates synthetic datasets that are similar to clients' training space in the server, 
it may violate the privacy regulations in FL.
Also, DFDG requires clients to upload the label statistics of the data, which also is at risk of compromising privacy.
However, according to our observation~(as shown in Fig.~\ref{raw_data_data_gen_mul:}), DFDG can only capture the shared attributes of the dataset without access to real data, and it is difficult to concretely and visually reveal the unique attributes of individual data.
See Fig.~\ref{sup:raw_data_data_gen_mul} in Appendix~\ref{sup:Add_Ex_Re} for more synthetic data.
Although our work does not address the privacy issue, in tasks with high privacy protection requirements, DFDG can incorporate some privacy protection techniques to enhance the reliability of one-shot FL systems in practice, such as differential privacy~(DP)~\cite{dwork2008differential, geyer2017differentially, cheng2022differentially}.
% Note that the combination of DFRD and encryption technologies~\cite{Ma2022Privacy, Zhang2022Homomorphic} may not be feasible, since {\it training generator} and {\it robust model distillation} in DFRD require knowledge of the structure of the local models.

\textbf{Computational Efficiency and Communication Cost.}
To achieve a robust global model, DFDG requires additional computation.
% Specifically, compared with DENSE and FedFTG, the training time of DFDG will be longer, as it needs to additionally train a generator on the server.
Specifically, compared with DENSE and FedFTG, the training time of DFDG will be longer, which yields a significant performance improvement of the global model, as it needs to additionally train a generator on the server.
In our experiments, DFDG runs about \textbf{one and half times} longer than they do in each communication round.
% \textbf{Notably, DFDG can be extended to train more generators by harnessing a one-against-the-rest strategy to frame cross-divergence loss, but it requires the server to allocate more computational resources and memory for training.}
\textbf{Of note, DFDG can be extended to train more generators, namely three or more generators, but it is also confronted with more challenges.}
For example, it requires the server to allocate more computational resources and memory for training.
Also, it is not trivial to frame cross-divergence loss to simultaneously guide multiple generators to extract more comprehensive and efficient knowledge from the local models.
We leave this for future work.
Moreover, compared to FedAvg and DENSE, DFDG requires an additional vector of label statistics to be transmitted.
However, the communication cost of this vector is negligible compared to that of the model.
\section{Conclusion}
In this paper, with the help of DFKD, we propose a novel data-free one-shot FL method DFDG, which can explore a broader local models' training space via training dual generators.
To be specific, DFDG is performed in an adversarial manner on the server and consists of two stages: \textit{dual-generator training} and \textit{dual-model distillation}.
In \textit{dual-generator training}, we dig deeply into each generator in terms of fidelity, transferability and diversity to ensure its utility on the one hand, and on the other hand, we carefully tailor a cross-divergence loss to guide the generators to explore different regions of the local models' training space, thereby reducing the overlap of the dual generators' output spaces.
In \textit{dual-model distillation}, the trained dual generators work together to provide the training data for updates of the global model.
Also, we conduct extensive experiments to verify the superiority of DFDG.
% In \textit{dual-generator training}, we dig into each generator concerning fidelity, transferability and diversity to ensure its utility on the one hand, and on the other hand, we tailor a cross-divergence loss to reduce the overlap of dual generators' output spaces.

\section{Acknowledgments}
\label{sec:acknowledge}
This work has been supported by the National Natural Science Foundation of China under Grant No.12471331.

{
    \small
    \bibliographystyle{ieeenat_fullname}
    \bibliography{main}
}
\clearpage
\appendix
\setcounter{page}{1}
% \maketitlesupplementary
\onecolumn

\section{Pseudocodes}
\label{sup_sec:pseudo}
We summarize the pseudocodes of DFDG in Algorithm~\ref{alg_all:}.

\begin{algorithm}[h]
  \caption{DFDG for one-shot FL.} %DFRD: Data-Free Robustness Distillation
  \label{alg_all:}
  \hspace*{0.02in} {\bf Input:}
  $N$: the number of clients; $B$: batch size.
  (\textbf{client side}) $f_i$: local model of client $i$ parametered by $\bm{\theta}_i$~($i \in [N]$); $I_c$: iteration of the training procedure in each client; $\eta_c$: learning rate for SGD.
  (\textbf{server side}) $f$: global model parameterized by $\bm{\theta}$; $G_1(\cdot)$~($G_2(\cdot)$): generator parameterized by $\bm{w}_1$~($\bm{w}_2$);
    $I$: iteration of the training procedure in server; ($I_g$, $I_d$): inner iterations of training the generator and the global model; ($\eta_g$, $b_1$, $b_2$): learning rate and momentums of Adam for the dual generators; $\eta_d$: learning rate of the global model; ($\beta_{tran}$, $\beta_{div}$, $\beta_{cd}$): hyper-parameters in \textit{training single generator}.
\begin{algorithmic}[1]
  \STATE Initial state global model $\bm{\theta}$, dual generators $\bm{w}_1$ and $\bm{w}_2$, local models $\bm{\theta}_i$ ($i\in[N]$).
  % \STATE Initial width capability $\{R_i\}_{i\in[N]}$. 
  \STATE Initial weighting counter $\Phi=\{\tau_{i, y}\}_{i\in [N], y \in [C]}\in \mathbf{R}^{N\times C}$, label distribution $P=\{p(y)\}_{y \in [C]}\in \mathbf{R}^{C}$ and label counter $LC=\{0\}_{i\in [N], y \in [C]}\in \mathbf{R}^{N\times C}$.
  \STATE \textbf{On Clients:}
  \FOR{$i \in [N]$ \textbf{parallel}}
        \STATE $\bm{\theta}_{i}, L_i \leftarrow$ Client\_Update($\bm{\theta}_{i}$, $\eta_c$, $I_c$, $B$), \# see Algorithm~\ref{client_update:}
        \STATE Send $\bm{\theta}_{i}$ and $L_i$ to server.
  \ENDFOR
  \STATE \textbf{On Server:}
  \STATE 
    Collecting local models $\{\bm{\theta}_{i}\}_{i\in [N]}$ from clients.
    \IF{$shape(\bm{\theta}_i)=shape(\bm{\theta}_j)$, for $\forall i, j \in [N]$ } 
        \STATE \# Note that $shape(\bm{\theta}
    )$ denotes the architecture of the model $\bm{\theta}$,
        \STATE $\bm{\theta} = \frac{1}{N} \sum \limits_{i=1}^{N} \bm{\theta}_i$.
    \ENDIF
    \FOR{$i \in [N]$}
        \STATE $LC[i,:] \leftarrow L_i$.
    \ENDFOR
  \FOR{$t=0,\cdots, I-1$}
    
    \STATE Compute $\Phi$ and $P$ according to $LC$, where $\tau_{i, y} = n_i^{y} / n^{y}$ and $p(y)=n^{y} / \sum_{y\in[C]}n^{y}$, where $n^{y} = \sum_{ j\in [N]}n_j^{y}$ and $n_i^{y}$ denotes the number of data points with label $y$ on the $i$-th client, % by using Eq.~(\ref{Dyn_wei:}) and~(\ref{lab_samp:}), respectively.
    \STATE  $\bm{\theta}, \bm{w}_1, \bm{w}_2 \leftarrow$ Server\_Update($\{\bm{\theta}_{i}\}_{i\in [N]}$, $\bm{\theta}$, $\bm{w}_1$,  $\bm{w}_2$, $\Phi$, $P$,\\ \quad \quad \quad \quad \quad \quad \quad \quad \quad \quad \quad \quad \quad \quad $I_g$, $I_d$, $\eta_g$, $b_1$, $b_2$, $\eta_d$, $\beta_{tran}$, $\beta_{div}$, $\beta_{cd}$, $B$). \# see Algorithm~\ref{server_update:}
    % \STATE $LC=\{0\}_{i\in [N], y \in [C]}$.
  \ENDFOR
\end{algorithmic}
\hspace*{0.02in} {\bf Output:} 
  $\bm{\theta}$ 
% \hspace*{0.02in} {\bf Note that}  
%   $\tau(t)=\max \left\{l:l\leq t \ {\rm and} \ {\rm mod} {(l,\tau)} = 0 \right\}$
\end{algorithm}

\begin{algorithm}[h]
  \caption{Client\_Update($\bm{\theta}$, $\eta$, $I$, $B$) \# Take client $i$ as an example}
  \label{client_update:}
\begin{algorithmic}[1]
    \STATE Set $\bm{\theta}_i = \bm{\theta}$ and $L_i=\{0\}_{y\in [C]} \in \mathbf{R}^{C}$.
    \STATE Cache data $Cache=[]$.
    \FOR{$e = 0,1,\ldots, I-1$}
        \STATE Sample $\mathcal{B} = \{(\bm{x}_i^b, y_i^b)\}_{b=1}^B$ from $\{(\bm{X}_i, \bm{Y}_i)\}$.
        \STATE $\bm{\theta}_{i} \leftarrow \bm{\theta}_{i}- \frac{\eta}{B}\sum_{b\in[B]} \nabla \mathcal{L}_{i}(f_i(\bm{x}_i^b,\bm{\theta}_{i}), y_i^b )$, where $\mathcal{L}_{i}$ denotes the cross-entropy loss.
        \FOR{$(\bm{x}, y)$ in $\mathcal{B}$}
            \IF{$(\bm{x}, y)$ not in $Cache$}
                \STATE $Cache\leftarrow Cache \cup \{(\bm{x}, y)\}$.
            \ENDIF
        \ENDFOR
        \FOR {$(\bm{x}, y)$ in $Cache$}
            \STATE $L_i^y\leftarrow L_i^y + 1$, where $L_i^y$ denotes the value of the $y^{th}$ element in $L_i$.
        \ENDFOR
    \ENDFOR
    % \STATE Sets $\bm{x}_{t+1}^{(i)}=\bm{x}_{t,I}^{(i)}$.
    \STATE {\bfseries Output:} $\bm{\theta}_{i}$, $L_i$.
\end{algorithmic}
\end{algorithm}

\begin{algorithm}[h]
  \caption{Server\_Update($\{\bm{\theta}_{i}\}_{i\in [N]}$, $\bm{\theta}$,  $\bm{w}_1$,  $\bm{w}_2$, $\Phi$ , $P$, $I_g$, $I_d$, $\eta_g$,  $b_1$, $b_2$, $\eta_d$, $\beta_{tran}$, $\beta_{div}$, $\beta_{cd}$, $B$)}
  \label{server_update:}
\begin{algorithmic}[1]
    \STATE Sample batches of $\{\bm{z}^b, y^b\}_{b=1}^B$ for $G_1(\cdot)$ and $G_2(\cdot)$, where $\bm{z}\sim \mathcal{N}(\bm{0}, \bm{I})$ and $y\sim p(y):=P[y]$. 
    \STATE \# \textbf{dual-generator training}
    \STATE $\bm{w}_1 \leftarrow$ Generator\_Update($\{\bm{\theta}_{i}\}_{i\in [N]}$, $\bm{\theta}$,  $\bm{w}_1$, $\{\bm{z}^b, y^b\}_{b=1}^B$, $\Phi$, $I_g$, $\eta_g$,  $b_1$, $b_2$, $\beta_{tran}$, $\beta_{div}$, $\beta_{cd}$). \# see Algprithm~\ref{generator_update:}
    \STATE $\bm{w}_2 \leftarrow$ Generator\_Update($\{\bm{\theta}_{i}\}_{i\in [N]}$, $\bm{\theta}$,  $\bm{w}_2$, $\{\bm{z}^b, y^b\}_{b=1}^B$, $\Phi$, $I_g$, $\eta_g$,  $b_1$, $b_2$, $\beta_{tran}$, $\beta_{div}$, $\beta_{cd}$). \# see Algprithm~\ref{generator_update:}
    \STATE \# \textbf{Dual-model distillation}
    \FOR{$e_d = 0,1,\ldots, I_d-1$}
        % \State Sample a batch $\mathcal{B}^e$ of $\bm{z}\sim \mathcal{N}(\bm{0}, \bm{I})$ and $y\sim p(y):=P[y]$
        \STATE Generate $\{\bm{s}_1^b\}_{b=1}^B$ with $\{\bm{z}^b, y^b\}_{b=1}^B$ and $G(\cdot)$ parametered by $\bm{w}_1$,
        \STATE Generate $\{\bm{s}_2^b\}_{b=1}^B$ with $\{\bm{z}^b, y^b\}_{b=1}^B$ and $G(\cdot)$ parametered by $\bm{w}_2$,
        \STATE Compute loss $\mathcal{L}_{dmd}$ by using $\{\bm{s}_1^b, y^b\}_{b=1}^B$, $\{\bm{s}_2^b, y^b\}_{b=1}^B$, $\Phi$ and Eq.~(\ref{L_dmd:}),
            \STATE $\bm{\theta} \leftarrow \bm{\theta}-\frac{\eta_d}{B}\sum_{b\in [B]}\nabla_{\bm{\theta}} \mathcal{L}_{dmd}$. 
    \ENDFOR
    % \STATE Sets $\bm{x}_{t+1}^{(i)}=\bm{x}_{t,I}^{(i)}$.
    \STATE {\bfseries Output:} $\bm{\theta}$, $\bm{w}_1$, $\bm{w}_2$.
\end{algorithmic}
\end{algorithm}

\begin{algorithm}[h]
  \caption{Generator\_Update($\{\bm{\theta}_{i}\}_{i\in [N]}$, $\bm{\theta}$,  $\bm{w}$, $\{\bm{z}^b, y^b\}_{b=1}^B$, $\Phi$, $I_g$, $\eta_g$,  $b_1$, $b_2$, $\beta_{tran}$, $\beta_{div}$, $\beta_{cd}$)}
  \label{generator_update:}
\begin{algorithmic}[1] 
    \STATE Set $\bm{m}=0$ and $\bm{v}=0$.
    \FOR{$e_g = 0,1,\ldots, I_g-1$}
        \STATE Generate $\{\bm{s}^b\}_{b=1}^B$ with $\{\bm{z}^b, y^b\}_{b=1}^B$ and $G(\cdot)$ parametered by $\bm{w}$,
        \STATE Compute generator loss $\mathcal{L}_{gen}=\mathcal{L}_{fid}+\beta_{tran}\cdot\mathcal{L}_{tran}+\beta_{div} \cdot\mathcal{L}_{div} +\beta_{cd} \cdot\mathcal{L}_{cd}$ by using $\{\bm{s}^b, y^b\}_{b=1}^B$, $\Phi$ and Eq.~(\ref{L_fidelity:})-(\ref{L_cd:}),
        \STATE $\bm{g} = \frac{1}{B}\sum_{b\in [B]}\nabla_{\bm{w}} \mathcal{L}_{gen}$, %(\bm{s}^b, y^b; \bm{w})
        \STATE $\bm{m} \leftarrow b_1 \cdot \bm{m} + (1-b_1)\cdot\bm{g}$, $\bm{v} \leftarrow b_2\cdot\bm{v} + (1-b_2)\cdot \bm{g}^2$,
        \STATE $\hat{\bm{m}}\leftarrow \bm{m}/ (1-b_1)$, $\hat{\bm{v}}\leftarrow \bm{v}/ (1-b_2)$,
        \STATE  $\bm{w}\leftarrow \bm{w}-\eta_g \hat{\bm{m}} / (\sqrt{\hat{\bm{v}}} + 10^{-8})$.
    \ENDFOR
    % \STATE Sets $\bm{x}_{t+1}^{(i)}=\bm{x}_{t,I}^{(i)}$.
    \STATE {\bfseries Output:} $\bm{w}$.
\end{algorithmic}
\end{algorithm}

\section{Datasets}
\label{sup_sec:dataset}
In Table~\ref{Dataset_table:}, we provide information about the image size, the number of classes~(\#class), and the number of training/test samples~(\#train/\#test) of the datasets used in our experiment. 
Of note, we have used Resize$()$ in PyTorch to scale up and down the image size of FMNIST and FOOD101, respectively.

\begin{table}[htbp]
  \centering
  \caption{Statistics of the datasets used in our experiments.}
    \begin{tabular}{c|c|c|c|c}
    \toprule
    Dataset & Image size & \#class & \#train & \#test \\
    \midrule
    FMNIST & 1*32*32 & 10    & 60000 & 10000 \\
    \midrule
    SVHN  & \multirow{4}[6]{*}{3*32*32} & 10    & 50000 & 10000 \\
\cmidrule{1-1}\cmidrule{3-5}    CIFAR-10 &       & 10    & 50000 & 10000 \\
\cmidrule{1-1}\cmidrule{3-5}    CINIC-10 &       & 10   & 90000 & 90000 \\
\cmidrule{1-1}\cmidrule{3-5}    CIFAR-100 &       & 100   & 73257 & 26032 \\
    \midrule
    Tiny-ImageNet  & \multirow{2}[6]{*}{3*64*64} & 200    & 100000 & 10000 \\
\cmidrule{1-1}\cmidrule{3-5}    FOOD101 &       & 101    & 75750 & 25250 \\
    \bottomrule
    \end{tabular}%
  \label{Dataset_table:}%
\end{table}%

\section{Budget Distribution}
\label{Budget_Distribution_app:}
In a real-world FL scenario, each client $i$ holds an on-demand local model $f_i$ parameterized by $\bm{\theta}_i$.  
Due to the difference in resource budgets, the model capacity of each client may vary, i.e., $|\bm{\theta}_i| \neq |\bm{\theta}_j|, \exists i\neq j, i,j \in [N]$. 
In our work, we define a confined width capability $R_i \in (0, 1]$ according to the resource budget of client $i$, which is the proportion of nodes extracted from each layer in the global model $f$.

In order to simulate model-heterogeneous scenarios and investigate the effect of the different model heterogeneity distributions. 
We formulate exponentially distributed budgets for a given $N$: $R_i = [\frac{1}{2}]^{\min\{\sigma, \lfloor\frac{\rho \cdot i}{N}\rfloor\}} ( i \in [N])$, where $\sigma$ and $\rho$ are both positive integers.
% , and control the lower bound and distribution of clients' width capacity, respectively.
Specifically, $\sigma$ controls the capacity of the smallest model, i.e., $\left[\frac{1}{2}\right]^\sigma$-width w.r.t. global model. 
Also, $\rho$ manipulates client budget distribution. The larger the $\rho$ value, the higher the proportion of models with the smallest capacity. 

We now present client budget distributions under several different settings about parameters $\sigma$ and $\rho$.
For example, we fix $\sigma=2$, that is, the smallest model is $\frac{1}{4}$-width w.r.t. global model. 
Assuming $10$ clients participate in federated learning, we have: For any $i \in [10]$,
\begin{align}
    R_i = [\frac{1}{2}]^{\min\{2, \lfloor\frac{\rho \cdot i}{10}\rfloor\}}.
\end{align}
\begin{itemize}
    \item If we set $\rho=0$, $\{R_i\}_{i\in[10]}=\{1, 1, 1, 1, 1, 1, 1, 1, 1, 1\}=\{1\}_{i=1}^{10}$;
    \item If we set $\rho=0$, $\{R_i\}_{i\in[50]}=\{1\}_{i=1}^{50}$;
    \item If we set $\rho=0$, $\{R_i\}_{i\in[100]}=\{1\}_{i=1}^{100}$;
    \item If we set $\rho=2$, $\{R_i\}_{i\in[10]}=\{1, 1, 1, 1, \frac{1}{2}, \frac{1}{2}, \frac{1}{2}, \frac{1}{2}, \frac{1}{2}, \frac{1}{4}\}$;
    \item If we set $\rho=3$, $\{R_i\}_{i\in[10]}=\{1, 1, 1, \frac{1}{2}, \frac{1}{2}, \frac{1}{2}, \frac{1}{4}, \frac{1}{4}, \frac{1}{4}, \frac{1}{4}\}$;
    \item If we set $\rho=4$, $\{R_i\}_{i\in[10]}=\{1, 1, \frac{1}{2}, \frac{1}{2}, \frac{1}{4}, \frac{1}{4}, \frac{1}{4}, \frac{1}{4}, \frac{1}{4}, \frac{1}{4}\}$. 
\end{itemize}
Note that if $\rho\geq4N$, each client holds $[\frac{1}{2}]^\sigma$-width w.r.t. global model. 
More budget distributions can refer to the literature~\cite{diao2020heterofl, hong2022efficient, alam2022fedrolex}.

\section{Complete Experimental Setup}
\label{Com_Exp_Setup:}
In this section, for the convenience of the reader, we review the experimental setup for all the implemented methods on top of FMNIST, CIFAR-10, SVHN, CINIC-10, CIFAR-100, Tiny-ImageNet, and FOOD101.

\textbf{Datasets and Baselines.} 
We evaluate different methods with multiple image classification task-related datasets, namely Fashion-MNIST~\cite{xiao2017fashion}~(FMNIST in short), CIFAR-10~\cite{Krizhevsky2009Learning}, SVHN~\cite{Netzer2011Reading}, CINIC-10~\cite{darlow2018cinic}, CIFAR-100~\cite{Krizhevsky2009Learning}, Tiny-ImageNet~\footnote{http://cs231n.stanford.edu/tiny-imagenet-200.zip} and FOOD101~\cite{bossard2014food}.
We detail the said datasets in Appendix~\ref{sup_sec:dataset}.
To gauge the effectiveness of DFDG with dual generators, we design a baseline~(marked as DFAD), which considers DFKD with a single generator based on Eq.~(\ref{L_fidelity:})-(\ref{L_div:}).
Also, we compare DFDG against FedAvg~\cite{McMahan2017Communication}, FedFTG~\cite{Zhang2022Fine} and DENSE~\cite{Zhang2022DENSE}, which are the most relevant methods to our work. 

\textbf{Configurations.} 
Unless otherwise stated, all experiments are performed on a centralized network with $N=10$ clients.
To mimic data heterogeneity across clients, as in previous works~\cite{yurochkin2019bayesian, luo2023gradma}, we use Dirichlet process $Dir(\omega)$ to partition the training set for each dataset, thereby allocating local training data for each client.  
It is worth noting that $\omega$ is the concentration parameter and smaller $\omega$ corresponds to stronger data heterogeneity. We set $\omega=0.5$ as default.
To simulate model-heterogeneous scenarios, we formulate exponentially distributed resource budgets for a given $N$: $R_i = [\frac{1}{2}]^{\min\{\sigma, \lfloor\frac{\rho \cdot i}{N}\rfloor\}} (i \in [N])$, where $\sigma$ and $\rho$ are both positive integers. 
See Appendix~\ref{Budget_Distribution_app:} for more details. 
We set both $\sigma$ and $\rho$ to $0$ by default.
%~\ref{Budget_Distribution_app:}
Unless otherwise specified, we set $\beta_{tran}$, $\beta_{div}$ and $\beta_{cd}$ to $1$ in \textit{dual-generator training}.
\textbf{Of note, the reader is referred to Algorithm~\ref{alg_all:} for the notations we use thereafter.}

On the server, 
we take the values  $20$ and $2$ for $I_g$ and $I_d$ respectively.
Also, SGD and Adam are applied to optimize the global model and generators, respectively.
The learning rate $\eta_d$ for SGD is $0.01$.
We set $\eta_g=0.0002$, $b_1=0.5$ and $b_2=0.999$ for Adam. 
Besides, for FMNIST~(CIFAR-10, SVHN, CINIC-10, CIFAR-100, Tiny-ImageNet, FOOD101), we set $I=500$~($800$, $800$, $800$, $800$, $800$, $800$).

For each client, we select $\eta_c$ from $\{0.001, 0.01, 0.1\}$ and the best one is picked. 
For fairness, the popular SGD procedure is employed to perform local update steps for each client.
Meanwhile, for FMNIST~(CIFAR-10, SVHN, CINIC-10, CIFAR-100, Tiny-ImageNet, FOOD101), we set $I_c=50$~($100$, $100$, $100$, $200$, $400$, $400$). 
Note that $I_c$ represents epochs during training.
For all update steps, we set batch size $B$ to $64$.

To verify the superiority of DFDG, we execute substantial comparative experiments in terms of data and model heterogeneity for one-shot FL, respectively.

\begin{figure*}[h]\captionsetup[subfigure]{font=scriptsize}
  \centering
  \begin{subfigure}{0.3\linewidth}
    \centering
    \includegraphics[width=1.0\linewidth]{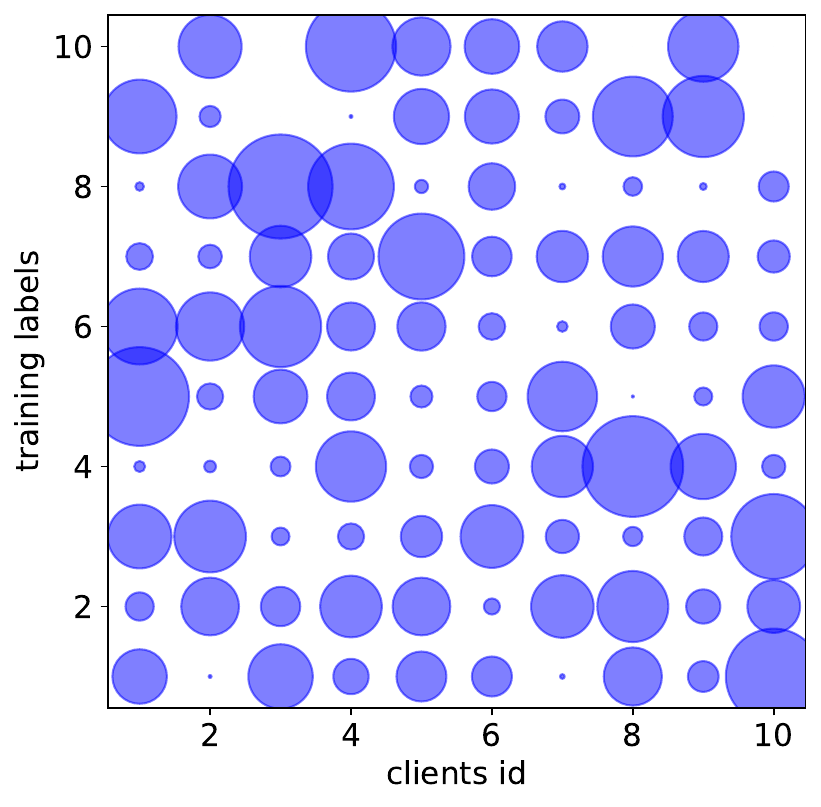}
        \caption{FMNIST, $\omega=1.0$}
  \end{subfigure}
  \centering
  \begin{subfigure}{0.3\linewidth}
    \centering
    \includegraphics[width=1.0\linewidth]{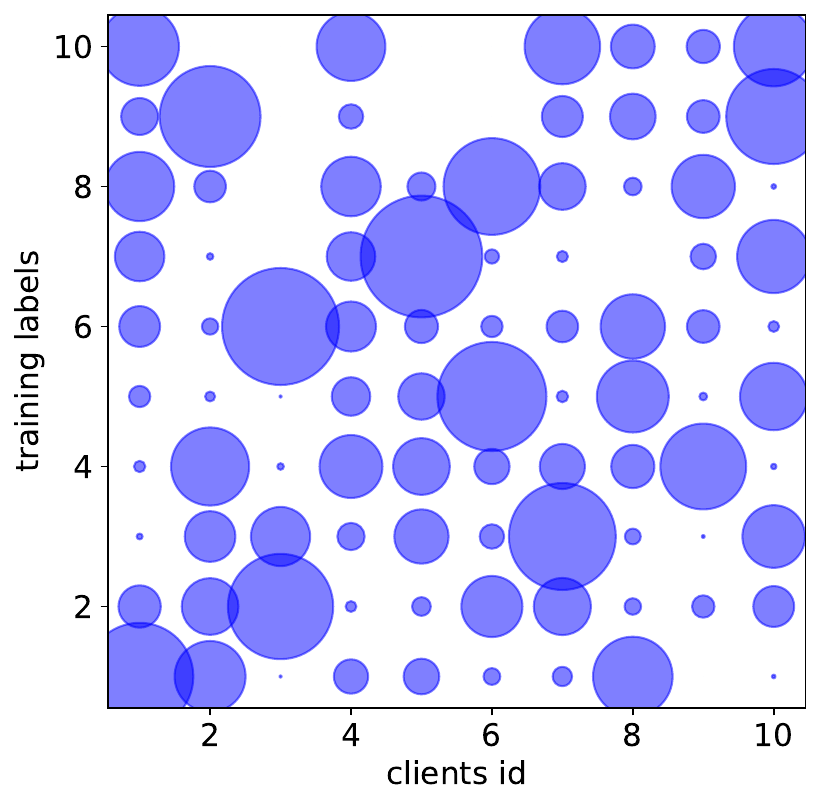}
        \caption{FMNIST, $\omega=0.5$}
  \end{subfigure}
  \centering
  \begin{subfigure}{0.3\linewidth}
    \centering
    \includegraphics[width=1.0\linewidth]{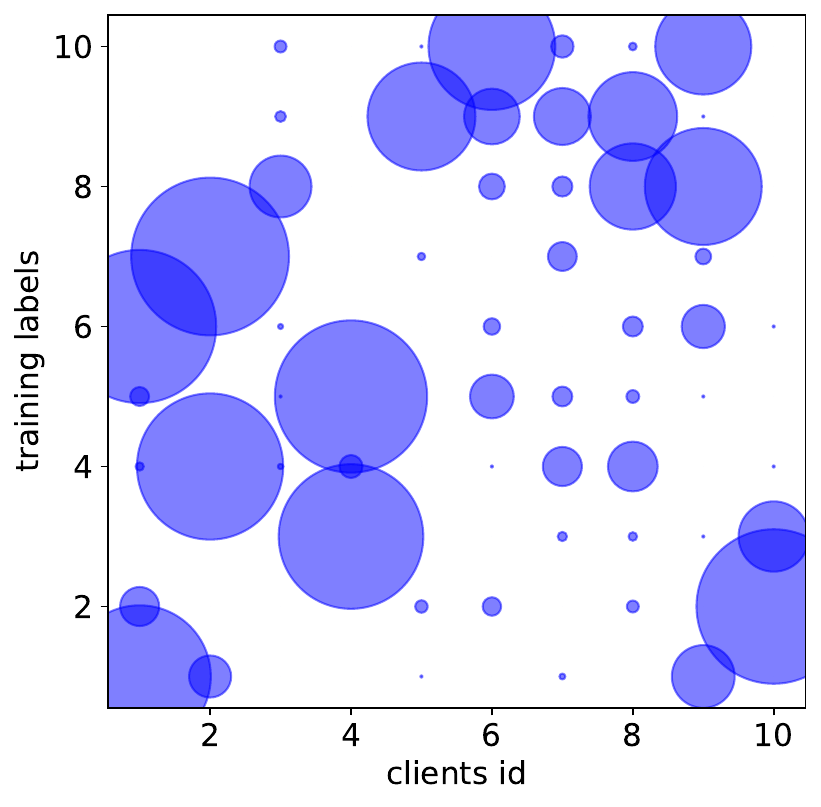}
        \caption{FMNIST, $\omega=0.1$}
  \end{subfigure}
  \centering
  \begin{subfigure}{0.3\linewidth}
    \centering
    \includegraphics[width=1.0\linewidth]{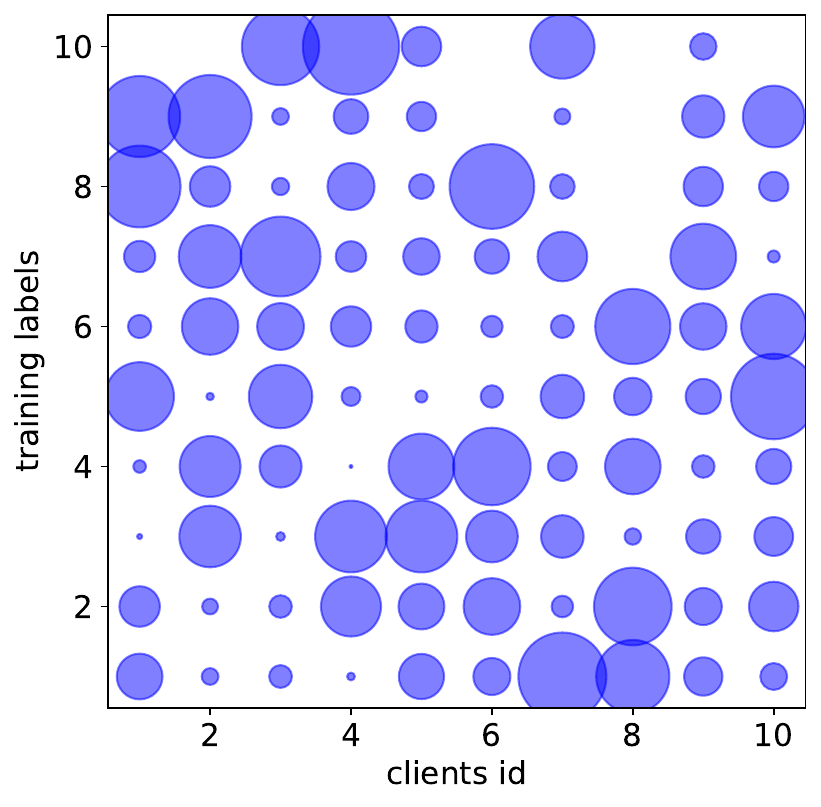}
        \caption{CIFAR-10, $\omega=1.0$}
  \end{subfigure}
  \centering
  \begin{subfigure}{0.3\linewidth}
    \centering
    \includegraphics[width=1.0\linewidth]{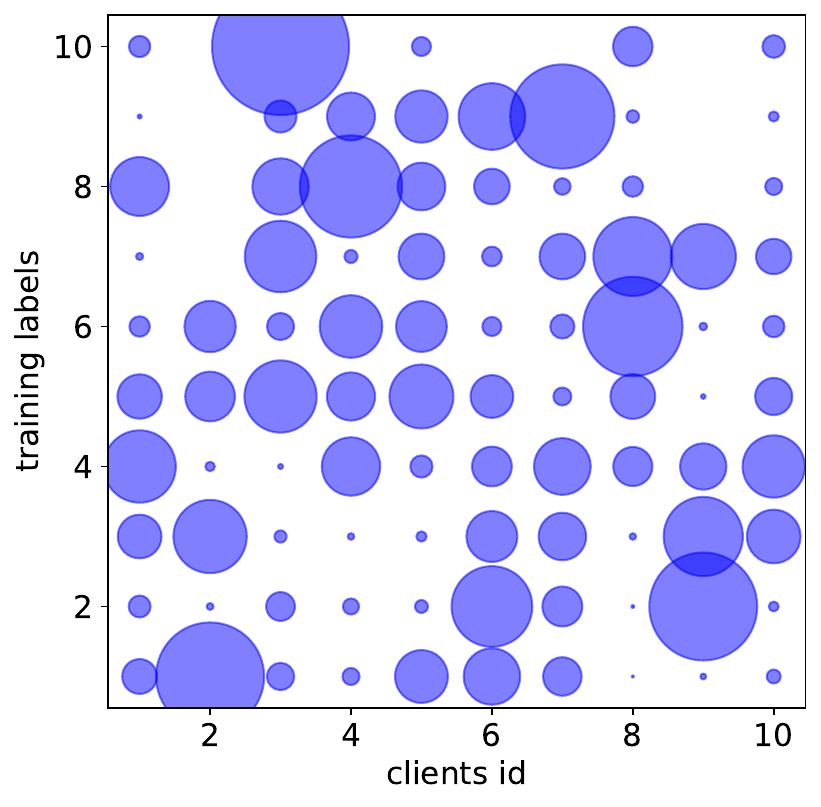}
        \caption{CIFAR-10, $\omega=0.5$}
  \end{subfigure}
  \centering
  \begin{subfigure}{0.3\linewidth}
    \centering
    \includegraphics[width=1.0\linewidth]{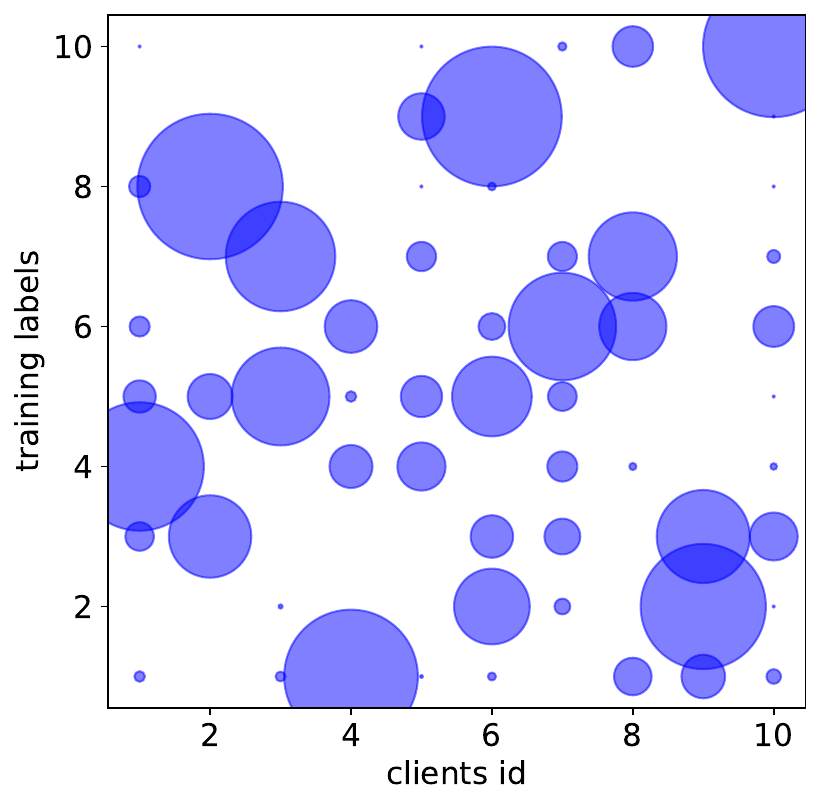}
        \caption{CIFAR-10, $\omega=0.1$}
  \end{subfigure}
  \centering
  \begin{subfigure}{0.3\linewidth}
    \centering
    \includegraphics[width=1.0\linewidth]{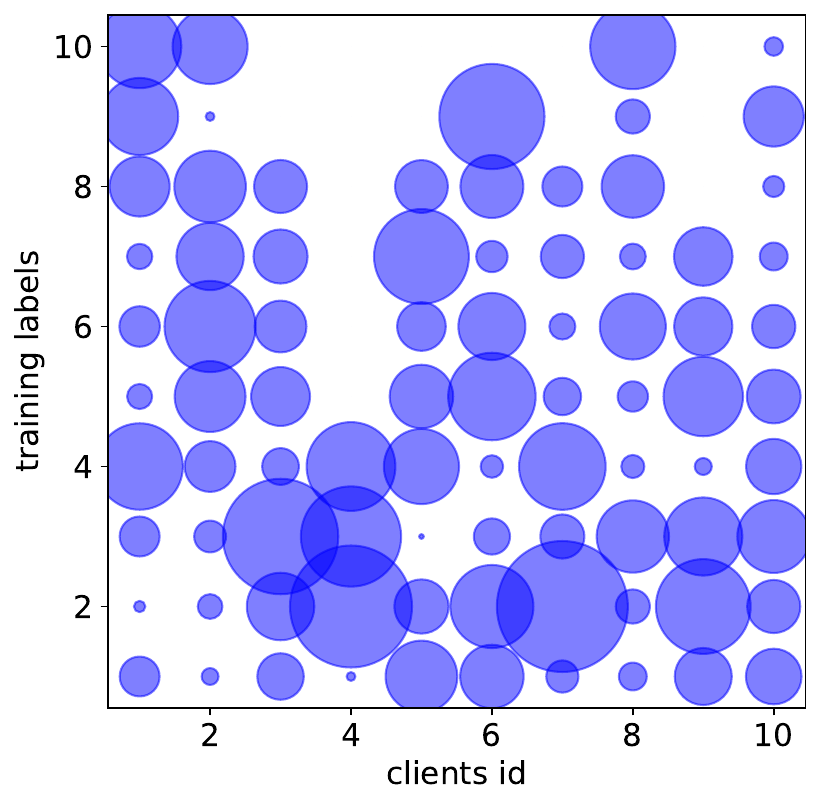}
        \caption{SVHN, $\omega=1.0$}
  \end{subfigure}
  \centering
  \begin{subfigure}{0.3\linewidth}
    \centering
    \includegraphics[width=1.0\linewidth]{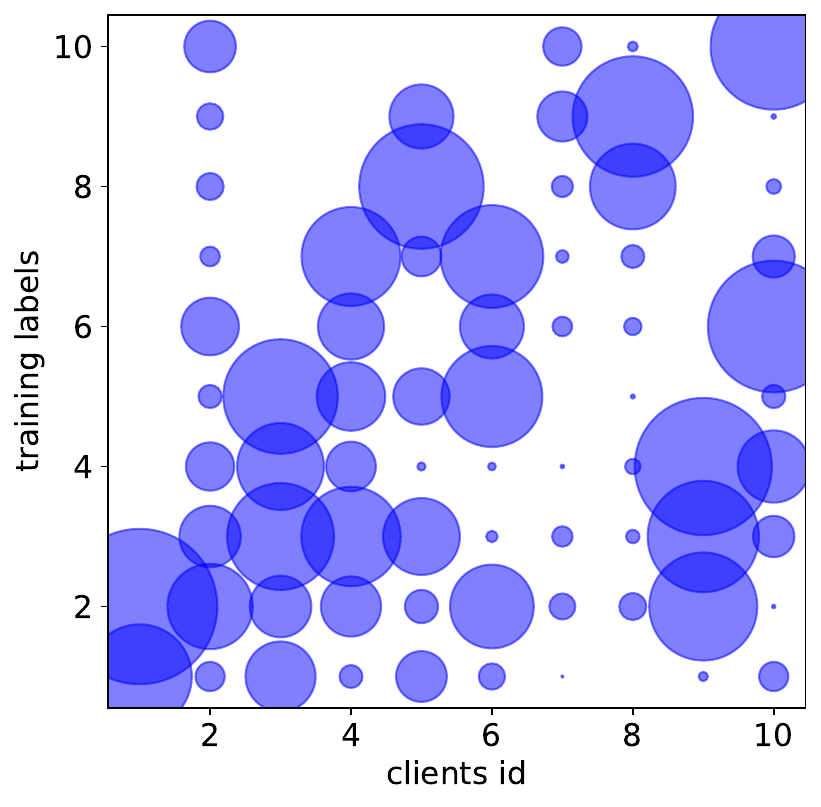}
        \caption{SVHN, $\omega=0.5$}
  \end{subfigure}
  \centering
  \begin{subfigure}{0.3\linewidth}
    \centering
    \includegraphics[width=1.0\linewidth]{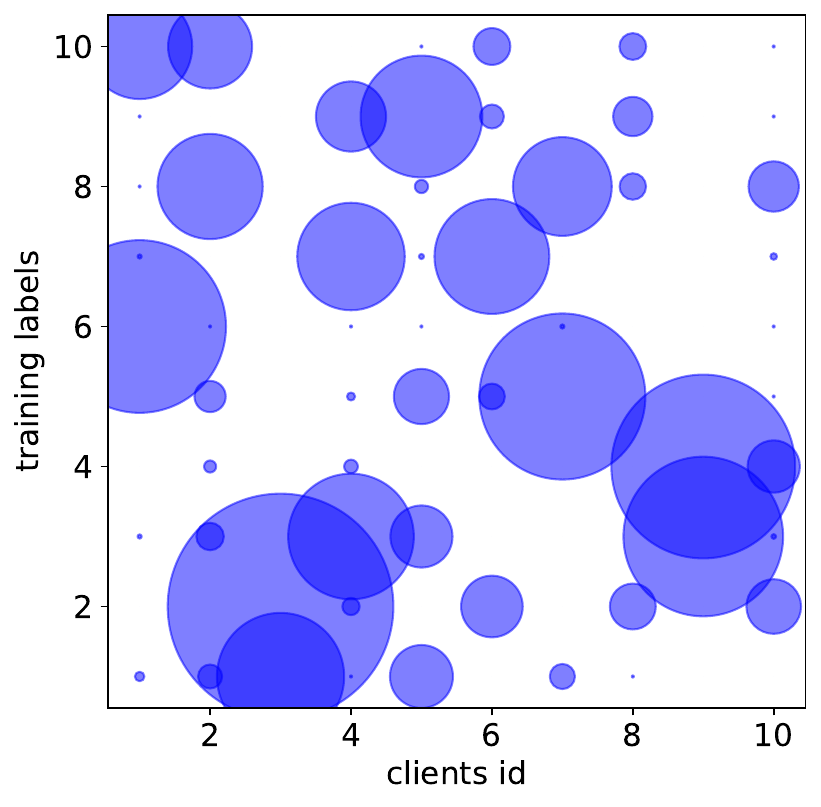}
        \caption{SVHN, $\omega=0.1$}
  \end{subfigure}
  \centering
  \begin{subfigure}{0.3\linewidth}
    \centering
    \includegraphics[width=1.0\linewidth]{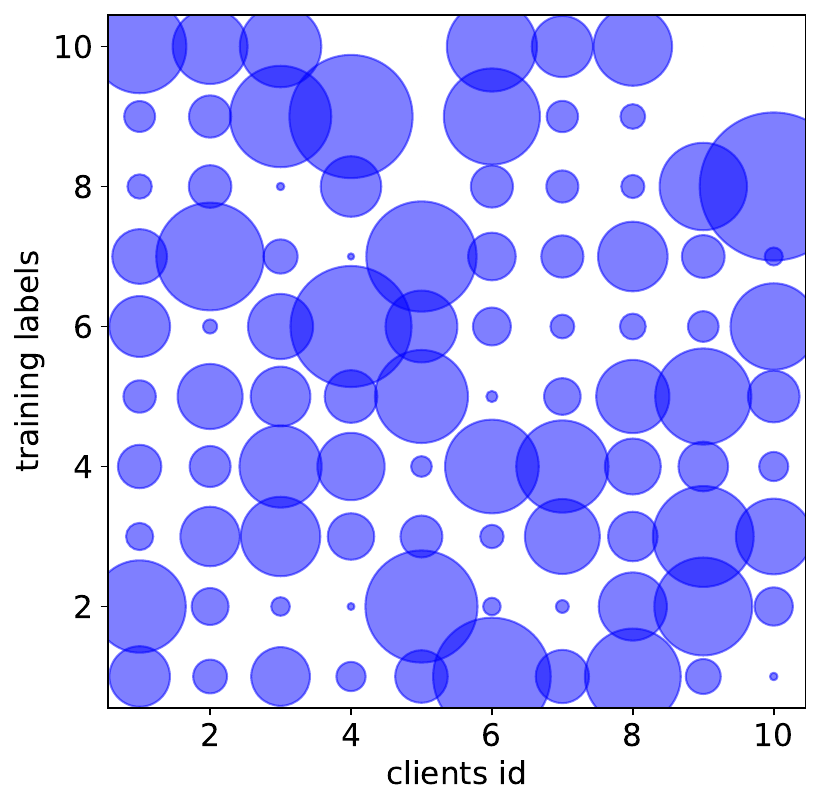}
        \caption{CINIC-10, $\omega=1.0$}
  \end{subfigure}
  \centering
  \begin{subfigure}{0.3\linewidth}
    \centering
    \includegraphics[width=1.0\linewidth]{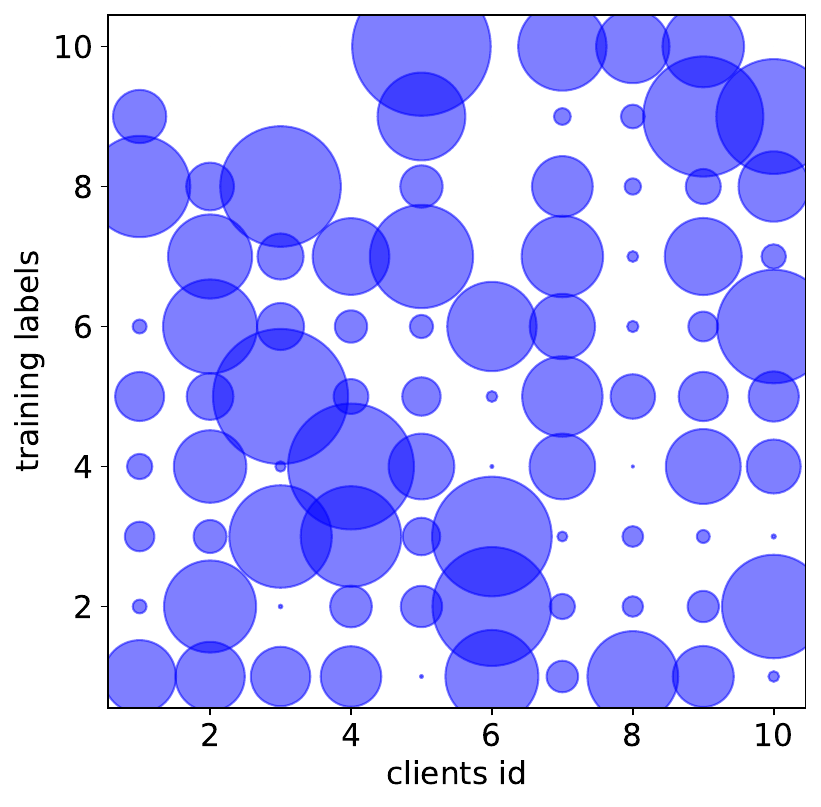}
        \caption{CINIC-10, $\omega=0.5$}
  \end{subfigure}
  \centering
  \begin{subfigure}{0.3\linewidth}
    \centering
    \includegraphics[width=1.0\linewidth]{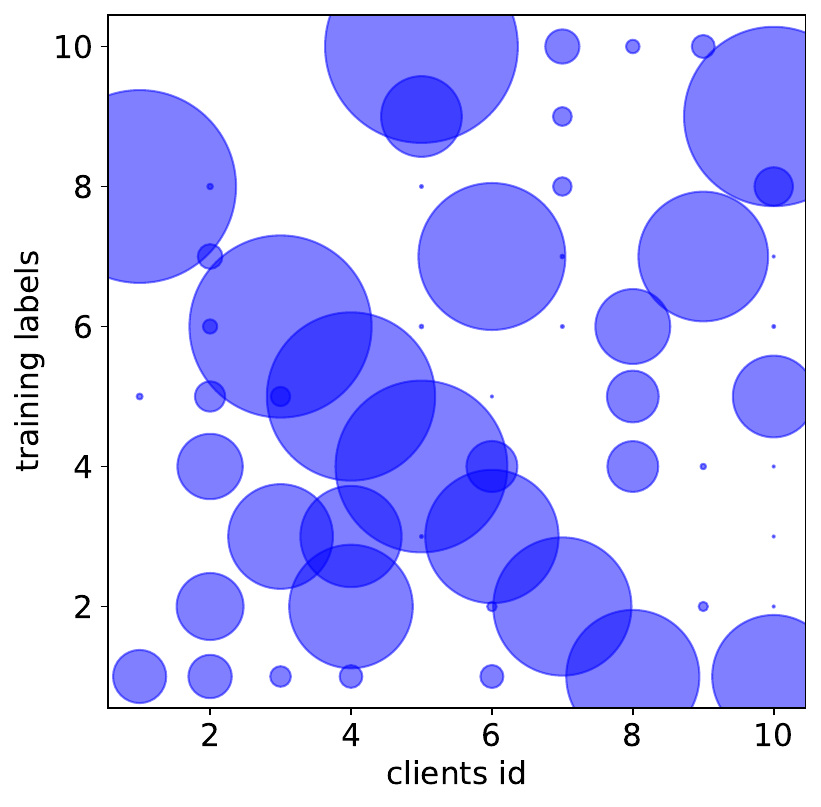}
        \caption{CINIC-10, $\omega=0.1$}
  \end{subfigure}
  \caption{Data heterogeneity among clients is visualized on four datasets~(FMNIST, CIFAR-10, SVHN and CINIC-10), where the $x$-axis represents the clients id, the $y$-axis represents the class labels on the training set, and the size of scattered points represents the number of training samples with available labels for that client.}
  \label{data_par_sum_appendix:}
\end{figure*}

Firstly, we investigate the performance of DFDG against that of baselines w.r.t. different levels of data heterogeneity in one-shot FL with homogeneous models~(i.e., $\rho=0$).
To be specific, we conduct experiments on varying levels of $\omega$, that is, $\omega$ is selected from $\{0.1, 0.5, 1.0\}$.
A visualization of the data partitions for the four datasets at varying $\omega$ values can be found in Fig.~\ref{data_par_sum_appendix:}.
Meanwhile, for FMNIST, a four-layer convolutional neural network with BatchNorm is implemented for each client. 
For CIFAR-10, SVHN and CINIC-10, each client implements a ResNet18~\cite{he2016deep} architecture.

Then, we look into the performance of DFDG against that of baselines in terms of different levels of model heterogeneity distribution in one-shot FL. % on FMNIST, CIFAR-10, SVHN and CINIC-10
Concretely, we fix $\sigma=2$ and select $\rho$ from $\{2, 3, 4\}$. 
We then extract the local model for each client from the global model that matches its computational budget based on $R_i = [\frac{1}{2}]^{\min\{2, \lfloor\frac{\rho \cdot i}{10}\rfloor\}}$, $i\in [10]$, $\rho \in \{2, 3, 4\}$.
Also, we set concentration parameter $\omega=0.5$. 
Additionally, for FMNIST, we equip the server with a four-layer convolutional neural network with BatchNorm as the global model, and extract suitable sub-models from the global model according to the width capability $R_i$ of each client. 
For CIFAR-10, SVHN and CINIC-10, a ResNet18~\cite{he2016deep} architecture is equipped on the server as the global model, and appropriate sub-models are extracted from the global model according to the width capability $R_i$ of each client.

\begin{figure*}[h]\captionsetup[subfigure]{font=scriptsize}
  \centering
  \begin{subfigure}{0.3\linewidth}
    \centering
    \includegraphics[width=1.0\linewidth]{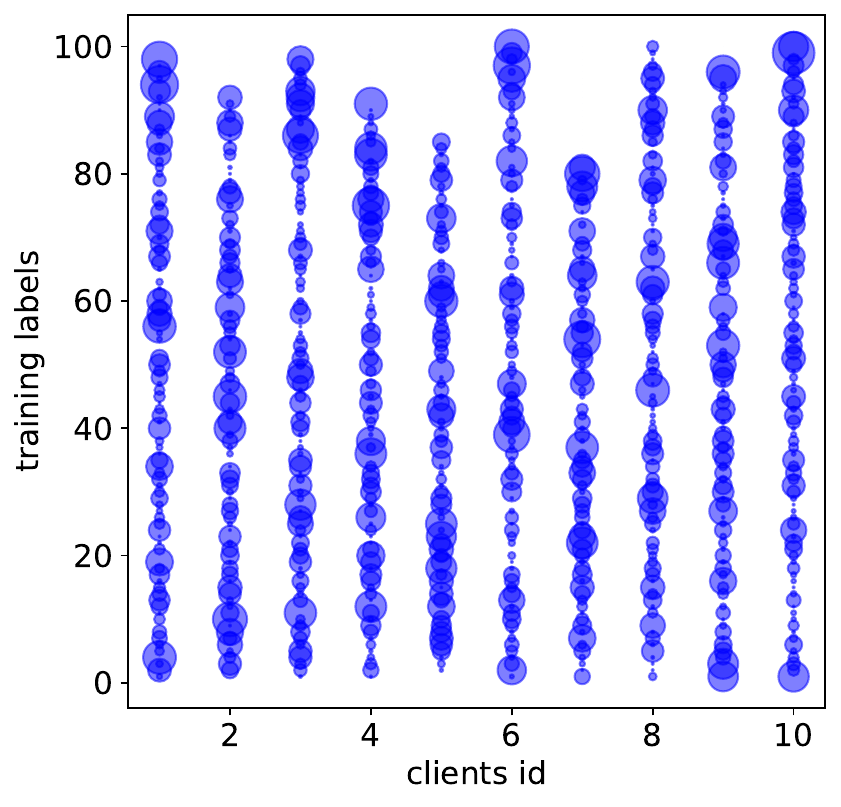}
        \caption{CIFAR-100, $N=10$}
  \end{subfigure}
  \centering
  \begin{subfigure}{0.3\linewidth}
    \centering
    \includegraphics[width=1.0\linewidth]{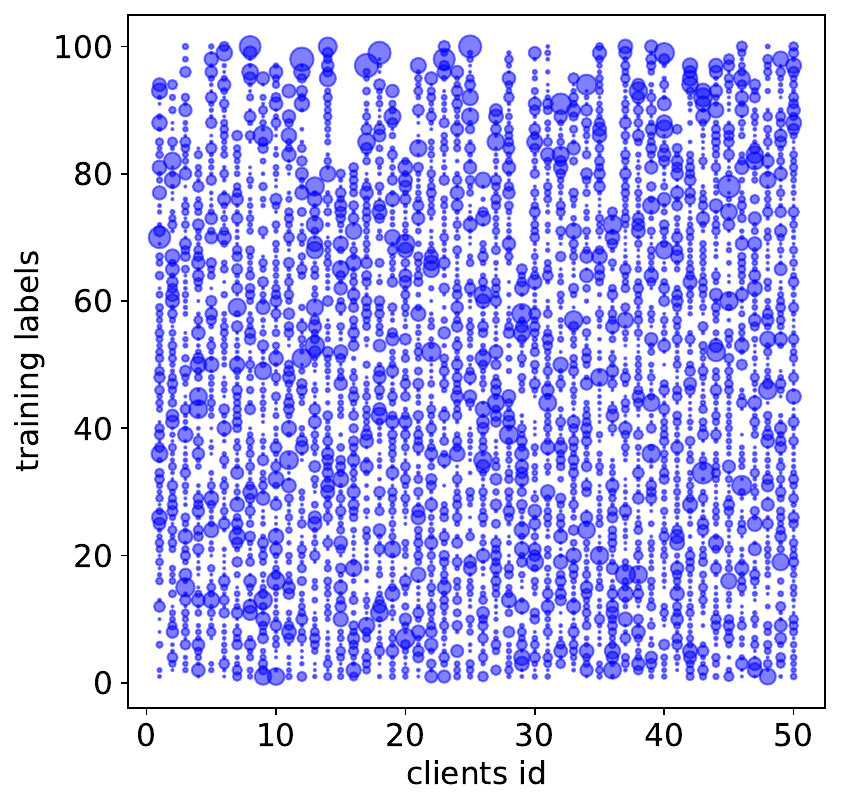}
        \caption{CIFAR-100, $N=50$}
  \end{subfigure}
  \centering
  \begin{subfigure}{0.3\linewidth}
    \centering
    \includegraphics[width=1.0\linewidth]{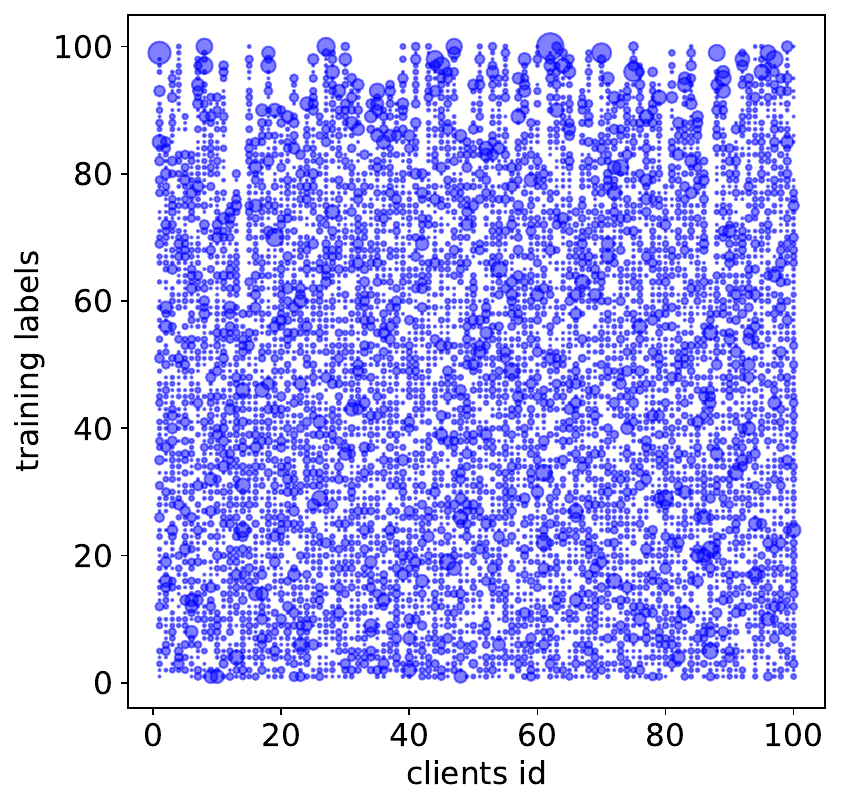}
        \caption{CIFAR-100, $N=100$}
  \end{subfigure}
  \centering
  \begin{subfigure}{0.3\linewidth}
    \centering
    \includegraphics[width=1.0\linewidth]{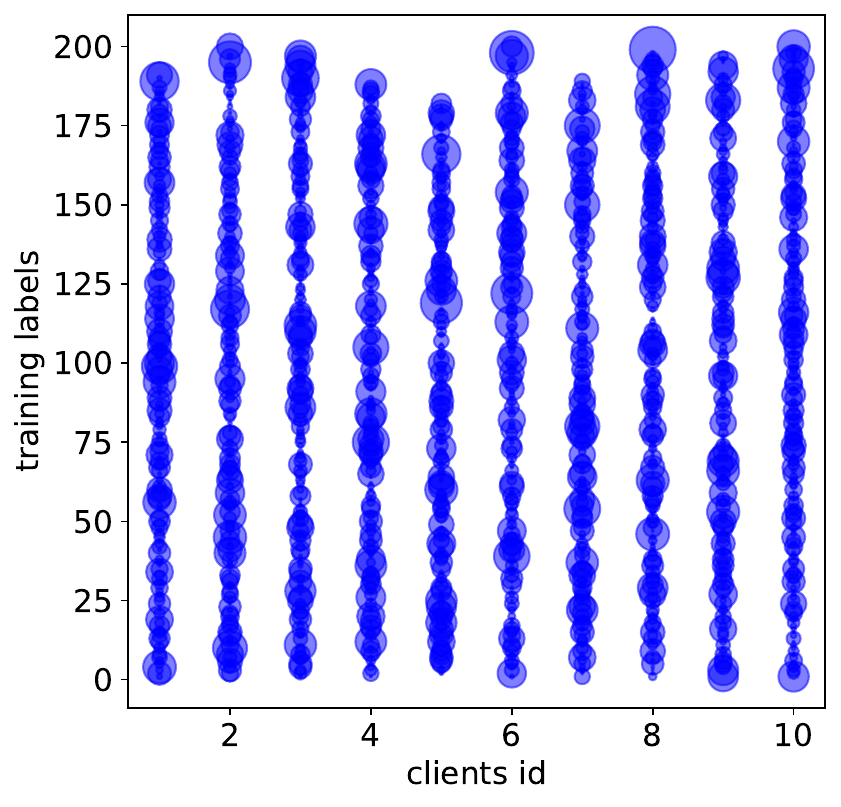}
        \caption{Tiny-ImageNet, $N=10$}
  \end{subfigure}
  \centering
  \begin{subfigure}{0.3\linewidth}
    \centering
    \includegraphics[width=1.0\linewidth]{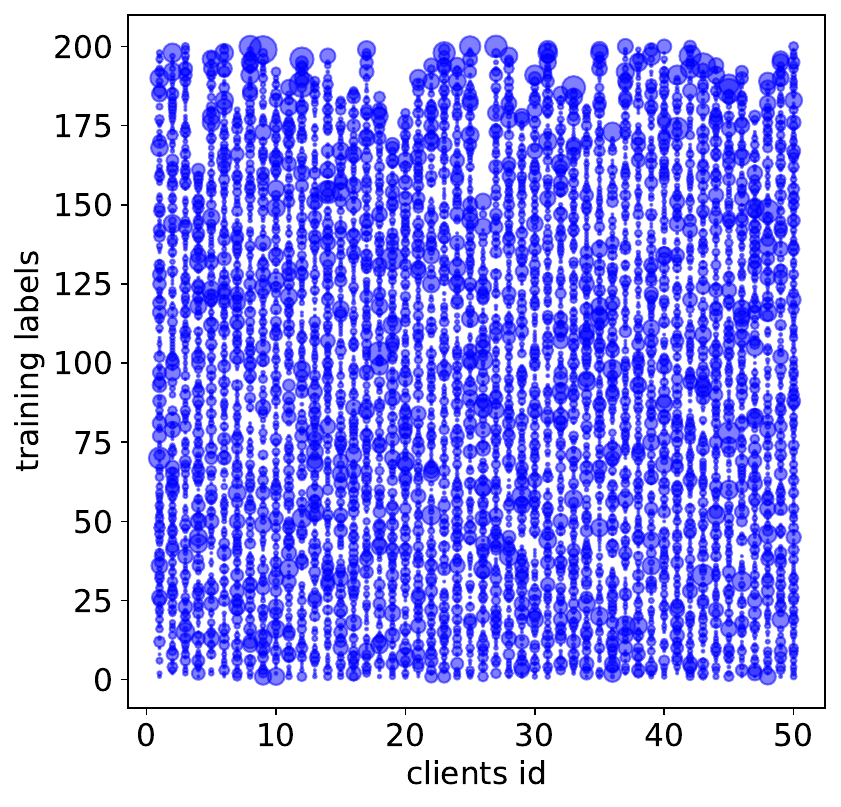}
        \caption{Tiny-ImageNet, $N=50$}
  \end{subfigure}
  \centering
  \begin{subfigure}{0.3\linewidth}
    \centering
    \includegraphics[width=1.0\linewidth]{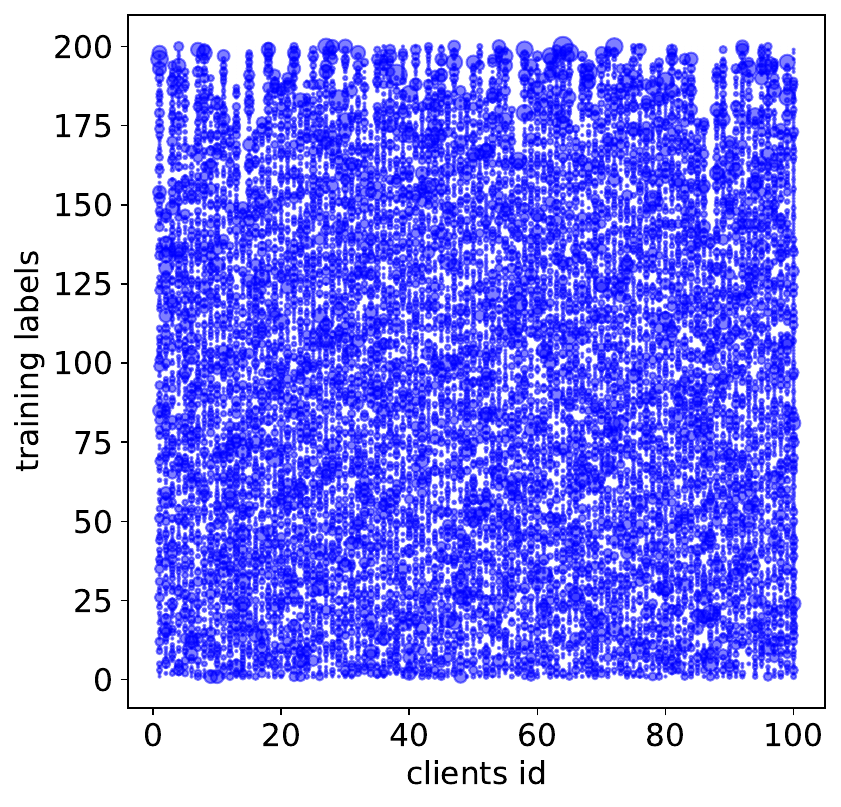}
        \caption{Tiny-ImageNet, $N=100$}
  \end{subfigure}
  \centering
  \begin{subfigure}{0.3\linewidth}
    \centering
    \includegraphics[width=1.0\linewidth]{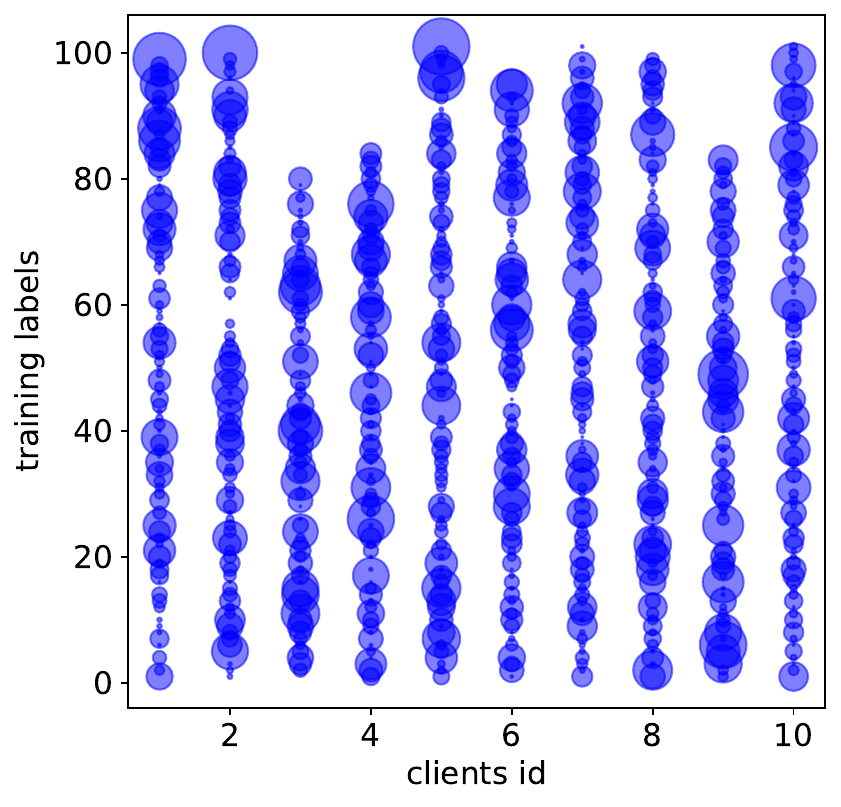}
        \caption{FOOD101, $N=10$}
  \end{subfigure}
  \centering
  \begin{subfigure}{0.3\linewidth}
    \centering
    \includegraphics[width=1.0\linewidth]{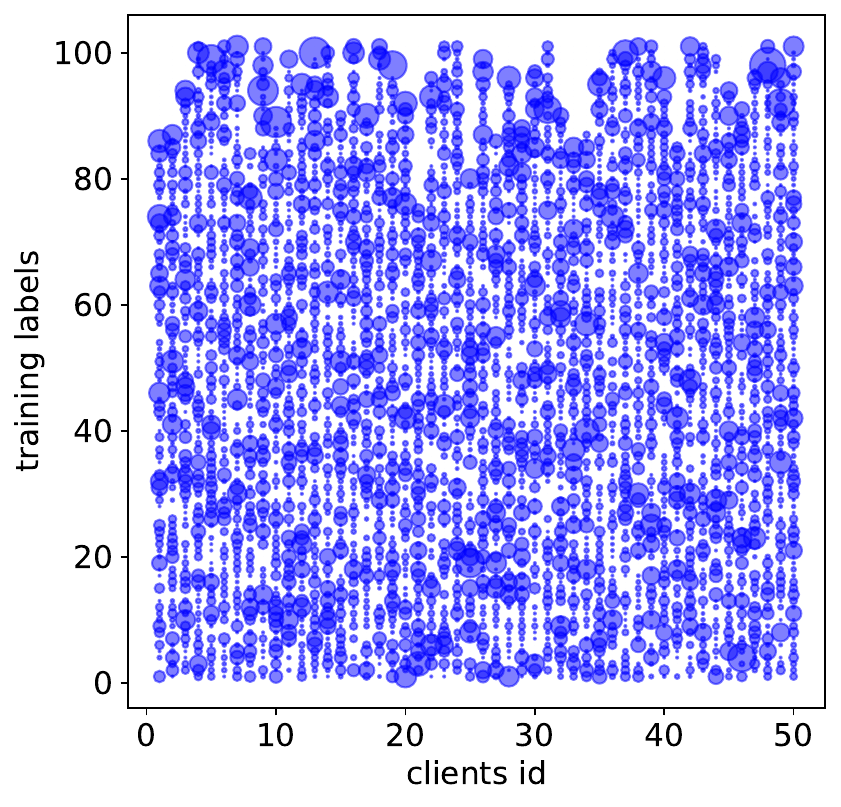}
        \caption{FOOD101, $N=50$}
  \end{subfigure}
  \centering
  \begin{subfigure}{0.3\linewidth}
    \centering
    \includegraphics[width=1.0\linewidth]{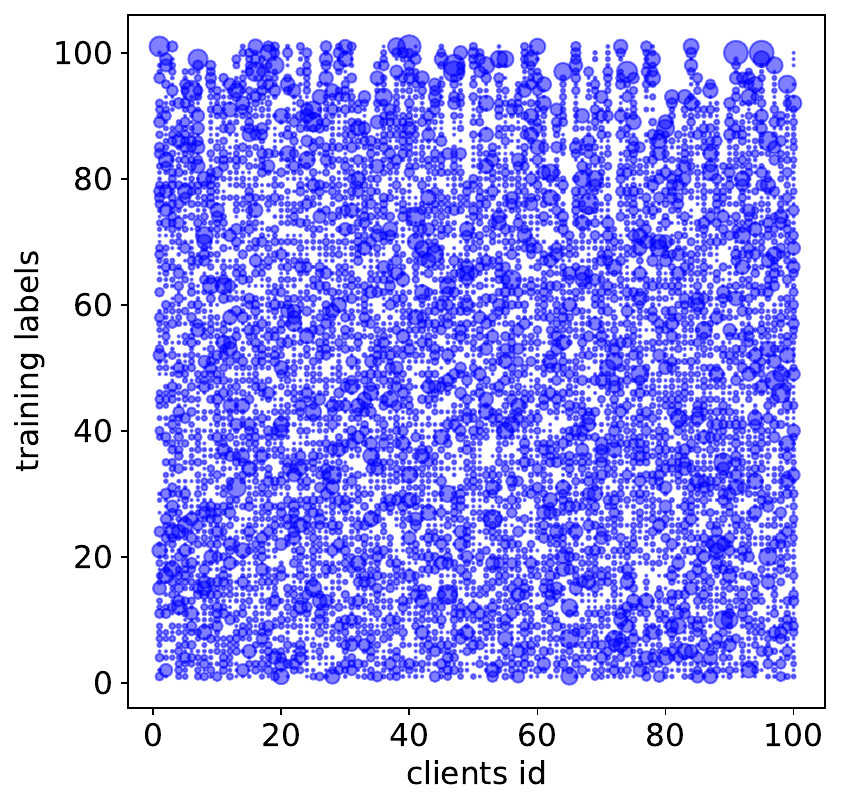}
        \caption{FOOD101, $N=100$}
  \end{subfigure}
  \caption{Data heterogeneity ($\omega=0.5$) among clients over varying $N$ is visualized on three datasets~(CIFAR-100, Tiny-ImageNet and FOOD101), where the $x$-axis represents the clients id, the $y$-axis represents the class labels on the training set, and the size of scattered points represents the number of training samples with available labels for that client.}
  \label{data_par_N_appendix:}
\end{figure*}

For difficult image classification tasks~(CIAFR-100, Tiny-ImageNet and FOOD101), we delve into the performance of DFDG against that of baselines w.r.t. different the number of clients in one-shot FL with homogeneous setting~(i.e., $\rho=0$). We set concentration parameter $\omega=0.5$ and select $N$ from $\{10, 50, 100\}$.
A visualization of the data partitions for the three datasets at varying $N$ values can be found in Fig.~\ref{data_par_N_appendix:}.
For CIFAR-100~(Tiny-ImageNet, FOOD101), each client implements a ResNet20~(ResNet34, ResNet34)~\cite{he2016deep} architecture. 

In addition, it is crucial to equip each dataset with a suitable generator on the server. 
The details of generator frameworks are shown in Tables~\ref{gen_FMNIST_tab:}-\ref{gen_tinyimage_food_tab:}.
Moreover, for FMNIST, we set the random noise dimension $d=100$. For CIFAR-10, SVHN and CINIC-10, we set the random noise dimension $d=200$. For difficult image classification tasks, i.e., CIFAR-100, Tiny-ImageNet and FOOD101, we set the random noise dimension $d=400$.
All baselines leverage the same setting as ours.

\begin{table}[htbp]
  \centering
  \caption{Generator for FMNIST}
    \begin{tabular}{c}
    \toprule
    $\bm{z}\in \mathbf{R}^d \sim \mathcal{N}(\bm{0}, \bm{I})$, $y\in[C]$ \\
    \midrule
    $\bm{h}=o(\bm{z}, y)\rightarrow {d}$ \\
    \midrule
    Reshape($\bm{h}$)$\rightarrow{ d\times1\times1}$ \\
    \midrule
    Relu(ConvTransposed2d($d$, $512$)) $\rightarrow{B\times 512\times4\times4}$ \\
    \midrule
    Relu(BN(ConvTransposed2d($512$, $256$))) $\rightarrow{ 256\times8\times8}$ \\
    \midrule
    Relu(BN(ConvTransposed2d($256$, $128$))) $\rightarrow{128\times16\times16}$ \\
    \midrule
    Tanh(ConvTransposed2d($128$, $1$)) $\rightarrow{ 1\times32\times32}$ \\
    \bottomrule
    \end{tabular}%
  \label{gen_FMNIST_tab:}%
\end{table}%

\begin{table}[htbp]
  \centering
  \caption{Generator for CIFAR-10, SVHN, CINIC-10 and CIFAR-100}
    \begin{tabular}{c}
    \toprule
    $\bm{z}\in \mathbf{R}^d \sim \mathcal{N}(\bm{0}, \bm{I})$, $y\in[C]$ \\
    \midrule
    $\bm{h}=o(\bm{z}, y)\rightarrow {d}$ \\
    \midrule
    Reshape($\bm{h}$)$\rightarrow{ d\times1\times1}$ \\
    \midrule
    Relu(ConvTransposed2d($d$, $512$)) $\rightarrow{B\times 512\times4\times4}$ \\
    \midrule
    Relu(BN(ConvTransposed2d($512$, $256$))) $\rightarrow{ 256\times8\times8}$ \\
    \midrule
    Relu(BN(ConvTransposed2d($256$, $128$))) $\rightarrow{128\times16\times16}$ \\
    \midrule
    Tanh(ConvTransposed2d($128$, $3$)) $\rightarrow{ 3\times32\times32}$ \\
    \bottomrule
    \end{tabular}%
  \label{gen_CIFAR10_SVHN_CINIC10_tab:}%
\end{table}%

\begin{table}[htbp]
  \centering
  \caption{Generator for Tiny-ImageNet and FOOD101}
    \begin{tabular}{c}
    \toprule
    $\bm{z}\in \mathbf{R}^d \sim \mathcal{N}(\bm{0}, \bm{I})$, $y\in[C]$ \\
    \midrule
    $\bm{h}=o(\bm{z}, y)\rightarrow {d}$ \\
    \midrule
    Reshape($\bm{h}$)$\rightarrow{ d\times1\times1}$ \\
    \midrule
    Relu(ConvTransposed2d($d$, $512$)) $\rightarrow{B\times 512\times4\times4}$ \\
    \midrule
    Relu(BN(ConvTransposed2d($512$, $256$))) $\rightarrow{ 256\times8\times8}$ \\
    \midrule
    Relu(BN(ConvTransposed2d($256$, $128$))) $\rightarrow{128\times16\times16}$ \\
    \midrule
    Relu(BN(ConvTransposed2d($128$, $64$))) $\rightarrow{64\times32\times32}$ \\
    \midrule
    Tanh(ConvTransposed2d($64$, $3$)) $\rightarrow{ 3\times64\times64}$ \\
    \bottomrule
    \end{tabular}%
  \label{gen_tinyimage_food_tab:}%
\end{table}%

% % \section{B \quad Full Experimental Results}

\section{Additional Experimental Results}
\label{sup:Add_Ex_Re}
% % In this section, we provide the clients' local learning curves over $\omega \in [0.1, 0.5, 1.0]$  on FMNIST, CIFAR-10, SVHN and CINIC-10 datasets~($\rho=0$).

\begin{figure*}[h]\captionsetup[subfigure]{font=scriptsize}
  \centering
  \begin{subfigure}{0.3\linewidth}
    \centering
    \includegraphics[width=1.0\linewidth]{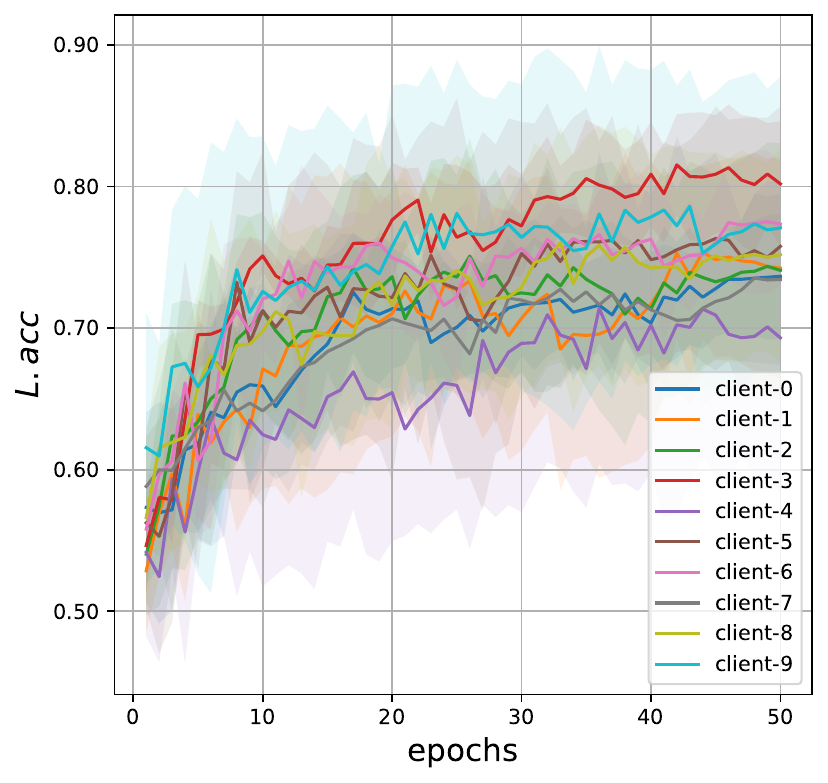}
        \caption{FMNIST, $\omega=1.0$}
  \end{subfigure}
  \centering
  \begin{subfigure}{0.3\linewidth}
    \centering
    \includegraphics[width=1.0\linewidth]{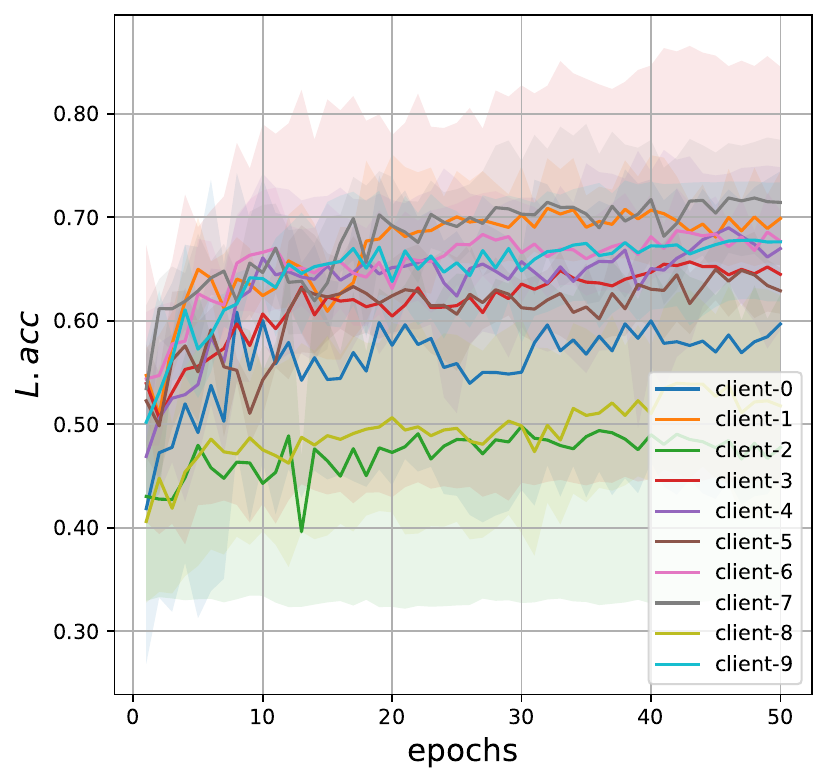}
        \caption{FMNIST, $\omega=0.5$}
  \end{subfigure}
  \centering
  \begin{subfigure}{0.3\linewidth}
    \centering
    \includegraphics[width=1.0\linewidth]{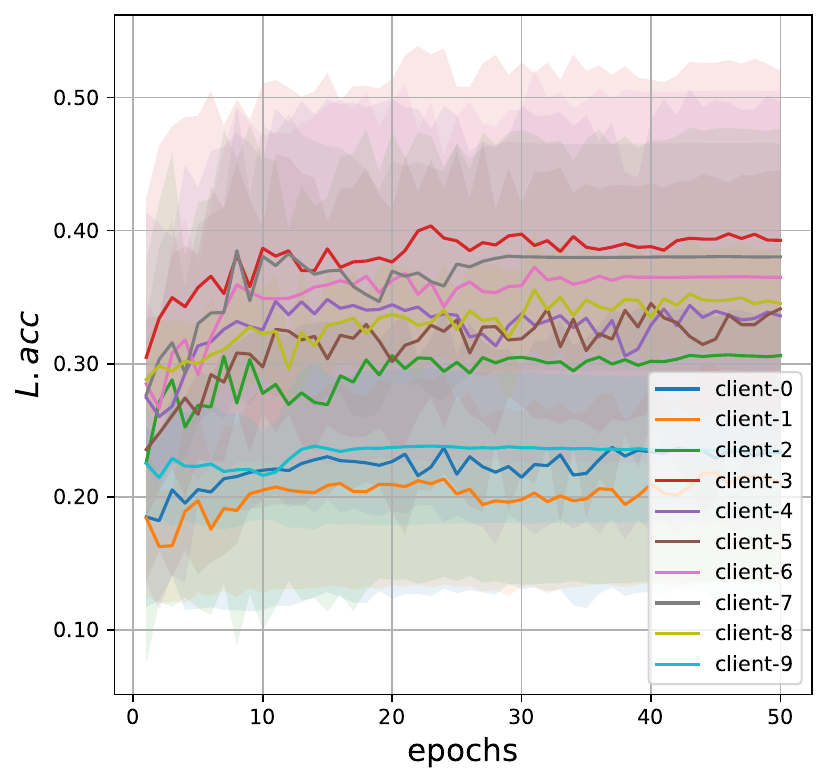}
        \caption{FMNIST, $\omega=0.1$}
  \end{subfigure}
  \centering
  \begin{subfigure}{0.3\linewidth}
    \centering
    \includegraphics[width=1.0\linewidth]{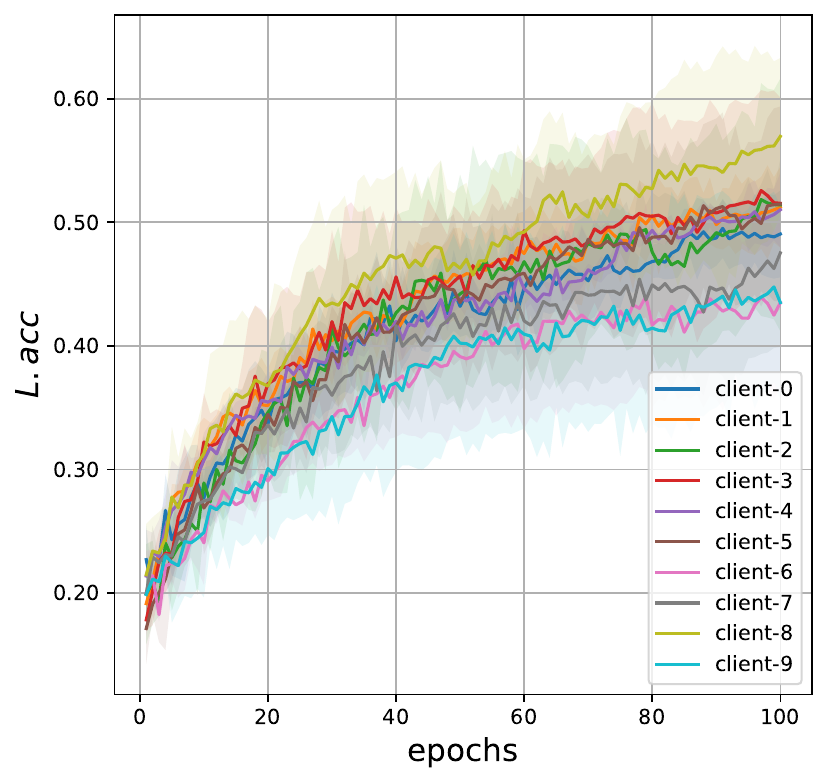}
        \caption{CIFAR-10, $\omega=1.0$}
  \end{subfigure}
  \centering
  \begin{subfigure}{0.3\linewidth}
    \centering
    \includegraphics[width=1.0\linewidth]{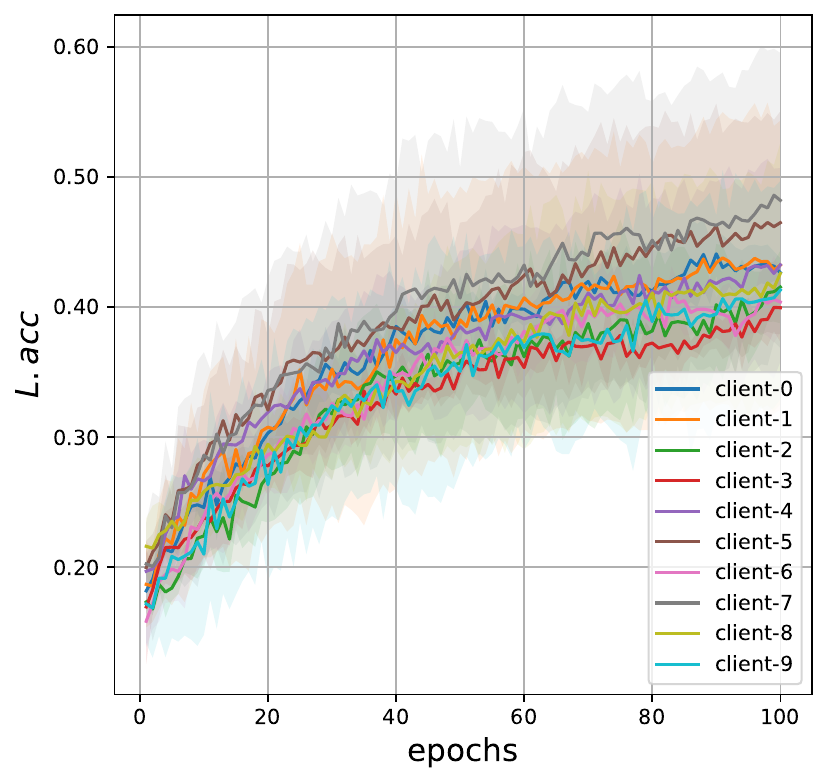}
        \caption{CIFAR-10, $\omega=0.5$}
  \end{subfigure}
  \centering
  \begin{subfigure}{0.3\linewidth}
    \centering
    \includegraphics[width=1.0\linewidth]{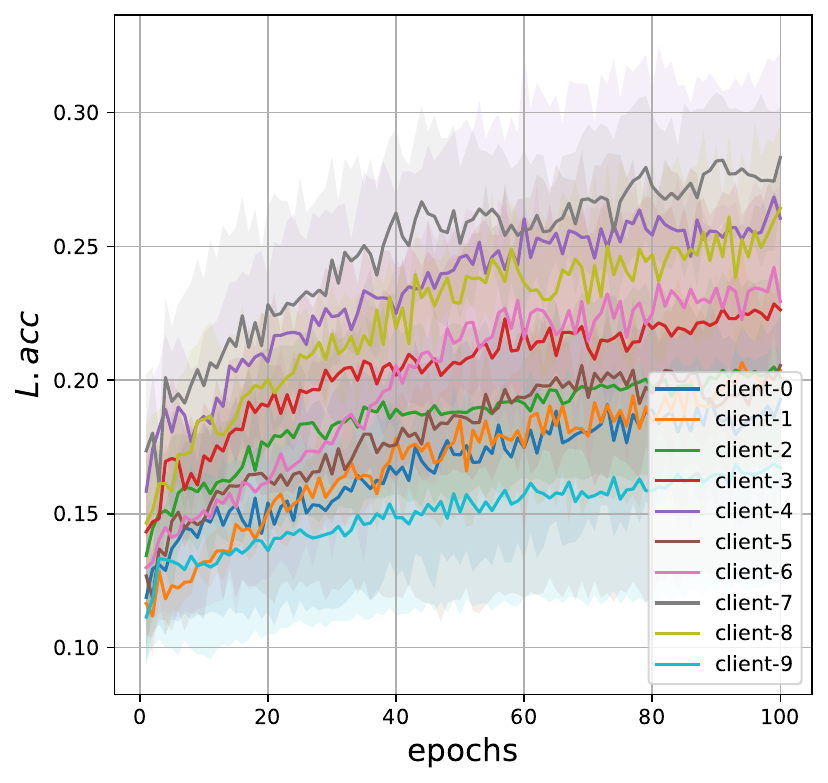}
        \caption{CIFAR-10, $\omega=0.1$}
  \end{subfigure}
  \centering
  \begin{subfigure}{0.3\linewidth}
    \centering
    \includegraphics[width=1.0\linewidth]{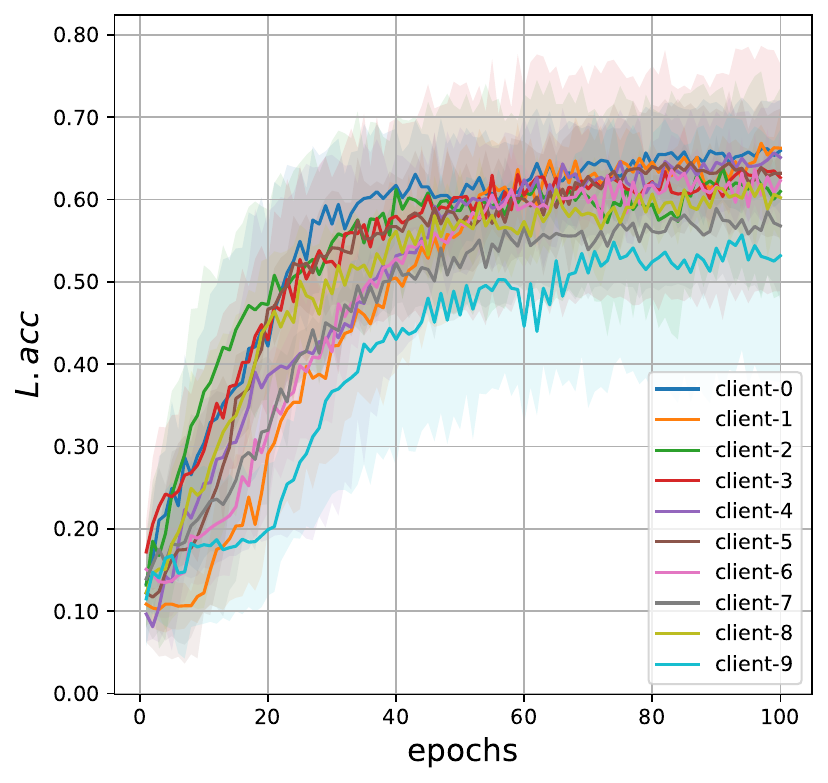}
        \caption{SVHN, $\omega=1.0$}
  \end{subfigure}
  \centering
  \begin{subfigure}{0.3\linewidth}
    \centering
    \includegraphics[width=1.0\linewidth]{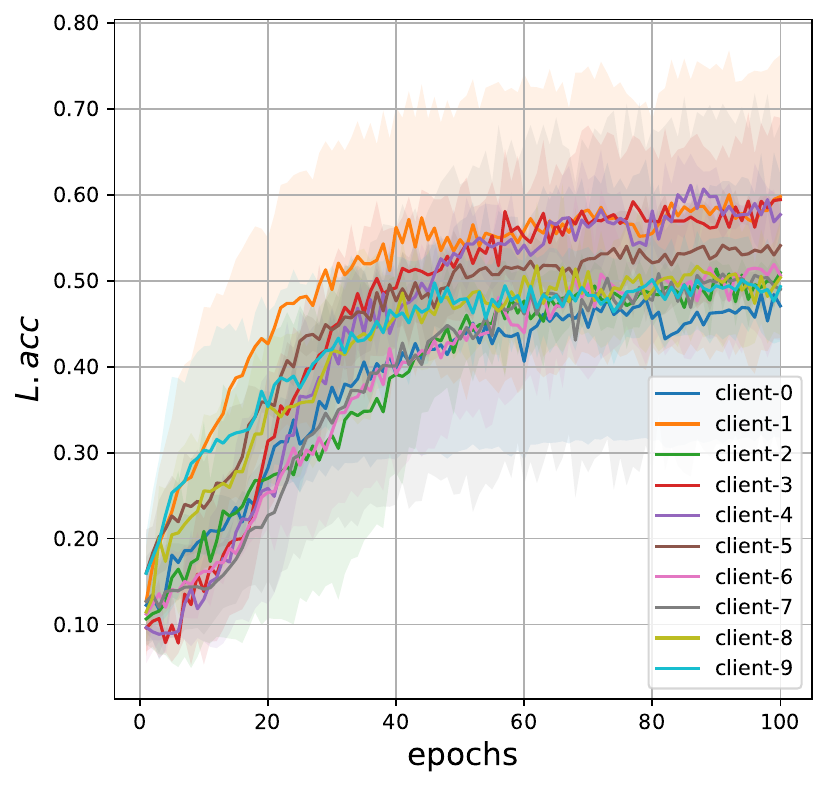}
        \caption{SVHN, $\omega=0.5$}
  \end{subfigure}
  \centering
  \begin{subfigure}{0.3\linewidth}
    \centering
    \includegraphics[width=1.0\linewidth]{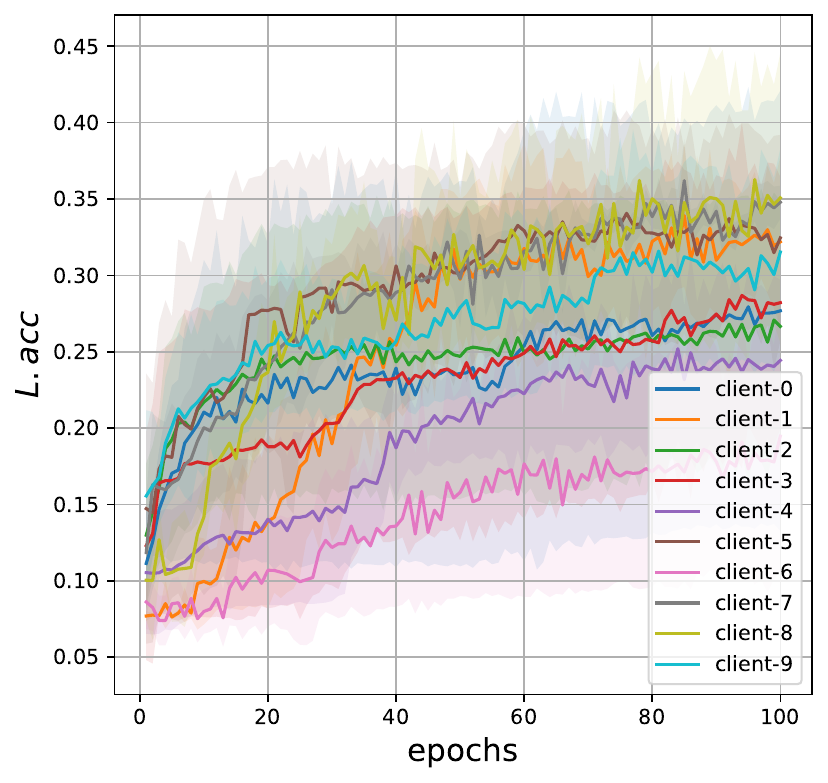}
        \caption{SVHN, $\omega=0.1$}
  \end{subfigure}
  \centering
  \begin{subfigure}{0.3\linewidth}
    \centering
    \includegraphics[width=1.0\linewidth]{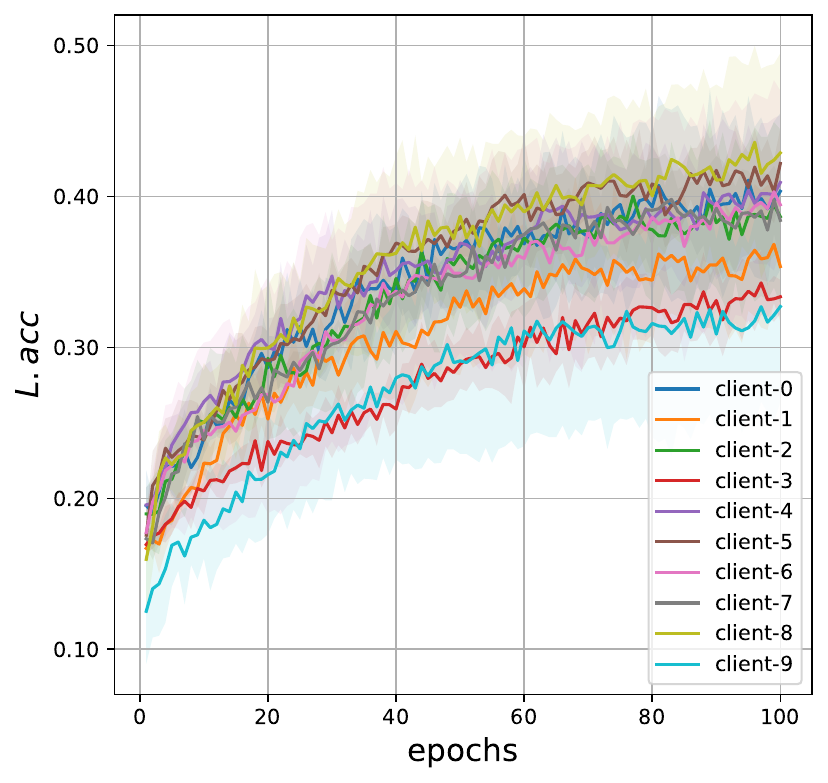}
        \caption{CINIC-10, $\omega=1.0$}
  \end{subfigure}
  \centering
  \begin{subfigure}{0.3\linewidth}
    \centering
    \includegraphics[width=1.0\linewidth]{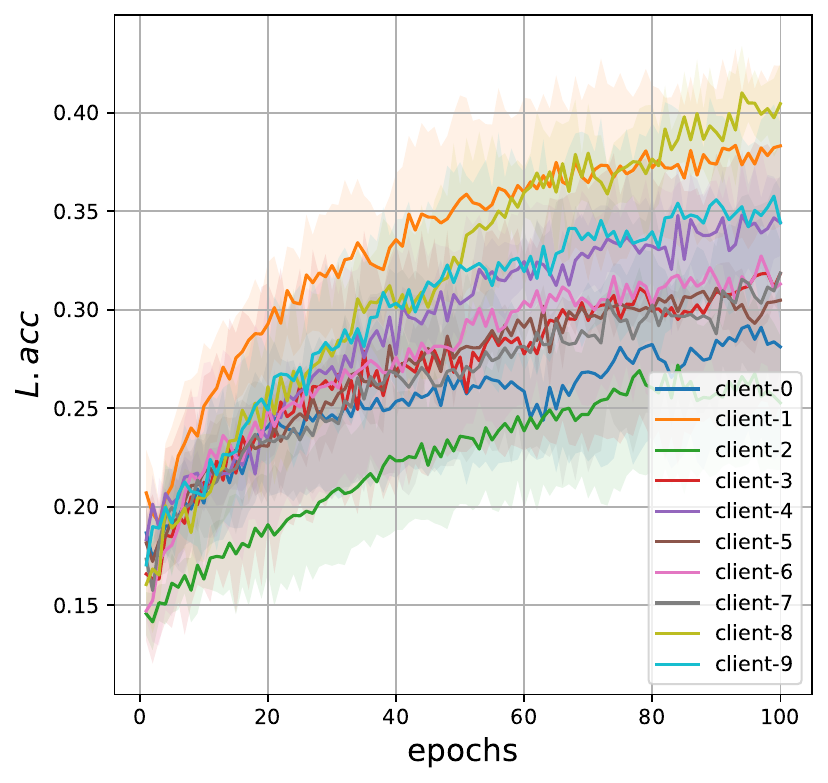}
        \caption{CINIC-10, $\omega=0.5$}
  \end{subfigure}
  \centering
  \begin{subfigure}{0.3\linewidth}
    \centering
    \includegraphics[width=1.0\linewidth]{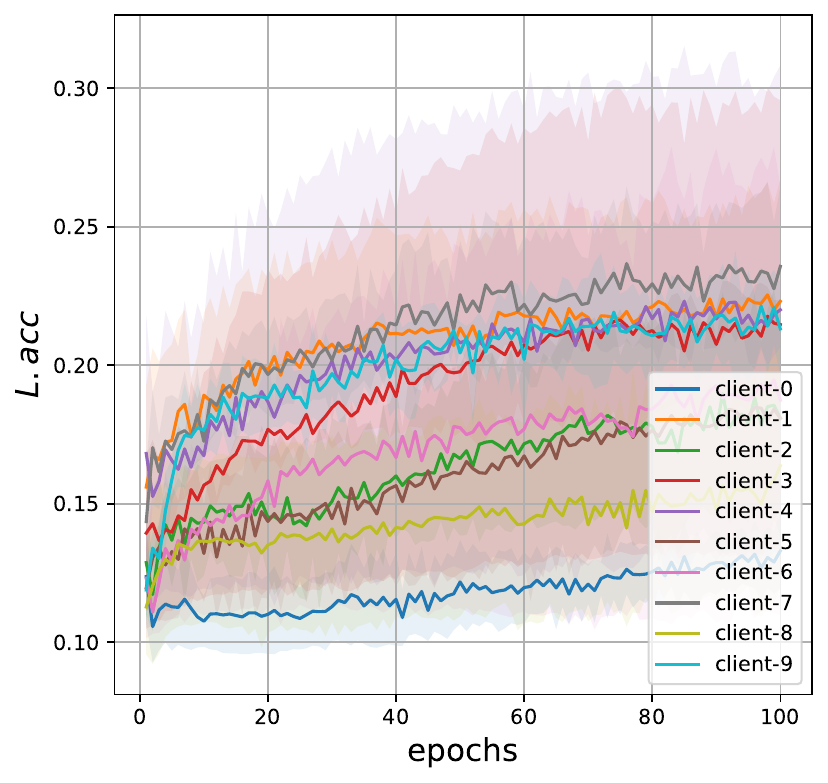}
        \caption{CINIC-10, $\omega=0.1$}
  \end{subfigure}
  \caption{Full local learning curves of clients across $\omega \in \{0.1, 0.5, 1.0\}$ on FMNIST, CIFAR-10, SVHN and CINIC-10 datasets~($\rho=0$), which are  averaged over $3$ random seeds.}
  \label{fig:data_full_local_res}
\end{figure*}

\begin{figure*}[h]\captionsetup[subfigure]{font=scriptsize}
  \centering
  \begin{subfigure}{0.3\linewidth}
    \centering
    \includegraphics[width=1.0\linewidth]{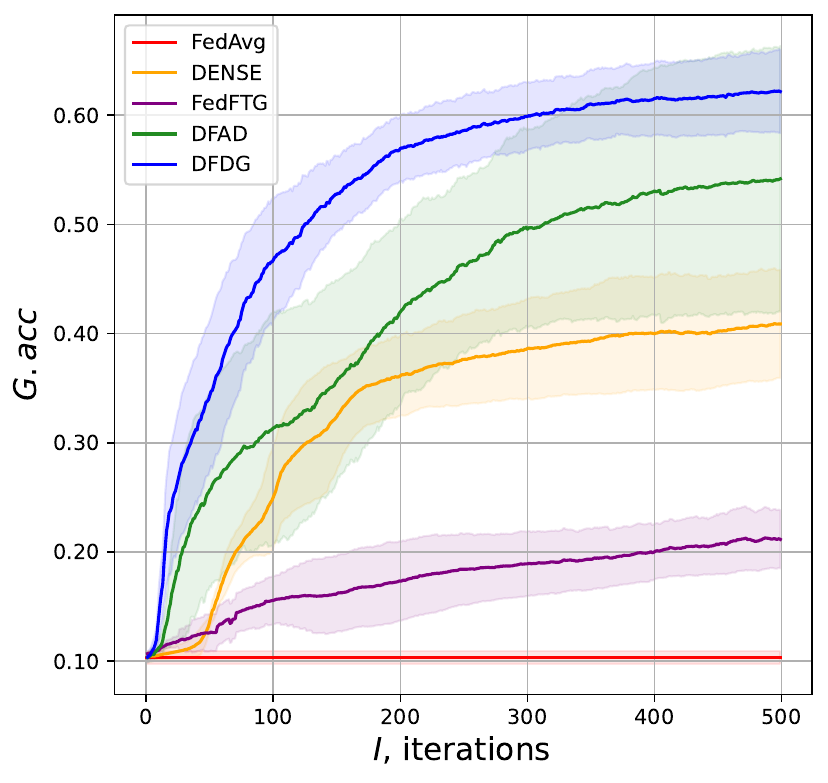}
        \caption{FMNIST, $\omega=1.0$}
  \end{subfigure}
  \centering
  \begin{subfigure}{0.3\linewidth}
    \centering
    \includegraphics[width=1.0\linewidth]{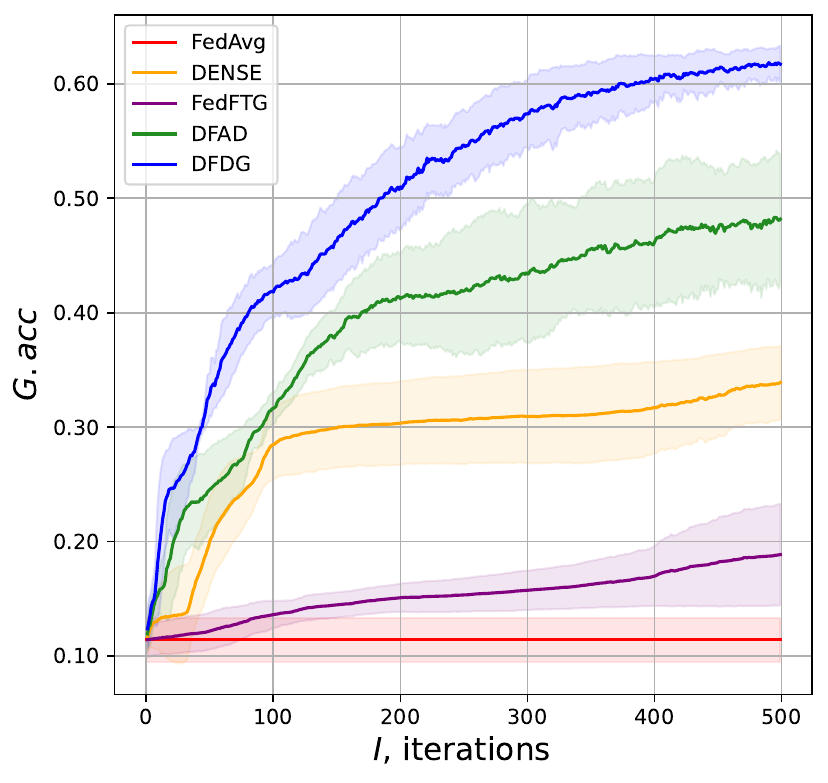}
        \caption{FMNIST, $\omega=0.5$}
  \end{subfigure}
  \centering
  \begin{subfigure}{0.3\linewidth}
    \centering
    \includegraphics[width=1.0\linewidth]{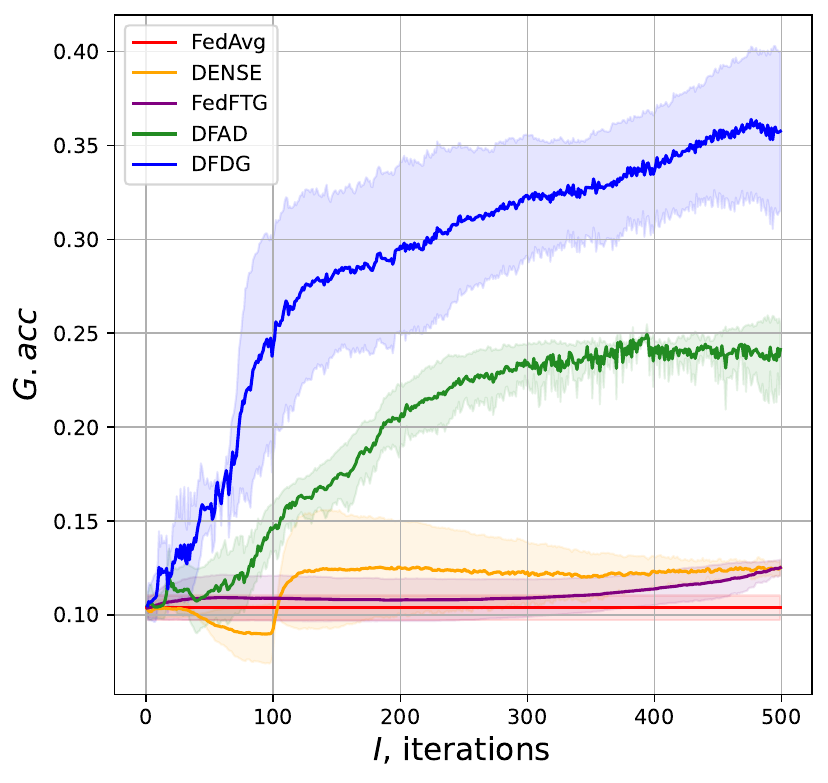}
        \caption{FMNIST, $\omega=0.1$}
  \end{subfigure}
  \centering
  \begin{subfigure}{0.3\linewidth}
    \centering
    \includegraphics[width=1.0\linewidth]{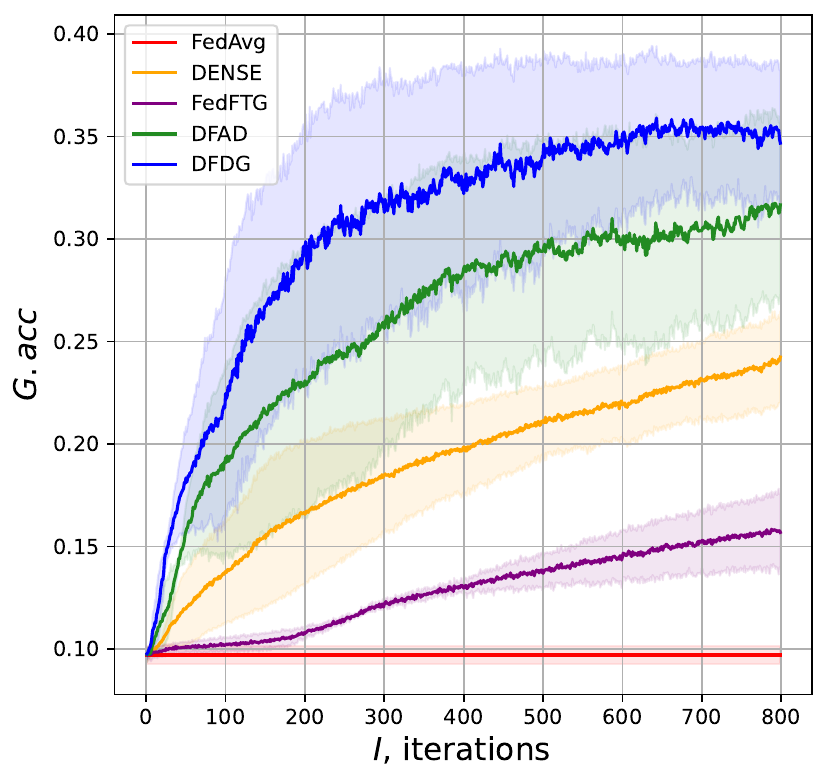}
        \caption{CIFAR-10, $\omega=1.0$}
  \end{subfigure}
  \centering
  \begin{subfigure}{0.3\linewidth}
    \centering
    \includegraphics[width=1.0\linewidth]{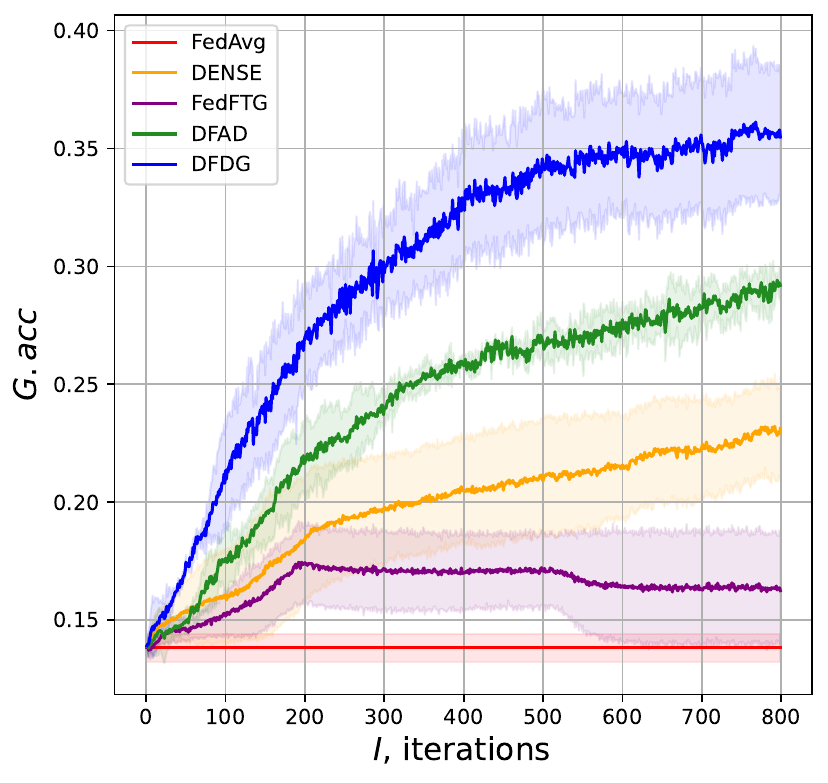}
        \caption{CIFAR-10, $\omega=0.5$}
  \end{subfigure}
  \centering
  \begin{subfigure}{0.3\linewidth}
    \centering
    \includegraphics[width=1.0\linewidth]{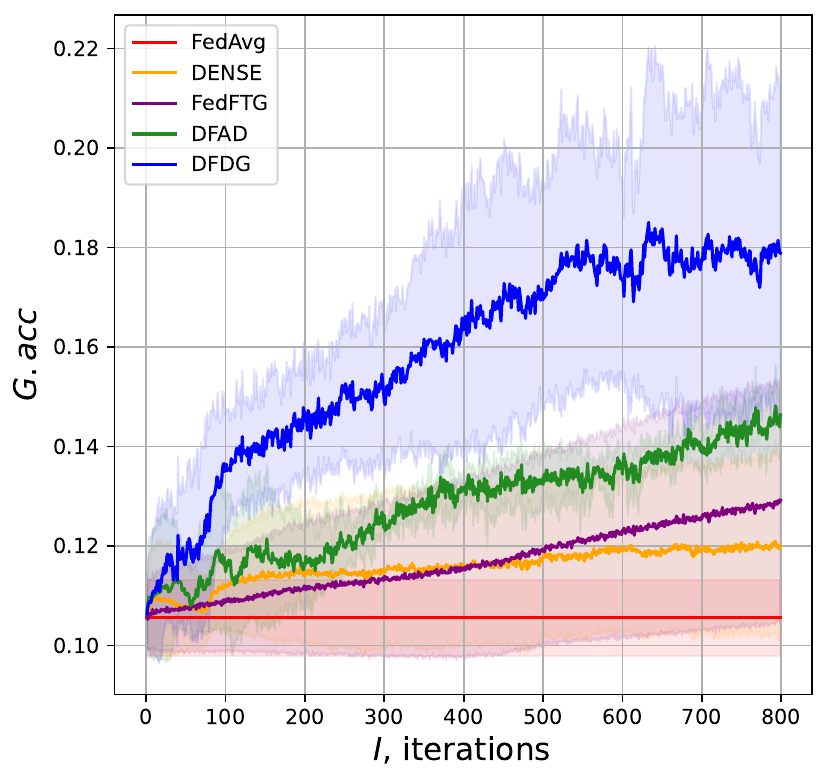}
        \caption{CIFAR-10, $\omega=0.1$}
  \end{subfigure}
  \centering
  \begin{subfigure}{0.3\linewidth}
    \centering
    \includegraphics[width=1.0\linewidth]{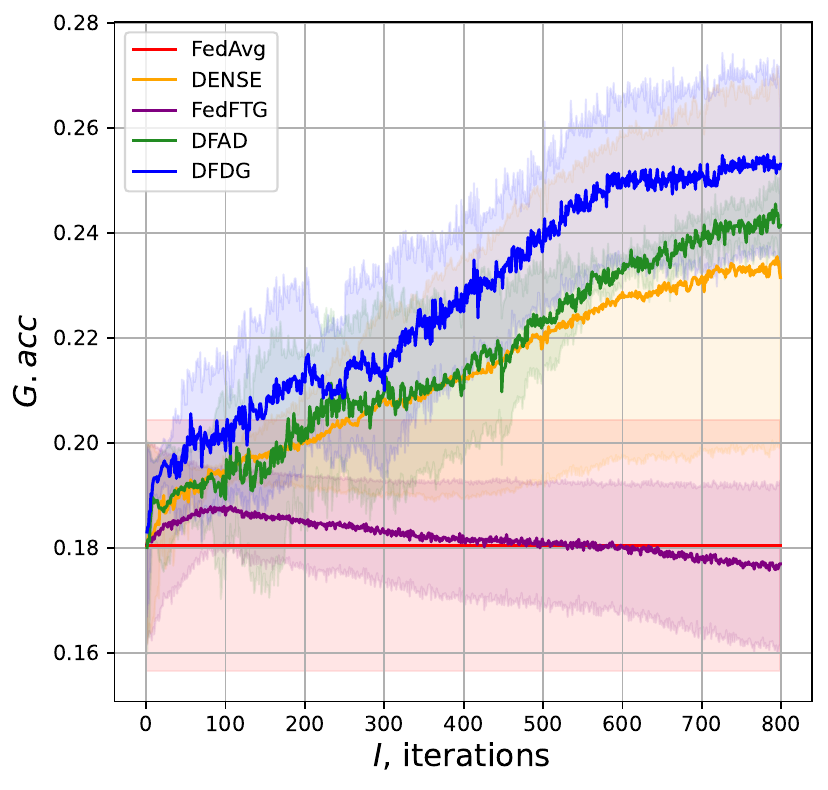}
        \caption{SVHN, $\omega=1.0$}
  \end{subfigure}
  \centering
  \begin{subfigure}{0.3\linewidth}
    \centering
    \includegraphics[width=1.0\linewidth]{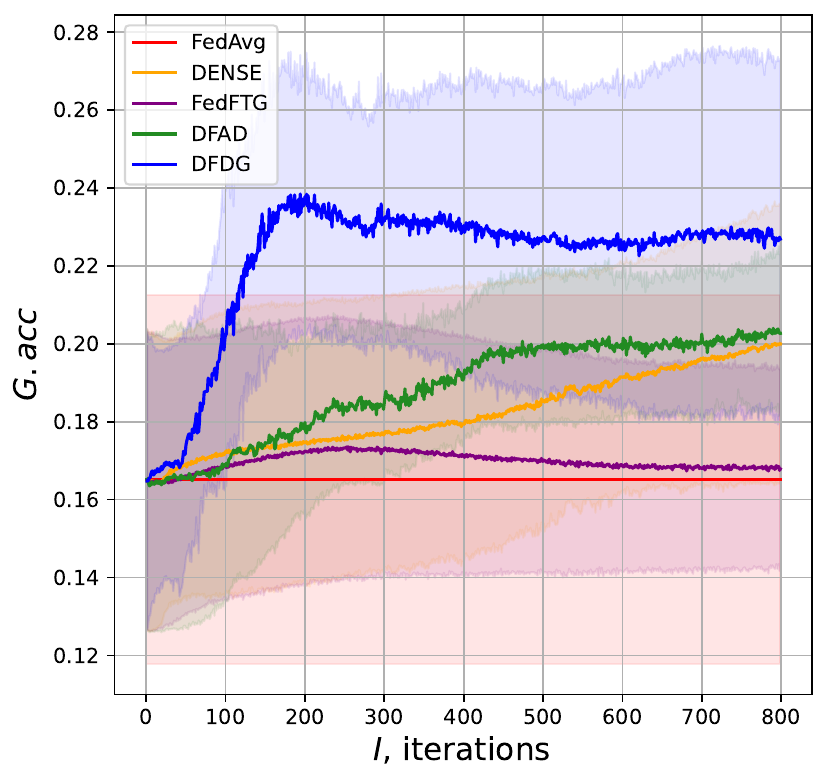}
        \caption{SVHN, $\omega=0.5$}
  \end{subfigure}
  \centering
  \begin{subfigure}{0.3\linewidth}
    \centering
    \includegraphics[width=1.0\linewidth]{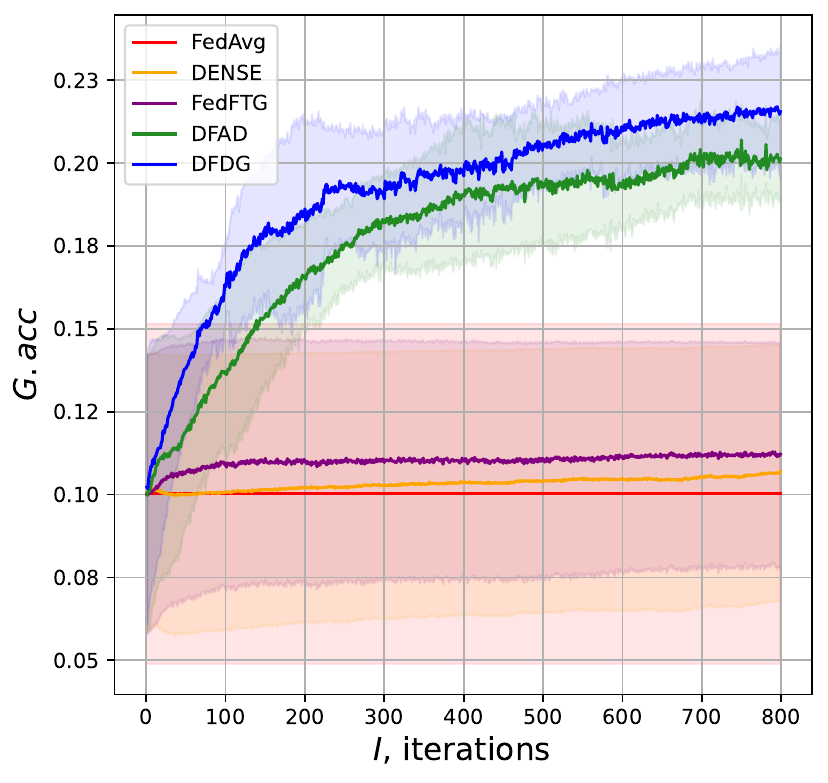}
        \caption{SVHN, $\omega=0.1$}
  \end{subfigure}
  \centering
  \begin{subfigure}{0.3\linewidth}
    \centering
    \includegraphics[width=1.0\linewidth]{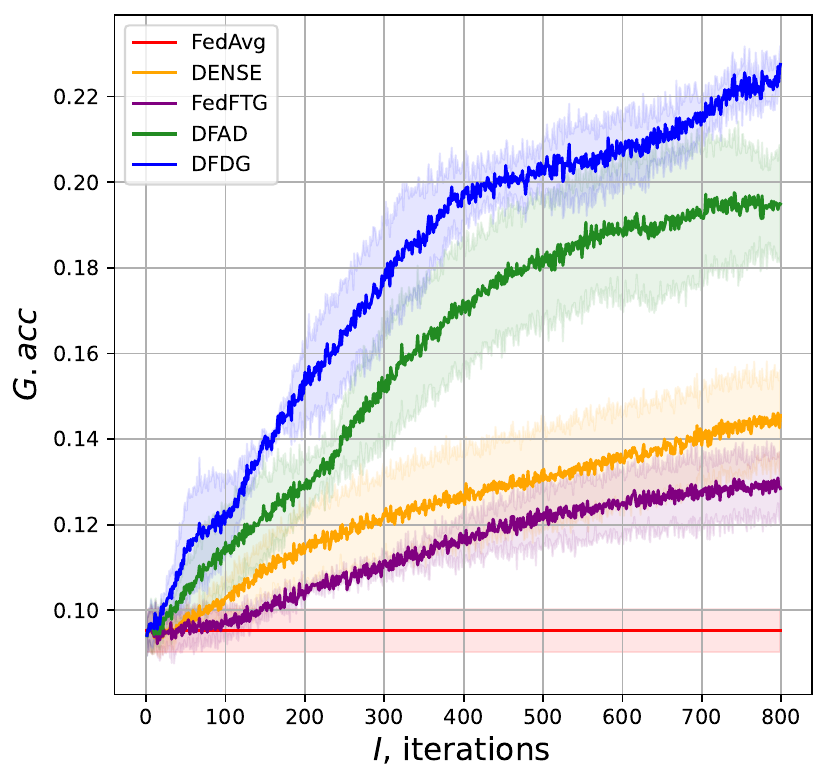}
        \caption{CINIC-10, $\omega=1.0$}
  \end{subfigure}
  \centering
  \begin{subfigure}{0.3\linewidth}
    \centering
    \includegraphics[width=1.0\linewidth]{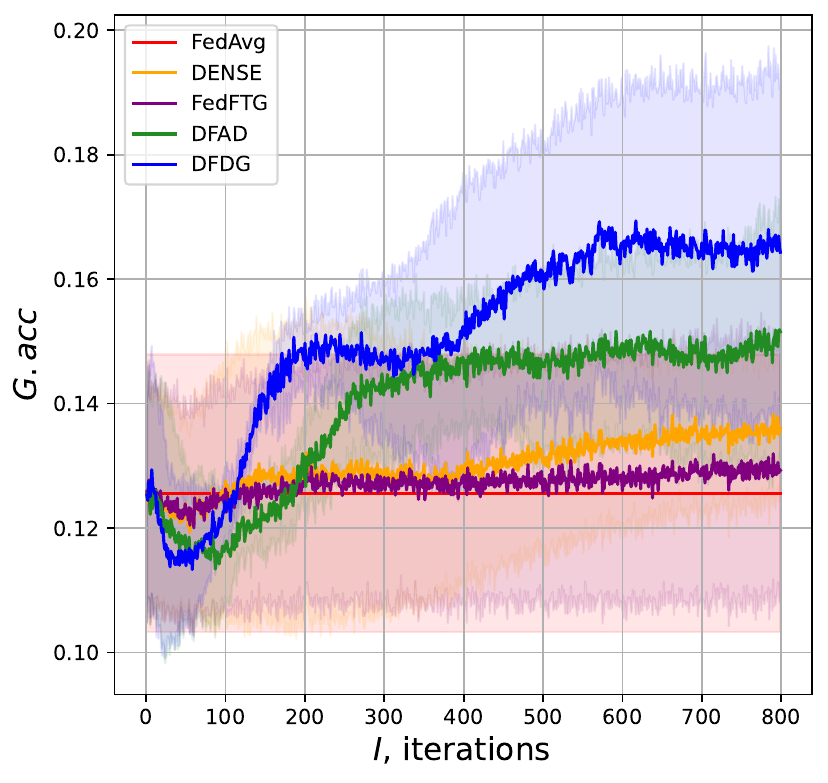}
        \caption{CINIC-10, $\omega=0.5$}
  \end{subfigure}
  \centering
  \begin{subfigure}{0.3\linewidth}
    \centering
    \includegraphics[width=1.0\linewidth]{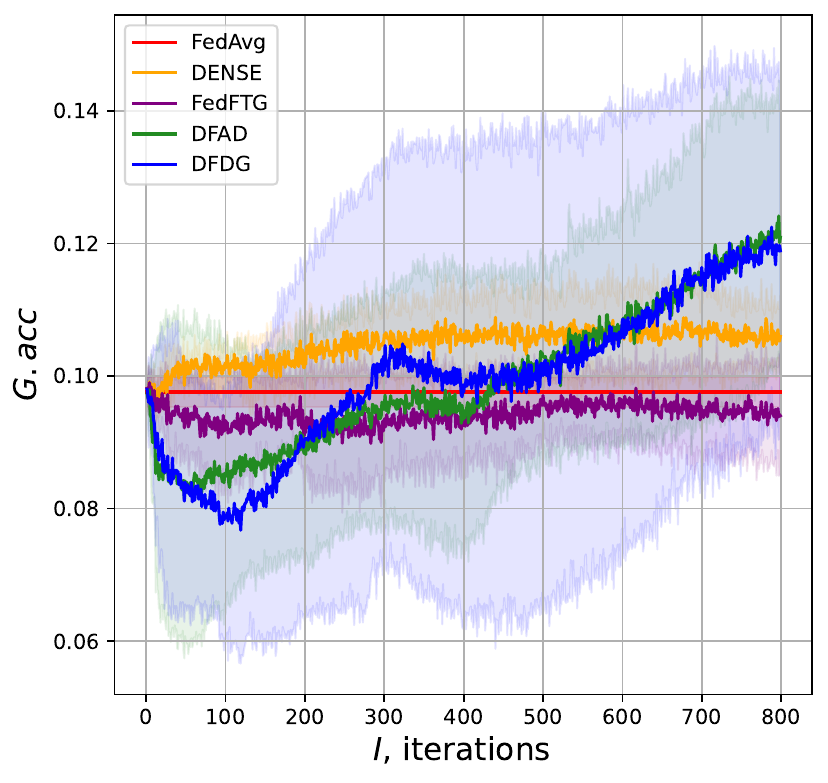}
        \caption{CINIC-10, $\omega=0.1$}
  \end{subfigure}
  \caption{Full learning curves of distinct methods across $\omega \in \{0.1, 0.5, 1.0\}$ on FMNIST, CIFAR-10, SVHN and CINIC-10 datasets~($\rho=0$), which are  averaged over $3$ random seeds.}
  \label{fig:data_full_res}
\end{figure*}

\begin{figure*}[h]\captionsetup[subfigure]{font=scriptsize}
  \centering
  \begin{subfigure}{0.3\linewidth}
    \centering
    \includegraphics[width=1.0\linewidth]{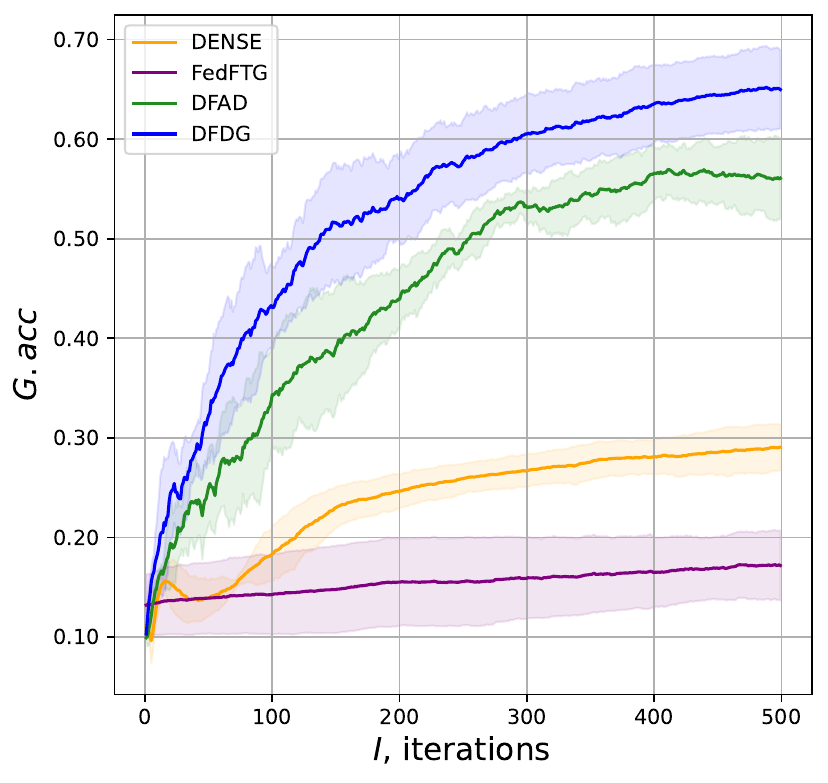}
        \caption{FMNIST, $\rho=2$}
  \end{subfigure}
  \centering
  \begin{subfigure}{0.3\linewidth}
    \centering
    \includegraphics[width=1.0\linewidth]{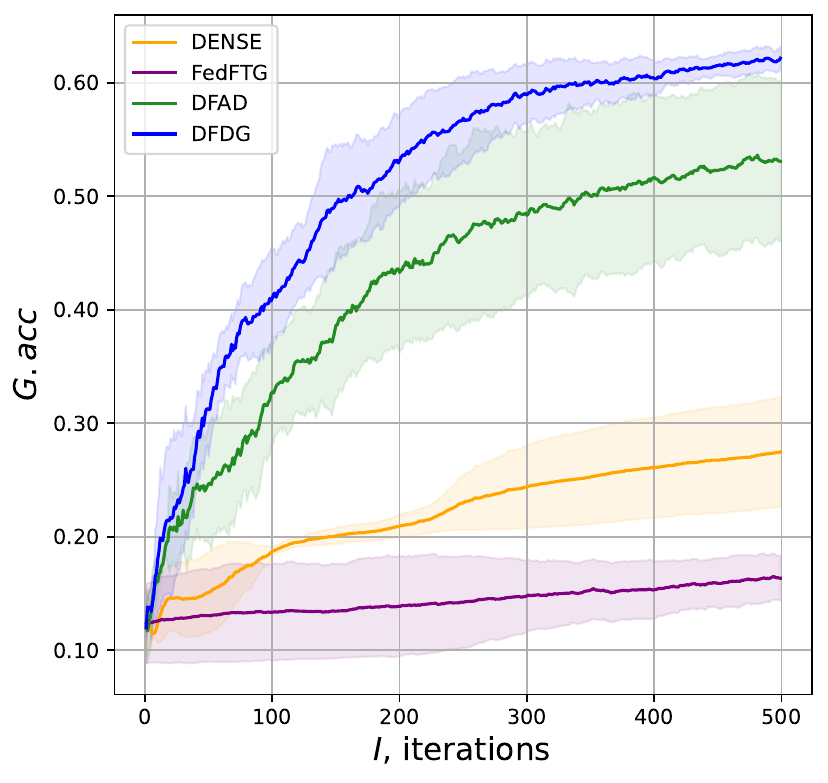}
        \caption{FMNIST, $\rho=3$}
  \end{subfigure}
  \centering
  \begin{subfigure}{0.3\linewidth}
    \centering
    \includegraphics[width=1.0\linewidth]{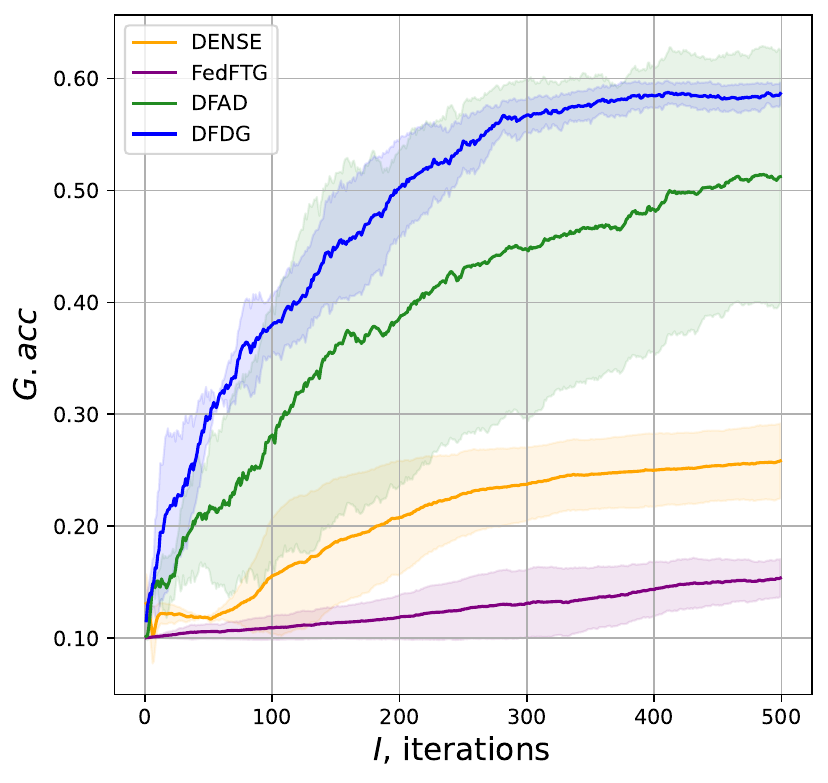}
        \caption{FMNIST, $\rho=4$}
  \end{subfigure}
  \centering
  \begin{subfigure}{0.3\linewidth}
    \centering
    \includegraphics[width=1.0\linewidth]{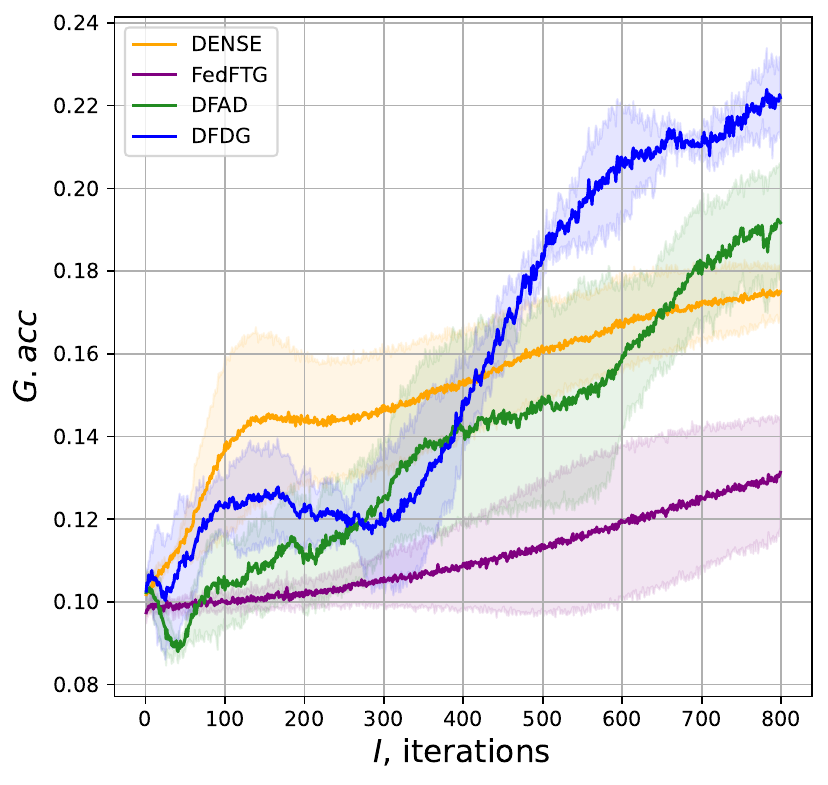}
        \caption{CIFAR-10, $\rho=2$}
  \end{subfigure}
  \centering
  \begin{subfigure}{0.3\linewidth}
    \centering
    \includegraphics[width=1.0\linewidth]{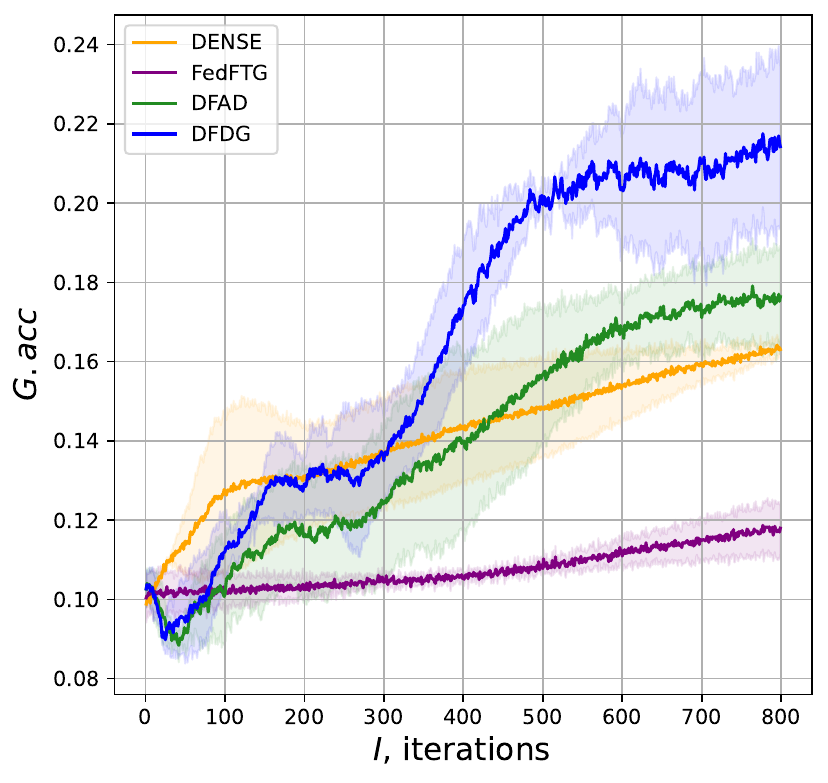}
        \caption{CIFAR-10, $\rho=3$}
  \end{subfigure}
  \centering
  \begin{subfigure}{0.3\linewidth}
    \centering
    \includegraphics[width=1.0\linewidth]{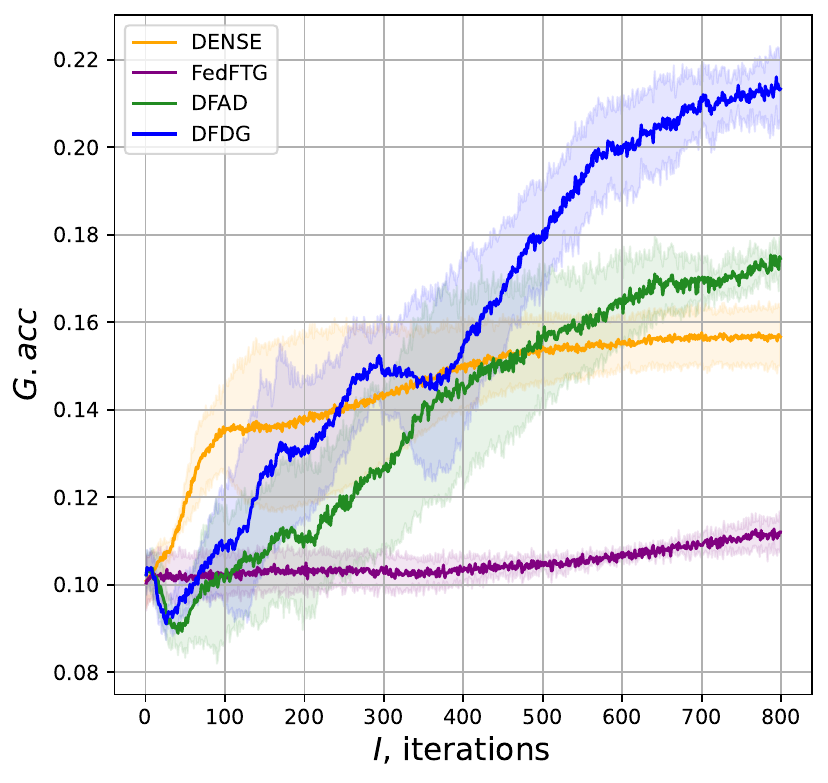}
        \caption{CIFAR-10, $\rho=4$}
  \end{subfigure}
  \centering
  \begin{subfigure}{0.3\linewidth}
    \centering
    \includegraphics[width=1.0\linewidth]{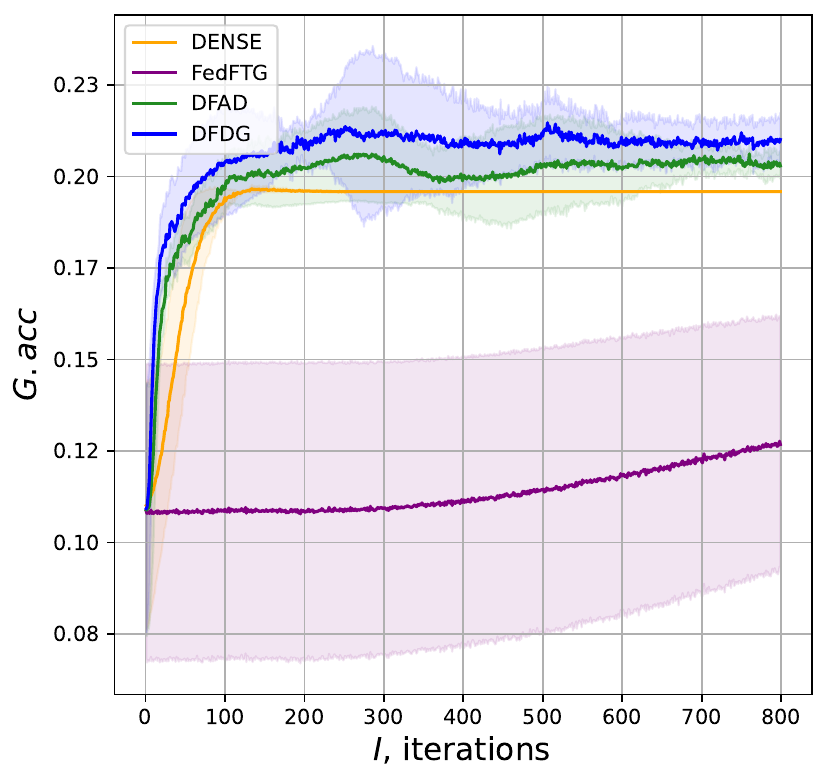}
        \caption{SVHN, $\rho=2$}
  \end{subfigure}
  \centering
  \begin{subfigure}{0.3\linewidth}
    \centering
    \includegraphics[width=1.0\linewidth]{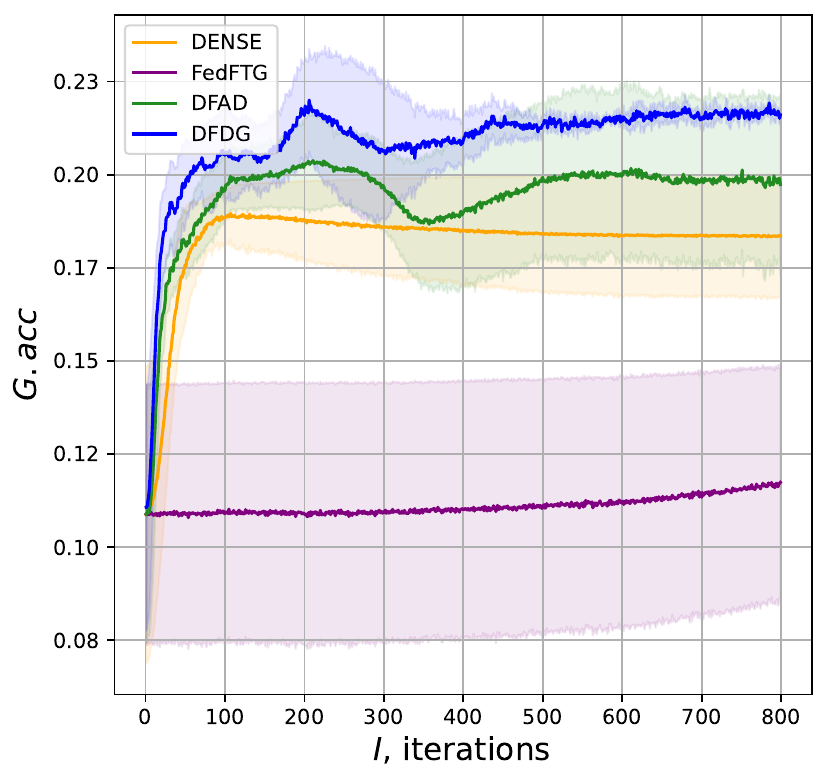}
        \caption{SVHN, $\rho=3$}
  \end{subfigure}
  \centering
  \begin{subfigure}{0.3\linewidth}
    \centering
    \includegraphics[width=1.0\linewidth]{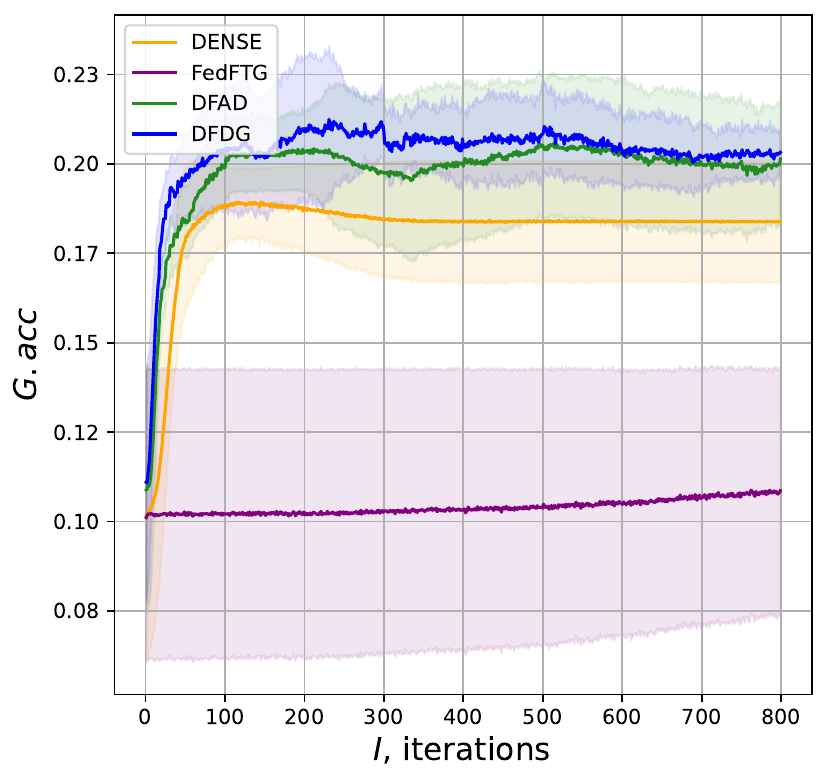}
        \caption{SVHN, $\rho=4$}
  \end{subfigure}
  \centering
  \begin{subfigure}{0.3\linewidth}
    \centering
    \includegraphics[width=1.0\linewidth]{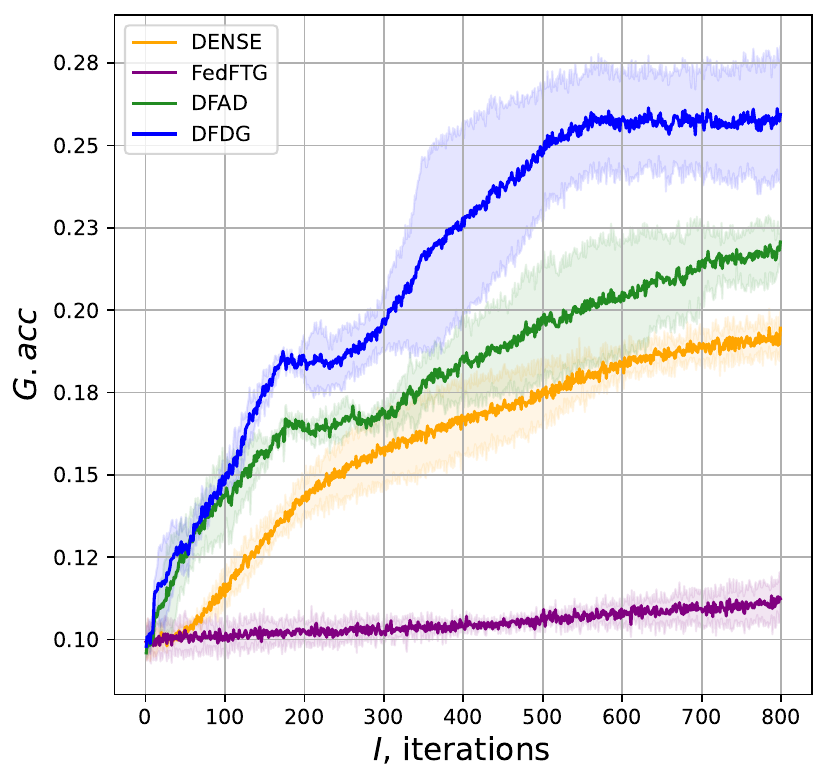}
        \caption{CINIC-10, $\rho=2$}
  \end{subfigure}
  \centering
  \begin{subfigure}{0.3\linewidth}
    \centering
    \includegraphics[width=1.0\linewidth]{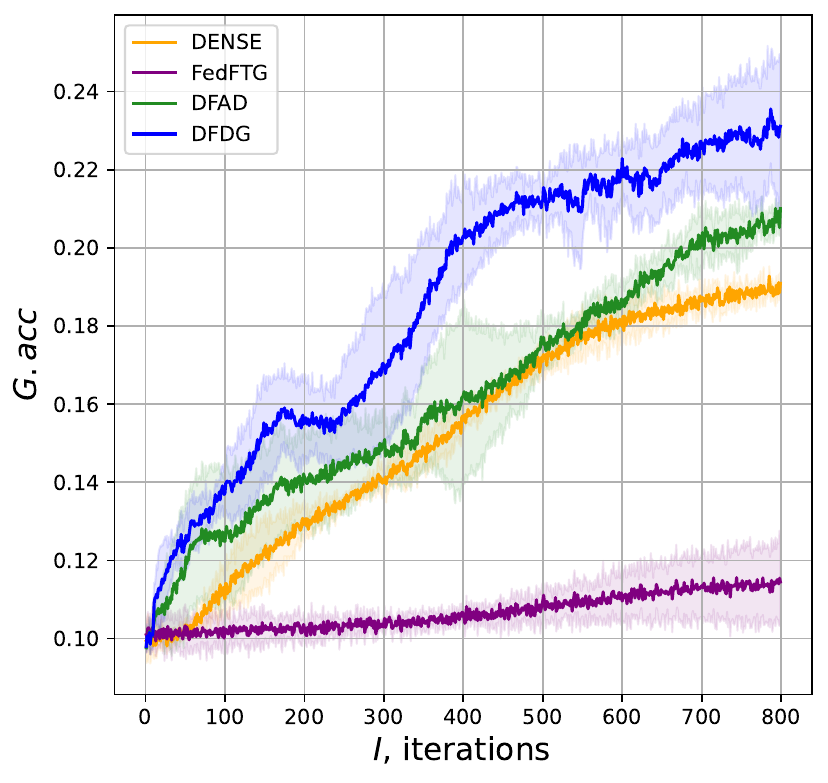}
        \caption{CINIC-10, $\omega=0.5$}
  \end{subfigure}
  \centering
  \begin{subfigure}{0.3\linewidth}
    \centering
    \includegraphics[width=1.0\linewidth]{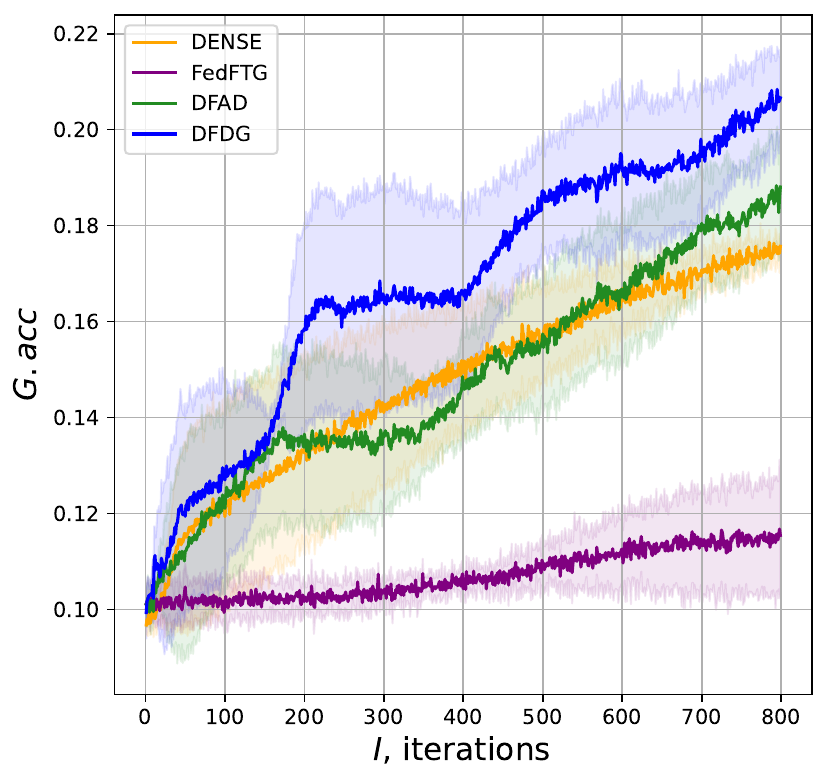}
        \caption{CINIC-10, $\omega=0.1$}
  \end{subfigure}
  \caption{Full learning curves of distinct methods across $\rho \in \{2, 3, 4\}$ on FMNIST, CIFAR-10, SVHN and CINIC-10 datasets~($\omega=0.5$), which are  averaged over $3$ random seeds.}
  \label{fig:model_full_res}
\end{figure*}

\begin{table}[htbp]
  \centering
  \caption{Top $G. acc$~(\%) of different methods over $N\in\{10, 50, 100\}$ on CIFAR-100, Tiny-ImageNet and FOOD101 datasets with $(\omega, \rho)=(0.5, 0)$.}
  \resizebox{1.0\columnwidth}{!}{
    \begin{tabular}{cccc|ccc|ccc}
    \toprule
    \multicolumn{1}{c}{\multirow{2}[4]{*}{Alg.s}} & \multicolumn{3}{c|}{CIFAR-100} & \multicolumn{3}{c|}{Tiny-ImageNet} & \multicolumn{3}{c}{FOOD101} \\
\cmidrule{2-10}          & $N=10$  & $N=50$  & $N=100$ & $N=10$  & $N=50$  & $N=100$ & $N=10$  & $N=50$  & $N=100$ \\
    \midrule
    FedAvg & 8.62$\pm$0.89 & 8.33$\pm$0.11 & 7.98$\pm$0.65 & 3.58$\pm$0.26 & 4.22$\pm$0.16 & 3.87$\pm$0.67 & 4.35$\pm$0.54 & 3.86$\pm$0.25 & 3.73$\pm$0.65 \\
    \midrule
    DENSE & 12.55$\pm$1.35 & 13.65$\pm$1.46 & 14.66$\pm$0.77 & 9.62$\pm$2.11 & 9.88$\pm$1.55 & 10.06$\pm$1.01 & 6.14$\pm$0.24 & 8.01$\pm$0.31 & 8.97$\pm$0.28 \\
    FedFTG & 10.92$\pm$0.74 & 9.55$\pm$0.26 & 9.01$\pm$0.38 & 9.06$\pm$1.88 & 8.87$\pm$1.53 & 8.24$\pm$0.78 & 7.16$\pm$0.36 & 7.04$\pm$0.21 & 6.33$\pm$0.47 \\
    DFAD  & 14.24$\pm$1.61 & 15.24$\pm$1.32 & 16.39$\pm$0.88 & 11.52$\pm$1.36 & 12.26$\pm$1.58 & 13.29$\pm$1.34 & 8.68$\pm$0.55 & 9.86$\pm$0.53 & 10.57$\pm$0.26 \\
    DFDG  & \textbf{17.05$\pm$1.62} & \textbf{18.33$\pm$0.98} & \textbf{18.98$\pm$0.66} & \textbf{13.97$\pm$1.75} & \textbf{14.79$\pm$1.43} & \textbf{15.99$\pm$1.32} & \textbf{10.98$\pm$0.67} & \textbf{11.59$\pm$0.29} & \textbf{12.88$\pm$0.36} \\
    \bottomrule
    \end{tabular} }%
  \label{full_di_im_task:}%
\end{table}%

\begin{figure*}[h]\captionsetup[subfigure]{font=scriptsize}
  \centering
  \begin{subfigure}{0.3\linewidth}
    \centering
    \includegraphics[width=1.0\linewidth]{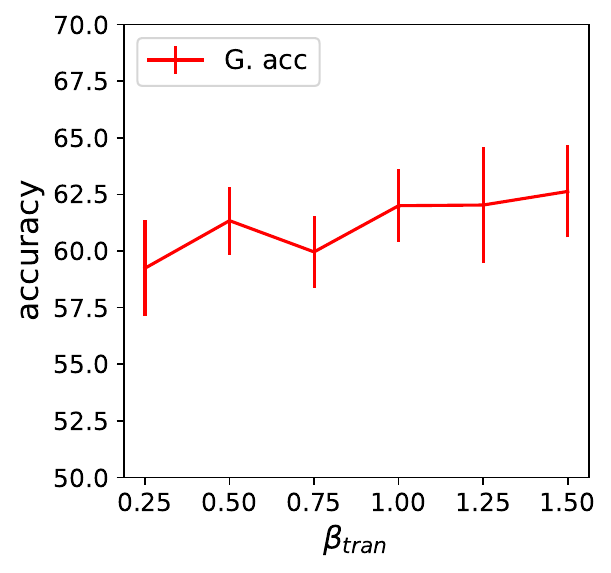}
        \caption{FMNIST}
  \end{subfigure}
  \centering
  \begin{subfigure}{0.3\linewidth}
    \centering
    \includegraphics[width=1.0\linewidth]{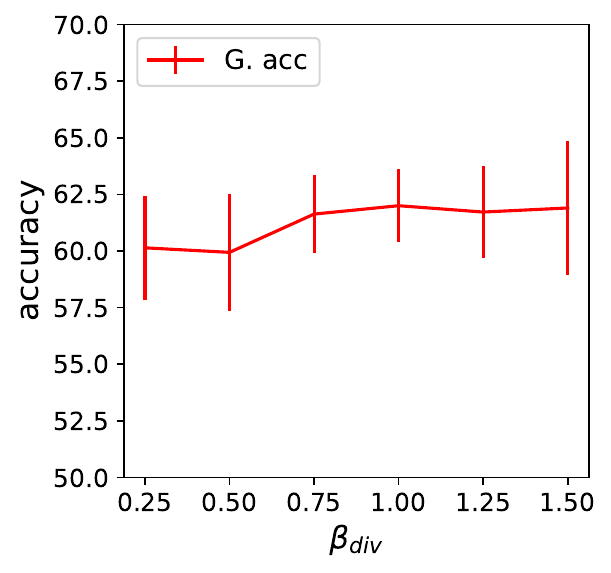}
        \caption{FMNIST}
  \end{subfigure}
  \centering
  \begin{subfigure}{0.3\linewidth}
    \centering
    \includegraphics[width=1.0\linewidth]{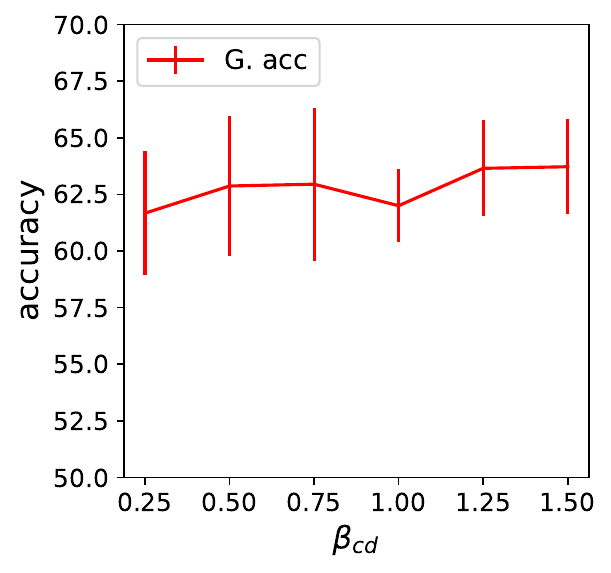}
        \caption{FMNIST}
  \end{subfigure}
  \centering
  \begin{subfigure}{0.3\linewidth}
    \centering
    \includegraphics[width=1.0\linewidth]{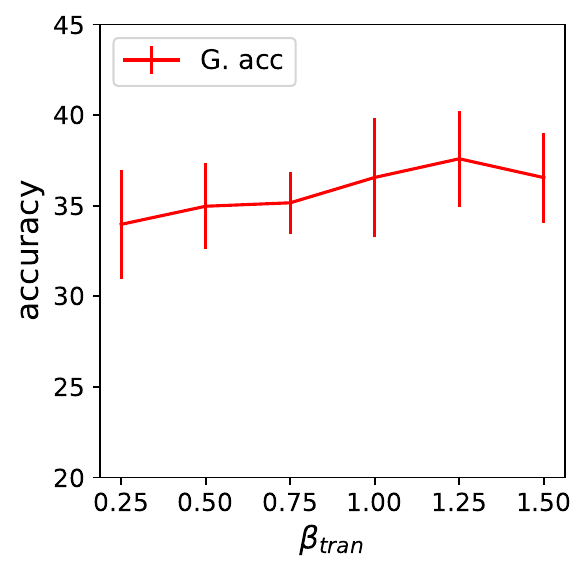}
        \caption{CIFAR-10}
  \end{subfigure}
  \centering
  \begin{subfigure}{0.3\linewidth}
    \centering
    \includegraphics[width=1.0\linewidth]{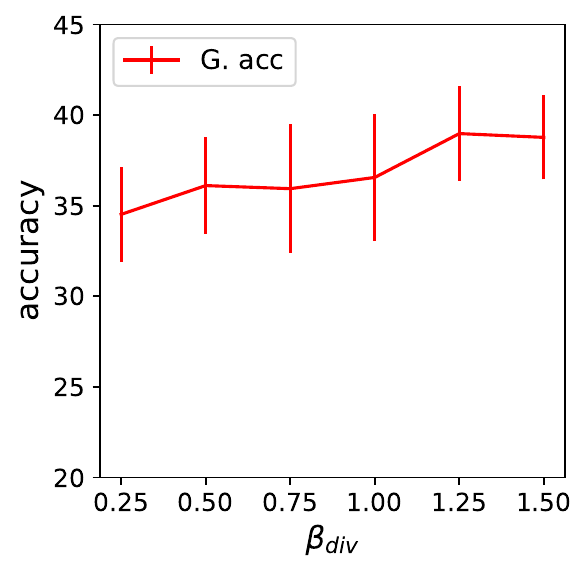}
        \caption{CIFAR-10}
  \end{subfigure}
  \centering
  \begin{subfigure}{0.3\linewidth}
    \centering
    \includegraphics[width=1.0\linewidth]{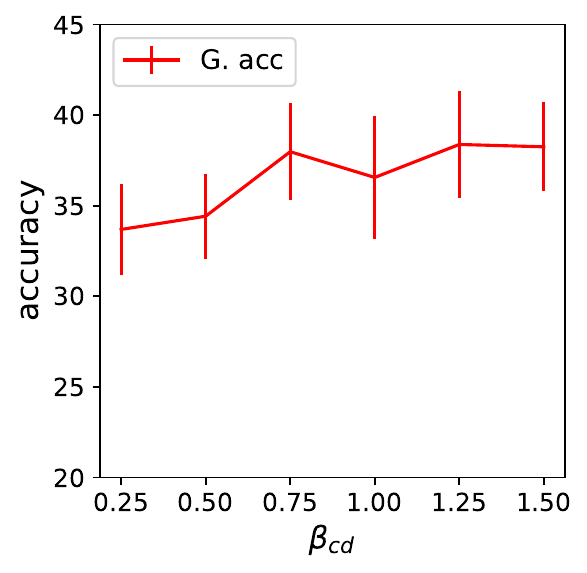}
        \caption{CIFAR-10}
  \end{subfigure}
  \caption{Test accuracy~(\%) with varying $\beta_{tran}$, $\beta_{div}$ and $\beta_{cd}$ over FMNIST and CIFAR-10.}
  \label{fig:model_full_res}
\end{figure*}

\begin{figure*}[h]\captionsetup[subfigure]{font=scriptsize}
  \centering
  \begin{subfigure}{0.4\linewidth}
    \centering
    \includegraphics[width=1.0\linewidth]{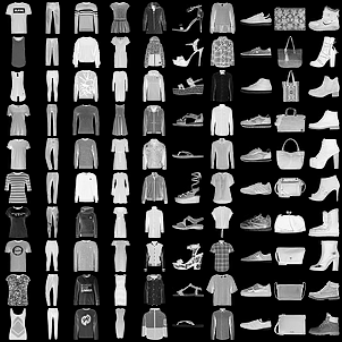}
        \caption{FMNIST}
  \end{subfigure}
  \centering
  \begin{subfigure}{0.4\linewidth}
    \centering
    \includegraphics[width=1.0\linewidth]{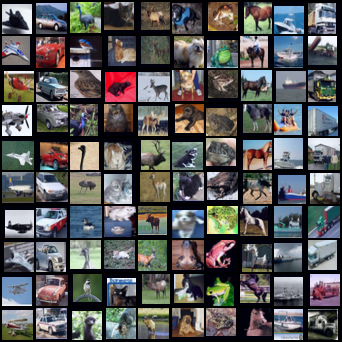}
        \caption{CIFAR-10}
  \end{subfigure}
  \centering
  \begin{subfigure}{0.4\linewidth}
    \centering
    \includegraphics[width=1.0\linewidth]{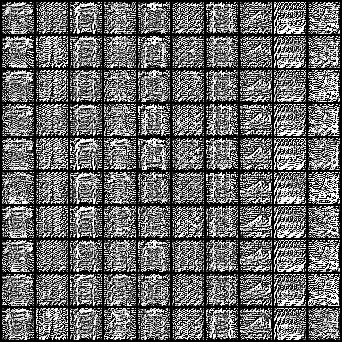}
        \caption{FMNIST, generator $G_1$}
  \end{subfigure}
  \centering
  \begin{subfigure}{0.4\linewidth}
    \centering
    \includegraphics[width=1.0\linewidth]{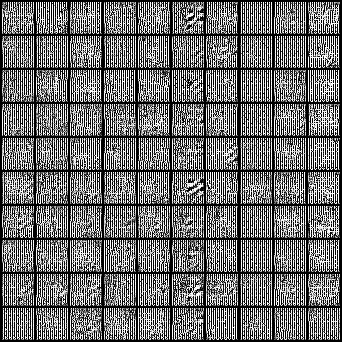}
        \caption{FMNIST, generator $G_2$}
  \end{subfigure}
  \centering
  \begin{subfigure}{0.4\linewidth}
    \centering
    \includegraphics[width=1.0\linewidth]{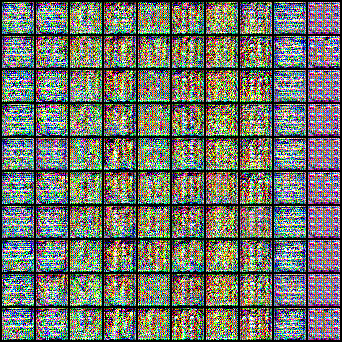}
        \caption{CIFAR-10, generator $G_1$}
  \end{subfigure}
  \centering
  \begin{subfigure}{0.4\linewidth}
    \centering
    \includegraphics[width=1.0\linewidth]{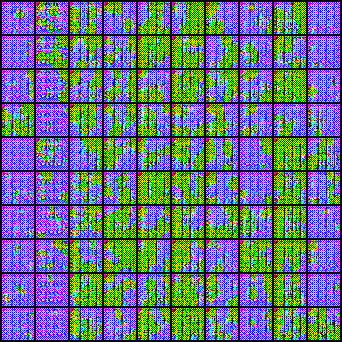}
        \caption{CIFAR-10, generator $G_2$}
  \end{subfigure}
  \caption{(a)-(b): Original data selected from FMNIST and CIFAR-10 datasets. We select $10$ images from each class in each dataset as a column. (c)-(f): Visualization of synthetic data on FMNIST and CIFAR-10 by using DFDG with diversity constraint based on \textit{mul}.}
  \label{sup:raw_data_data_gen_mul}
\end{figure*}

\end{document}